\documentclass[final]{jpp}
\usepackage{graphicx}
\usepackage[T1]{fontenc}
\usepackage{amsmath}
\usepackage{bm}
\usepackage{natbib}

\usepackage[colorlinks=true,
  linkcolor=blue, citecolor=blue, urlcolor=blue]{hyperref}   

\usepackage[capitalize]{cleveref}
\usepackage{threeparttable} 

\renewcommand{\vec}[1]{\bm{#1}}
\newcommand{\E}{\vec{E}}
\newcommand{\B}{\vec{B}}
\newcommand{\hB}{\hat{\B}}

\newcommand{\x}{\bm{x}}
\newcommand{\vi}{\bm{v}}

\newcommand{\z}{\bm{z}}
\newcommand{\A}{\bm{A}}
\newcommand{\e}{\bm{e}}
\newcommand{\R}{\bm{R}}
\newcommand{\U}{\bm{U}}
\renewcommand{\e}{\bm{e}}
\renewcommand{\c}{\bm{c}}
\renewcommand{\a}{\bm{a}}
\renewcommand{\b}{\vec{b}}
\newcommand{\hb}{\hat{\b}}

\newcommand{\rhoa}{\bm{\rho}_a}
\renewcommand{\u}{\vec{u}}
\newcommand{\cperp}{\c'_{\perp}}
\newcommand{\g}{\bm{g}}
\newcommand{\Z}{\bm{Z}} 
\newcommand{\grad}{\nabla}
\newcommand{\ptheta}{\frac{ \partial }{\partial \theta}}
\newcommand{\pmu}{\frac{ \partial }{\partial \mu}}
\newcommand{\pvparallel}{\frac{\partial }{ \partial v_{\parallel}}}
\renewcommand{\L}{\mathcal{L}}
\newcommand{\gyptheta}{\frac{ \partial }{\partial \overline{\theta}}}

\newcommand{\gypvparallel}{\frac{ \partial }{\partial \overline{v}_{\parallel}}}
\newcommand{\gypmu}{\frac{ \partial }{\partial \overline{\mu}}}
\newcommand{\gygrad}{\overline{\grad}}
\newcommand{\curvature}{\bm{\kappa}}
\newcommand{\kernel}[1]{\mathcal{K}_{#1}}
\newcommand{\norm}[1]{\left\lVert#1\right\rVert}
\newcommand{\gyaver}[1]{\left<  #1 \right>}
\newcommand{\sparallel}{\overline{s}_{\parallel a}}
\newcommand{\sperp}{\overline{s}_{\perp a}}
\newcommand{\uparallel}{\overline{u}_{\parallel a}}
\newcommand{\vthparallel}{\overline{v}_{th\parallel a }}
\newcommand{\vthperp}{\overline{v}_{th \perp a}}
\newcommand{\Tperp}{\overline{T}_{\perp a}}
\newcommand{\Tparallel}{\overline{T}_{ \parallel a }}
\newcommand{\Pperp}{\overline{P}_{\perp a}}
\newcommand{\Pparallel}{\overline{P}_{ \parallel a }}
\newcommand{\gyrhoa}{\overline{\bm{\rho}}_a}
\newcommand{\vparallel}{v_{\parallel}}
\newcommand{\vperp}{v_{\perp}}

\newcommand{\Qparallel}{\overline{Q}_{\parallel  a}}
\newcommand{\Qperp}{\overline{Q}_{\perp a}}
\newcommand{\gyU}{\overline{\U}}
\newcommand{\gyuparallel}{\overline{u}_{\parallel a}}
\newcommand{\gyvthparallel}{\overline{v}_{th \parallel a}}
\newcommand{\gyTperp}{\overline{T}_{\perp a}}
\newcommand{\gyTparallel}{\overline{T}_{\parallel a}}
\newcommand{\gyR}{\overline{\R}}
\newcommand{\gymu}{\overline{\mu}}
\newcommand{\gytheta}{\overline{\theta}}

\newcommand{\gyvparallel}{\overline{v}_{\parallel}}

\newcommand{\gyvperp}{\overline{v}_\perp}
\newcommand{\gyZ}{ \overline{\Z}}
\newcommand{\momentstar}[2]{ \norm{#2}^{*#1}_a}
\newcommand{\moment}[2]{ \norm{#2}^{#1}_a}
\newcommand{\momenta}[1]{\norm{#1}_a}
\newcommand{\momentastar}[1]{\norm{#1}^*_a}
\newcommand{\phaseV}{\mathcal{V}}
\newcommand{\phaseM}{\mathcal{M}}
         
\newcommand{\gyN}{\overline{N}_a}
\newcommand{\gyFa}{\overline{F_a}}   
\newcommand{\gyFe}{\overline{F_e}} 
\newcommand{\gyFi}{\overline{F_i}} 
\newcommand{\gR}{\bm{g}^{\R}}
\newcommand{\gtheta}{g^{\theta}}
\newcommand{\gmu}{g^{\mu}}

\newcommand{\gbR}{\overline{\bm{g}}^{\R}}
\newcommand{\gbtheta}{\overline{g}^{\theta}}
\newcommand{\gbmu}{\overline{g}^{\mu}}
\newcommand{\gbparallel}{\overline{g}^{\parallel}}
\newcommand{\bB}{\overline{\B}}

\newcommand{\corr}[1]{{\color{black} #1}} 
\newcommand{\corrs}[1]{{\color{black} #1}} 
\newcommand{\corrIII}[1]{{\color{black} #1}}

\newcommand{\twolinetabular}[2]{\begin{tabular}{@{}c@{}} #1  \\ #2\end{tabular}}

\shorttitle{A gyrokinetic model for the plasma periphery}
\shortauthor{B. J. Frei, R. Jorge and P. Ricci}

\title{A gyrokinetic model for the plasma periphery of tokamak devices}

\author{B. J. Frei\aff{1}
  \corresp{\email{baptiste.frei@epfl.ch}},
  R. Jorge\aff{1,2}
 \and P. Ricci\aff{1}}
 
\affiliation{\aff{1} \'Ecole Polytechnique F\'ed\'erale de Lausanne (EPFL),
  Swiss Plasma Center (SPC),
CH-1015 Lausanne, Switzerland,
\aff{2} Instituto de Plasmas e Fus\~{a}o Nuclear, Instituto Superior T\'ecnico, Universidade de Lisboa, 1049-001 Lisboa, Portugal}

\begin{document}

\maketitle

\begin{abstract}
A gyrokinetic model is presented that can properly describe large and small amplitude electromagnetic fluctuations occurring on scale lengths ranging from the electron Larmor radius to the equilibrium perpendicular pressure gradient scale length, and arbitrary large deviations from thermal equilibrium that are present in the plasma periphery of tokamak devices. The formulation of the gyrokinetic model is based on a second order accurate description of the single charged particle dynamics, derived from Lie perturbation theory, where the fast particle gyromotion is decoupled from the slow drifts assuming that the ratio of the ion sound Larmor radius to the perpendicular equilibrium pressure scale length is small. The collective behaviour of the plasma is obtained by a gyrokinetic Boltzmann equation that describes the evolution of the gyroaveraged distribution function. The collisional effects are included by  a nonlinear gyrokinetic Dougherty collision operator. The gyrokinetic model is then developed into a set of coupled fluid equations referred to as the gyrokinetic moment hierarchy. To obtain this hierarchy, the gyroaveraged distribution function is expanded onto a Hermite-Laguerre velocity-space polynomial basis. Then, the gyrokinetic equation is projected onto the same basis obtaining the spatial and temporal evolution of the Hermite-Laguerre expansion coefficients. A closed set of fluid equations for the lowest order coefficients are presented. The Hermite-Laguerre projection is performed accurately at arbitrary perpendicular wavenumber values. Finally, the self-consistent evolution of the electromagnetic fields is described by a set of gyrokinetic Maxwell's equations derived from a variational principle where the velocity integrals are explicitly evaluated.
\end{abstract}

\section{Introduction}

The plasma periphery in a tokamak, extends from the external part of the closed flux surface region, typically referred to as the edge, to the scrape-off layer (SOL) region where the magnetic field lines intercept the machine vessel walls. These two regions are separated by the last closed flux surface (LCFS). Understanding the plasma dynamics in the periphery is necessary to address some the most crucial problems fusion is facing today, still undermining our capabilities to make reliable predictions of the performances of future tokamak, such as ITER \citep{Shimada2007}. Indeed, the plasma dynamics in the periphery largely controls the plasma heat exhaust, the refuelling, and \corr{helium removal}. In addition, it governs the overall confinement performance of a tokamak. In fact, a low-to-high (L-H) confinement mode transition can be triggered if a heat power threshold is exceeded \citep{Wagner2007}. This results from the formation of a transport barrier, which reduces considerably the turbulent transport, while a pressure pedestal appears in the proximity of the LCFS \citep{Gohil1994}. This barrier is characterized by strong radial electric fields yielding strong $\E \times \B$ sheared flows \citep{Burrell1994,Xia2006,Schirmer2006}, which are thought to be responsible for the observed reduction of the H-mode turbulence level \citep{Petty1998,Naulin2007}. Once formed, the H-mode pedestal can periodically relax because of the presence of edge-localized modes (ELMs) releasing large amplitude bursts of particle and heat into the SOL \citep{Zohm1996,Connor2000,Kirk2005}. ELM events are a major concern on the way to fusion energy.

The phenomena at play in the plasma periphery are mostly a consequence of the presence of electromagnetic fluctuations. These fluctuations determine the anomalous transport level \citep{Stoneking1994,Scott2003,LaBombard2005,Snyder2012,Battaglia2014}, the dynamics behind the L-H mode transition \citep{Connor2000}, they might have a role in the physics of the density limit \citep{Rogers1997,Rogers1998,LaBombard2001,Scott2003} and, furthermore, they are expected to limit the pedestal pressure gradient \citep{Snyder2011,Dickinson2012} and to set the SOL width \citep{Scott1997,DIppolito2002,Scott2003,Halpern2014,Mosetto2015,Militello2016}. Turbulent modes can develop into coherent filamentary structures with important perpendicular extension carrying a large amount of particles outwards in the SOL, strongly affecting the heat load on the plasma-facing components \citep{Zweben2007,Agostini2011,Zweben2015,Garcia2015}.These fluctuations occur on perpendicular scale lengths ranging from the particle Larmor radius, $\rho_a = v_{tha} / \Omega_a$, to a scale length of the order of the typical equilibrium (in the sense of time-averaged) total pressure and potential gradient scale lengths, namely $L_P$ and $L_\phi$ \citep{Ritz1987,Endler1995,LaBombard2001,DIppolito2002,Garcia2007a,Nespoli2017}. Here, $a$ is the species subscript (i.e., for electrons $a=e$ and for the ions $a=i$), $v_{tha}^2 = 2 T_a /m_a$ is the particle thermal velocity, $T_a$ is the particle temperature, and $\Omega_a = q_a B / m_a$ is the particle gyrofrequency, with $q_a$ and $m_a$ the particle charge and mass, respectively. The turbulent dynamics is thought to be the result of the complex nonlinear development of electromagnetic ion-temperature-gradient (ITG) modes, drift waves, Kelvin-Helmholtz instabilities, and ballooning modes \citep{Scott1997,Zeiler1998,Rogers2005,Mosetto2013}. While these modes occur on the $\rho_i$ and larger spatial scales \citep{Mazzucato2008,Guttenfelder2011,Guttenfelder2013}, there are evidences that also turbulent modes, occurring on the electron gyroscale with perpendicular wavenumber, $k_\perp$, satisfying $k_\perp \rho_e \sim 1$, can contribute significantly to the large scale heat flux and interact with modes at larger scales \citep{Dorland2000,Jenko2000,Neiser2018}. Moreover, finite Larmor radius (FLR) effects might be important in the description of the pedestal \citep{Snyder2002}. While the turbulent modes yield small amplitude fluctuations in the edge region \citep{Brower1987,Ritz1987}, order unity fluctuation levels can be found in the SOL region with $k_\perp L_P \sim 1$ \citep{Ritz1987,Wootton1990,Garcia2007,Xu2009,Zweben2015,Nespoli2017}. In general, these turbulent modes display elongated structures in the direction parallel to the equilibrium magnetic field \citep{Winslow1997,Zeiler1998,Thomsen2002,Halpern2013,Grulke2014}, and have typical frequencies considerably smaller than $\Omega_{i}$ \citep{Levinson1984,Ritz1987,Zweben2007}. 

Over the last decades, successful and significant progress has allowed important advances in the simulation of the turbulent plasma dynamics in the periphery. Both fluid \citep{Dudson2009,Tamain2009,Ricci2012,Easy2014,Ricci2015,Halpern2016,Paruta2018} and \corr{gyrokinetic simulations} \citep{Xu2007,Cohen2008,Ku2009,Shi2015,Hakim2016,Pan2016,Chang2017,Pan2018} have been used to achieve this goal. Fluid models are usually based on the drift-reduced Braginskii equations \citep{Braginskii1965,Zeiler1997,Simakov2005,Scott2007}, which evolve the three-lowest order fluid moments of the kinetic equation, that are the particle density, velocity, and temperature. These fluid equations rely on high plasma collisionality as the fluid closure assumes that the particle distribution function is locally close to thermal equilibrium, and on the assumption that the perpendicular scale length of the fluctuations is large compared to $\rho_i$ \citep{Hazeltine1973,Zeiler1997,Catto2004}. The assumption of local thermal equilibrium is hard to justify in the edge while it appears more reasonable in the SOL, where the particles temperatures are lower than in the core and at the top of the pedestal \citep{Kocan2008}. However, significant deviations from thermal equilibrium can occur in the SOL. These deviations from a Maxwellian distribution function were numerically investigated by kinetic simulations \citep{Lonnroth2006,Tskhakaya2012,Battaglia2014}, and are due to the presence of a tail of suprathermal particles, yielding important modifications to the parallel heat flux, of transient processes (e.g., ELM bursts influencing locally the particle collisionality), and of the sheath \citep{Sigmar1996,Batishcheva1996,Batishchev1997,Kirk2005}. These effects might be enhanced in typical ITER operating H-mode scenarios \citep{Martin2008}. To include kinetic modifications covering different range of collisional regimes in a fluid description and non-Maxwellian physics, an extension of fluid models based on higher order moments has been recently addressed by \citet{Jorge2017} within a drift-kinetic regime. \corr{This model considers potentially large electrostatic fluctuations by truncating an expansion of the electrostatic potential around the guiding-center position, i.e., the center of the particle gyration. Therefore, it is not able to properly describe fluctuations that occur on the particle gyroscale.} Such fluctuations are particularly important in the description of anomalous transport in the edge region \citep{Scott2003}. 

To describe fluctuations on the ion Larmor radius scale, gyrokinetic theories - pioneered by \citet{Catto1978,Frieman1982} - were successfully used in numerical and theoretical investigations of microturbulences \citep{Furnish1999,Heikkinen2006,Brizard2007,Idomura2009,Wang2010,Hatch2011,Krommes2012}. Most of the gyrokinetic models are derived within the standard gyrokinetic ordering \citep{Dubin1983,Hahm1988,Parra2014}, i.e.

\begin{equation} \label{eq:stdgyrokinetic}
\frac{\omega}{\Omega_i} \sim  \frac{e \phi_1 }{T_e} \sim \frac{k_\parallel }{k_\perp} \ll 1,
\end{equation}
\\
and 
\begin{equation}
k_\perp \rho_i \sim 1,
\end{equation}
\\
where $\phi_1$ is small amplitude fluctuating component of the electrostatic potential $\phi$, and $e = - q_e$. While allowing for $k_\perp \rho_i \sim 1$ and being appropriate for core plasma conditions, the ordering given in \cref{eq:stdgyrokinetic} breaks down in the presence of large scale and amplitude fluctuations, such as the ones present in the SOL. The analytical treatment of large scale and amplitude electromagnetic fluctuations was addressed by \citet{Dimits1992} in a generalized gyrokinetic ordering, assuming that $ k_\perp \rho_i e(  \phi - v_\parallel A_{\parallel}  )/ T_e \ll 1$ where $A_\parallel$ is the parallel component of the fluctuating magnetic vector potential. This ordering is less restrictive than \cref{eq:stdgyrokinetic}. Indeed, it allows for order unity fluctuations, $e  \phi / T_e \sim 1$, on long scale lengths $k_\perp \rho_i \ll  1$, but also for $e\phi/ T_e \ll 1$ on $k_\perp \rho_i \sim 1$ scales. Gyrokinetic theories have also been developed to retain the presence of strong $\E \times \B$ flows driven by large radial sheared electric fields in the internal and edge transport barrier regions \citep[see, e.g., ][]{Berntstein1985,Brizard1995,Hahm1996,Hahm2009,Dimits2012}. However, despite significant progress, the present gyrokinetic formulations do not provide yet an efficient framework for the description of the turbulent plasma periphery dynamics. \corr{The limitations of the present models are due to the lack of a proper description of arbitrarily far from equilibrium distribution functions in the presence of both large and small scale fluctuating electromagnetic fields and to the lack of a nonlinear gyrokinetic collision operator in a form that allows for an efficient numerical implementation, overcoming the extreme computational resources needed for the present edge gyrokinetic codes \citep{Chang2017,Pan2018}}. In this regard, we note that gyrofluid models were developed to limit the computational cost of gyrokinetic simulations while overcoming the limitations of fluid models. These models aim to incorporate FLR and kinetic effects in a fluid-based description by taking a finite number of moments of the gyrokinetic equation \citep[see, e.g.,][]{Brizard1992,Dorland1993,Waltz1994,Beer1996,Snyder2001,Ribeiro2008,Madsen2013,Held2016}. The truncation schemes for closing the fluid models including collisionless Landau damping and perpendicular coupling between moments of the distribution function (associated with parallel streaming, magnetic gradient drifts, and FLR effects) are usually obtained by techniques pioneered by \citet{Hammett1990,Hammett1992a,Dorland1993} and by methods developed by \citet{Waltz1994,Beer1996}. Unfortunately, the limitations of these closure schemes undermine the use of gyrofluid models to evolve the plasma periphery dynamics due to the wide range of collisionality and the possible large deviations from equilibrium. Recently, an extension of the gyrofluid model to an arbitrary number of moments has been obtained by \citet{Mandell2018} in an Hermite-Laguerre pseudo-spectral velocity formulation of delta-F gyrokinetic theory for electrostatic perturbations in core conditions. This model allows for a dynamical refinement of previous gyrofluid models with a tuneable accuracy, and provide an ideal framework to approach the development of a gyrokinetic model for the plasma periphery.

In the present paper, leveraging previous works \citep{Hammett1993,Beer1996,Sugama2000,Hahm2009,Zocco2011,Madsen2013,Omotani2015,Zocco2015,Schekochihin2016,Loureiro2016,Jorge2017,Mandell2018}, we derive a gyrokinetic model that we develop into a gyrokinetic moment (gyro-moment) hierarchy able to evolve the turbulent dynamics of the plasma in the periphery of tokamak devices. Our model allows to describe far from equilibrium distribution functions, linear and nonlinear FLR effects driven by small scale electromagnetic fluctuations, and effects associated with time-dependent background electromagnetic fields in the presence of a strong sheared radial electric field. To construct our model, we introduce the small expansion parameters $\epsilon = \rho_s/L_P \ll 1$ and $\epsilon_\delta = e \phi_1/T_e \ll 1$, with $\rho_s = c_s/ \Omega_i$ the sound Larmor radius and $c_s^2 = T_e / m_i$ the sound speed. Collisional effects are treated by assuming the plasma strongly magnetized, and by introducing the small parameter $\epsilon_\nu = \nu_i / \Omega_i \ll 1$ being $\nu_i \equiv \nu_{ii}$ the ion-ion collision frequency. The evolution equation of the distribution function, that is the gyrokinetic Boltzmann equation, is then obtained from fully nonlinear electromagnetic gyrokinetic equations of motion that are second order accurate in $\epsilon$ and $\epsilon_\delta$. These equations are derived within a perturbation theory known as Lie-transform perturbation theory, pioneered by \citet{Littlejohn1982} and \citet{Cary1983}. Within this formulation, two successive noncanonical phase-space coordinates transformations are performed to decouple and remove the gyrophase dependent part of the single particle motion from the slow drifts. We expand the full gyroaveraged distribution function onto a complete set of velocity-space Hermite-Laguerre polynomials. Projecting the gyrokinetic Boltzmann equation onto the Hermite-Laguerre basis yields an infinite set of coupled fluid equations describing the evolution of the expansion coefficients, termed gyro-moments. Collisional effects are introduced in our model by a nonlinear gyrokinetic Dougherty collision operator \citep{Dougherty1964,Abel2008}. Finally, the gyro-moment hierarchy is coupled to a set of gyrokinetic Maxwell's equations derived self-consistently from a gyrokinetic variational principle. This set includes two coupled gyrokinetic Poisson's equation and two coupled gyrokinetic Ampere's laws. Within this framework, polarization and magnetization effects appear from the phase-space coordinate transformations carried out in the single particle dynamics. These polarization and magnetization terms are expressed as functions of gyro-moments, yielding complete and closed analytical expressions of the gyrokinetic Maxwell's equations.

The rest of the paper is organized as follows. In \cref{PeripheryOrdering}, we order the main spatial and temporal scales at play in the plasma periphery and the relative amplitude of the electromagnetic fluctuations. In \Cref{Singleparticledynamicsinthetokamakperiphery}, we describe the single particle dynamics by deriving a set of second order accurate nonlinear electromagnetic gyrokinetic equations of motion by performing two phase-space coordinate transformations. In \cref{GyrokineticBoltzmannEquation}, we derive the gyrokinetic Boltzmann equation. The expansion of the distribution function into a Hermite-Laguerre polynomial basis and the derivation of the gyro-moment hierarchy is presented in \cref{GyroMomentHierarchy}, while the nonlinear gyrokinetic Dougherty collision operator is obtained and expanded in gyro-moments in \cref{GyrokineticCollisionOperator}. Then, the gyro-moment hierarchy is self-consistently coupled to a set of gyrokinetic Maxwell's equations deduced from a gyrokinetic variational principle in \cref{GyrokineticMaxwellEquations}. Finally, our main results are outlined in \cref{conclusion}. In particular, the improvements over previous gyrokinetic theories are summarized.

\section{Ordering Assumptions for the Plasma Periphery}
\label{PeripheryOrdering}
In this section, we order the main spatial and temporal scales at play in the plasma periphery of tokamak devices. We first consider the scale lengths of the time-averaged profiles. We define the time-averaged perpendicular scale lengths of the total pressure $P = \sum_a n_a T_a$, with $n_a$ and $T_a$ the density and temperature of species $a$, and of the electrostatic potential $\phi$, namely $L_P$ and $L_\phi$, as $ L_P \sim \lvert \grad_\perp \ln \left <P\right>_\tau \rvert^{-1}$ and $L_\phi \sim \lvert \grad_\perp \ln \left< \phi \right>_\tau \rvert^{-1}$, with $\grad_\perp$ the gradient in the direction perpendicular to the magnetic field. The time-average operator, $\left< \cdot \right>_\tau$, acts over a time $\tau$ such that $ \tau \omega \gg 1$, where $\omega$ is the typical frequency of the fluctuations, as defined by $\omega \sim \lvert \partial_t \ln n_a \rvert \sim \lvert \partial_t \ln T_a \rvert \sim \lvert \partial_t \ln \phi \rvert$.

The definitions of $L_P$ and $L_\phi$ allow us to introduce the small parameter $\epsilon$ as the ratio of the ion sound Larmor radius $\rho_{s}$ to the equilibrium total pressure scale length $L_P$ \citep{Frieman1982},

\begin{equation} \label{eq:defepsilon} 
     \epsilon =  \frac{\rho_s}{L_P}  \sim \frac{\rho_s}{L_\phi}  \ll 1,
 \end{equation}
 \\
being $L_P \sim L_\phi$. We remark that $\rho_s  \sim \rho_i$, whereas $\rho_i / \rho_e  \sim \sqrt{  m_i / m_e}  \simeq 60$ for a deuterium plasma, assuming $ T_i / T_e \sim 1$. Indeed, $1 \lesssim T_i/T_e \lesssim 4$ in the plasma periphery \citep{Kocan2008,Elmore2012}. \corr{While an alternative small parameter, $\rho_i / L_P \ll 1$, can also be proposed (instead of $\rho_s$ in \cref{eq:defepsilon}), the presence of turbulent modes in the plasma periphery with turbulent frequency $\omega/ \Omega_i \sim \rho_s / L_P $ \citep{Mosetto2013}, that persist in the cold ion limit, motivates the definition in \cref{eq:defepsilon}.}

The spatial ordering in \cref{eq:defepsilon} is justified in many tokamaks and a wide range of experimental conditions. For example, in the edge region of the TEXT tokamak with plasma parameters $B = 2 $ T, $I_p = 200 $ kA, $n_e \simeq 3 \times 10^{19} $ m$^{-3}$, equilibrium scale lengths $L_n \sim L_{T_e} \sim  L_{T_i} \sim L_P \sim 1.5 $ cm and $\rho_s \sim \rho_i \simeq 0.02 $ cm, one has $\epsilon \sim 0.0167$ \citep{Ritz1987}. For a medium size tokamak, such as TCV, with $B = 1.5$ T, $T_e \sim T_i \sim 40$ eV and $n_e \simeq 6 \times 10^{18}$ m$^3$, it is possible to estimate $\epsilon \sim 0.043$ \citep{Rossel2012,Halpern2016}. For the edge transport barrier, typical values of $\epsilon \sim 1/30 $ can be found in H-mode discharges in the JT60-U tokamak \citep{Gohil1994}. Additionally, \cref{eq:defepsilon} is also in agreement with detailed experimental measurements of the pedestal equilibrium scale lengths and scaling investigations of the H-mode pedestal structure \citep{Burrell1994,Hubbard2000,Wagner2007,Zweben2015}. \corr{In general, in H-mode experimental conditions of large aspect ratio tokamaks, the equilibrium pedestal pressure scale length is often found to be of the order of the ion poloidal gyroradius, $\rho_{ \theta i} = \rho_i B_T / B_\theta$ (where $B_T$ and $B_\theta$ are the toroidal and poloidal components of the equilibrium magnetic field, respectively) \citep{Snyder2009}, that is $L_P \sim \rho_{ \theta i} $, yielding $\epsilon \sim B_\theta / B_T \sim 1/10$.} \corr{Although the latter estimate may break down in the edge of low aspect ratio devices such as MAST and NSTX \citep{Ono2000}, the parameter $\epsilon $ remains small also in this kind of devices \citep{kirk2004,Militello2013}. Finally, our ordering \corr{is valid in} the operational conditions of ITER, as expected from predictions of its pedestal height and width \citep{Snyder2009b,Beurskens2011}.} 

 The equilibrium magnetic scale length $L_B  \sim \lvert \grad_\perp \ln \hat{B} \rvert^{-1}$ of the equilibrium magnetic field, $\hat{B}$, is ordered by the small parameter 

\begin{equation} \label{eq:epsilonB}
 \epsilon_B =\frac{ \rho_s}{  L_B}  \sim \epsilon^3
 \end{equation}
 \\
 since $L_B \sim R_0$, being $R_0$ the major radius of the tokamak device. We remark that, while one expects a maximal ordering $\epsilon_B \sim \epsilon$ for typical core conditions, in the plasma periphery where $L_P$ is steeper compared to the core, the scale length separation obeys $ L_P / R_0 < \rho_s/ L_P$ \citep{Burrell1994,Zweben2007} \corr{such that $\epsilon_B < \epsilon^2$, an assumption used in previous gyrokinetic edge model \citep[see, e.g.,][]{Hahm2009}.} 

To describe the perpendicular scale length of the fluctuations, we introduce the perpendicular wavenumber, 

 \begin{equation} \label{eq:kperp}
 k_\perp \sim \lvert \grad_\perp \ln n_a \rvert \sim  \lvert \grad_\perp \ln T_a \rvert \sim  \lvert \grad_\perp \ln P \rvert \sim  \lvert \grad_\perp \ln \phi \rvert,
 \end{equation}
 \\
and the parameter 

\begin{equation} \label{eq:epsilonperp}
\epsilon_\perp = k_\perp \rho_{s}.
 \end{equation}
 \\
 Our model addresses the presence of turbulent fluctuations on scale lengths ranging from $L_P$ to the $\rho_e$ scale. More precisely, we consider large scale $\epsilon_\perp \sim \epsilon$ fluctuations, typically present in the SOL, where they appear with $k_\perp L_P \sim 1$, and may have large amplitude \citep{Zweben2015,Nespoli2017}. At the same time, we include $\epsilon_\perp \gtrsim 1$ fluctuations \citep{Wootton1990,Shats2005}. These fluctuations have scale length of the order or smaller than $\rho_s$, and include electron gyroscale fluctuations ($k_\perp \rho_e \sim 1$), which are thought to be important in ETG driven turbulent transport experiments \citep{Dorland2000,Colyer2017}. 
 
 \corr{In order to describe the possible coexistence of electromagnetic fluctuations at amplitudes incompatible with the standard gyrokinetic ordering, in \cref{eq:stdgyrokinetic}, and at scales ranging from the $L_P$ scale to $\rho_s$ (or smaller), we order the $\E \times \B$ drift, $\u_E = \E \times \B/B^2$, with $\E = - \grad \phi$, to be small with respect to the sound speed $c_s$, by introducing the parameter,

\begin{equation} \label{eq:uEoverCs}
\epsilon_E = \frac{|\u_E|}{c_s}  \sim \epsilon_\perp \frac{e \phi}{T_e}  \ll 1.
\end{equation}
\\
The ordering in \cref{eq:uEoverCs} allows for fluctuations of the order of the thermal energy on large scale, i.e. $e \phi /T_e \sim 1$ for $\epsilon_\perp \sim \epsilon_E$ scales, and for small amplitude fluctuations, i.e. $e \phi / T_e \sim \epsilon_E$ for $\epsilon_\perp \sim 1$ scales. Also, it allows the presence of fluctuations occurring at intermediate scales between $L_P$ and $\rho_s$. While a sonic ordering, $\epsilon_E \sim 1$ yielding $e \phi /T_e \sim  1 /\epsilon_E$ for $\epsilon_\perp \sim \epsilon_E$, has been considered in, e.g., \citet{Artun1994,Brizard1995,Hahm1996,Qin2007}, here we use the fact that fluctuations in the SOL satisfy $e \phi / T_e \sim 1$ on $\epsilon_\perp  \sim \epsilon$ scales \citep{Zweben2015}, while in the edge $e \phi/ T_e \sim \epsilon_\delta \sim \epsilon$ on $\epsilon_\perp \sim 1$ scales \citep{Brower1987}, to find from \cref{eq:uEoverCs} that the $\epsilon_E$ and $\epsilon$ parameters can have similar size.} \\
The $\E \times \B$ ordering in \cref{eq:uEoverCs} is valid in both H- and L-mode, even when strongly equilibrium sheared radial electric fields and steep pressure gradients are present. In the L-H mode experimental conditions observed in the DIII-D tokamak, we observe $\E \times \B$ drift to the sound speed of the order of $|\u_E| / c_s \sim 1/10$, estimating $T_i \simeq 0.3$ KeV and $B \simeq 2 $ T  \citep{Doyle1991,Gohil1994}. Similar values can be inferred from measurements of the long wavelength $\E \times \B$ flows in the edge of the HT-$7$ tokamak \citep{xu2003}. The ordering in \cref{eq:uEoverCs} is also appropriate to describe the broadband turbulent power spectrum of the plasma periphery \citep{Levinson1984,Ritz1987,Shats2005} and the corresponding fluctuation levels. Indeed, the turbulent dynamics in the SOL is typically characterized by coherent filamentary structures with perpendicular gradient scale lengths $L_P \sim L_\phi$ ($k_\perp \rho_s \sim 0.1$) and large amplitude fluctuations in the density and in the electrostatic potential, that are of the order of $T_e/e$, \citep{Garcia2007a,Xu2009,Zweben2015,Nespoli2017}. As observed in experimental characterizations of edge plasma turbulence \citep{Ritz1987,Wootton1990,Fonck1993,Wolf2002,Wang2004,Xia2006}, the amplitude of these large fluctuations present in the SOL decreases inside the LCFS to residual levels, while the turbulent power spectrum in the edge peaks at $k_\perp \rho_i \simeq 0.6$ with $\rho_i \sim \rho_s$ \citep{Brower1987,Ritz1987}. Thus, ion FLR effects might become non-negligible in this region \citep{Snyder2002,Dickinson2012}. On the other hand, we remark that the particle parallel streaming velocity, $v_\parallel$, can be comparable to the sound speed. Values of the dimensionless parallel Mach number of order of unity have been reported by experimental investigations of parallel flows \citep{Wang2004,Pitts2007}.

\corr{For our model to allow for fluctuations at scales from $L_P$ to $\rho_s$, we separate the potential $\phi$ into two components, i.e. a drift-kinetic and a gyrokinetic component, $\phi_0$ and $\phi_1$, respectively (see \cref{Singleparticledynamicsinthetokamakperiphery} for their rigorous definitions). Thus, we write \citep{Dimits1992} 

\begin{equation} \label{eq:phi}
\phi = \phi_0 + \phi_1,
\end{equation}
\\
and order the size of $\phi_1$ to be small, at all scales from $L_P$ to $\rho_s$, compared to the thermal energy by introducing the small parameter 

\begin{equation} \label{eq:epsilondelta}
\epsilon_\delta =  \frac{e \phi_1}{T_e}  \ll 1.
\end{equation}
\\
By imposing that both $\phi_0$ and $\phi_1$ provide a similar contribution to $\u_E$, we order their gradients such that $\lvert \grad_\perp \phi_0 \rvert \sim \lvert \grad_\perp \phi_1 \rvert$. \Cref{eq:uEoverCs,eq:epsilondelta} imply $\lvert \rho_s \grad_\perp \ln \phi_1 \rvert \sim 1$ such that $\phi_1$ is allowed to vary on all scales (from $L_P$ to $\rho_s$), while $\phi_0$ is potentially large on large scale if $\epsilon_\perp \sim \epsilon$ (comparable to $L_P$) since $ \epsilon_\perp e \phi_0 / T_e \sim \epsilon_E \sim \epsilon $.}

Following the critical balance conjecture \citep{Goldreich1995,Schekochihin2008,Schekochihin2009,Schekochihin2016}, we order the parallel scale length $ k_\parallel  \sim \lvert \grad_\parallel \ln P \rvert \sim  \lvert \grad_\parallel \ln \phi \rvert $ by assuming that $ c_s k_\parallel \sim k_\perp |\u_E|$ where $k_\perp |\u_E|$ is the frequency of the $\E \times \B$ flow. Following \cref{eq:uEoverCs}, this implies that $\rho_s k_\parallel \sim \epsilon_\perp \epsilon $, i.e.

\begin{equation} \label{eq:kparallel}
\frac{k_\parallel}{ k_\perp }\sim   \epsilon.
\end{equation}
\\
\Cref{eq:kparallel} is compatible with experimental evidences of turbulent structures elongated in the parallel direction in the tokamak periphery and theoretical expectations of ballooning modes \citep{Winslow1997,Bleuel2002,Garcia2007a,Naulin2007,Halpern2013,Grulke2014}. \corr{We remark  that \cref{eq:epsilonB,eq:kparallel} implies that $k_\parallel L_B \sim 1/\epsilon$ on the $L_P$ scales. This encompasses, e.g., ballooning mode structures for which the parallel wavenumber is of the order of the connection length, i.e. $L_c \simeq 2 \pi q R_0$ with $R_0\sim L_B$ \citep{Halpern2013}, and drift waves turbulent modes which can be characterized by large $k_\parallel$ values \citep{Rogers2010,Mosetto2013}.}

The typical turbulent frequency $\omega$ in the plasma periphery is much smaller than the ion gyrofrequency $\Omega_{i}$. More precisely, we assume that the turbulent frequency is comparable to the frequency of the $\E \times \B$ flow, i.e. $\omega \sim k_\perp | \u_E|$ \citep{Dimits2012}, yielding

\begin{equation} \label{eq:lowfrequency}
\frac{\omega}{\Omega_i} \sim \epsilon_\perp \epsilon,
\end{equation}
\\
being $\omega / \Omega_i \sim \epsilon^2$ for large scale fluctuations \citep{Zeiler1997,Simakov2005} while $\omega/\Omega_i \sim \epsilon$ on the $\epsilon_\perp \sim 1$ scale length \citep{Dubin1983,Hahm1988,Hahm2009}. As a result of the orderings in \cref{eq:kparallel,eq:lowfrequency}, the parallel advection at the sound speed occurs on the same time scale as the turbulent fluctuations, i.e. $k_\parallel c_s \sim \omega$, in agreement with the critical balance that sets the parallel correlation length $L_\parallel \sim 1/ k_\parallel$ of the perpendicular turbulent structure. 

To describe magnetic fluctuations, we introduce a fluctuating magnetic vector potential $\delta \bm  A$. We assume that the fluctuating magnetic field, $\delta \bm B = \grad \times \delta \A$, is small compared to the large scale equilibrium magnetic field, $\hB = \grad \times \hat{\bm A}$ where $\hat{\bm{A}}$ is the magnetic vector potential associated with the large scale equilibrium magnetic field. More precisely, we impose 

\begin{equation} \label{eq:deltaBoverB}
\frac{|\delta \bm  B|}{ \hat{ B}} \sim \epsilon,
\end{equation}
\\
with $\hat{B} = |\hB|$, consistently with experimental measurements of edge magnetic fluctuations \citep{Stockel1999,Graessle1991}. \corrs{ We remark that both $\hat{\bm{B}}$ and $\hat{\A}$ vary on the $L_B$ scale, and that the temporal scale of $\hat{\A}$ is ordered as 

\begin{equation} \label{eq:dAdt}
\frac{\left| \partial_t \hat{\bm{A}} \right|}{c_s B} \sim \epsilon.
\end{equation}
\\
This implies that $\hB$ evolves on time scales such that $|\partial_t \ln \hat{B}|/\Omega_i \sim \epsilon \epsilon_B$ \citep{Hazeltine2003,Green2003,Wagner2009}. \corrIII{We remark that \cref{eq:dAdt} allows us to retain the possible effects due to the change of the equilibrium magnetic field (e.g., for plasma shaping) that can occur on a time scale faster that the current diffusion time scale, but smaller that the typical frequency of the fluctuations.} As a consequence of the ordering in \cref{eq:dAdt}, the inductive $\E \times \B$ velocity, i.e. $\b \times \partial_t \hat{\bm A}/B$, remains small compared to $c_s$ (see \cref{eq:uEoverCs}).} \corrs{The total fluctuating magnetic field can be written as

\begin{equation}
\delta \B = \grad \times \delta \A_\perp + \grad \times \left( \delta A_{\parallel } \hb\right),
\end{equation}
\\
where $\delta A_{\parallel } \equiv \hb \cdot \delta \A$ and $\delta \A_\perp \equiv \delta \A - \delta A_\parallel \hb$ being the parallel and perpendicular components of the fluctuating magnetic vector potential, respectively. Since compressional waves, associated with $\delta B_\parallel \equiv \hb \cdot \delta \B$, are much faster than the typical turbulent time scale considered on the scales ranging from $L_P$ to $\rho_s$ in low-$\beta$ plasmas (with $\beta = 8 \pi P / B^2$ the total thermal beta plasma) - as inferred from experimental measurements \citep{Petty1998,Doyle2007,Saibene2007,Zweben2007}-, we systematically neglect $\delta B_\parallel$. Thus, we describe $\delta \B_\perp$, at the lowest order, by a fluctuating magnetic vector potential along the equilibrium magnetic field, i.e. 

\begin{equation} \label{eq:deltaB}
    \delta \B = \delta \B_\perp \simeq \grad_\perp \delta A_{\parallel} \times \hb,
\end{equation}
\\
where we neglect the parallel component, $\delta B_\parallel = \delta A_{\parallel} \hb \cdot \grad \times \hb $ since it is of higher order and $\delta \bm A_\perp$. This removes completely the compressional Alfv\'en waves in our model. We remark that this approximation is consistent with low-$\beta$ plasma models currently implemented in global and edge gyrokinetic codes (see, e.g, \citet{Goerler2011,Lanti2019,Mandell2019}), and fulfills the plasma periphery conditions. We also show that our low-$\beta$ approximation is in fact compatible with the presence of steep equilibrium pressure gradients in \cref{LeadingOrderEquilibriumPressureBalanceEquation}. Finally, we deduce that $\phi \sim c_{s} \delta A_{\parallel}$ by comparing the $\E \times \B$ ordering in \cref{eq:uEoverCs} with \cref{eq:deltaBoverB} \citep{Brizard2007,Dimits2012,Tronko2017}. This allows a general treatment of magnetic fluctuations in a maximal ordering.}



Similarly to \cref{eq:phi}, the parallel component of the fluctuating magnetic potential, $\delta A_{\parallel}$, can be written as

\begin{equation}\label{Aparallel}
\delta A_\parallel = A_{\parallel 0} +  A_{\parallel 1},
\end{equation}
\\
\corr{with $A_{\parallel 0}$ and $A_{\parallel 1}$ the drift-kinetic and gyrokinetic parts of the fluctuating magnetic potential, respectively.} The time variation of $A_{\parallel 0}$ and $A_{\parallel 1} $ obeys \cref{eq:lowfrequency}. \corr{By imposing that both $A_{\parallel 0}$ and $A_{\parallel 1} $ provide a similar contribution to $\delta \B_\perp$, we order $\lvert \grad_\perp A_{\parallel 0} \rvert \sim \lvert \grad_\perp A_{\parallel 1}\rvert$. In addition, the ordering given in \cref{eq:epsilondelta} with $\phi_1 \sim c_s A_{\parallel 1}$ yields $c_s e A_{\parallel 1} / T_e \sim \epsilon_\delta$ such that $A_{\parallel 1}$ is small at all scales, while $A_{\parallel 0}$ can be potentially large at large scales since $\epsilon_\perp e c_s A_{\parallel 0}/ T_e \sim \epsilon_E$.} Thus, the magnetic field from $A_{\parallel 0}$ and the large scale component, $\hB$, is written as

\begin{equation} \label{eq:totalB}
\B = \hB + \grad_\perp A_{\parallel 0} \times \hb,
\end{equation}
\\
while the magnetic field generated by the small amplitude component $A_{\parallel 1}$ is $\B_1 = \grad_\perp  A_{\parallel 1} \times \hb $, such the total magnetic field is given by $\B + \B_1$. We also remark that \cref{eq:lowfrequency} gives the ordering of the inductive part of the parallel electric field, being $\partial_t A_{\parallel 0} \sim \epsilon^2 \Omega_i A_{\parallel 0}$ and $\partial_t A_{\parallel 1} \sim \epsilon \Omega_i A_{\parallel 1}$ yielding

\begin{equation} \label{eq:EparaEperp}
    \frac{E_{\parallel}}{|\E_\perp|} \sim \epsilon.
\end{equation}

In order to describe the different collisional regimes in the plasma periphery, we introduce the small parameter $\epsilon_\nu = \nu_i /\Omega_i \ll 1$ where $\nu_i \equiv \nu_{ii}$ is the ion-ion collision frequency. By imposing that the plasma remains strongly magnetized, we require that

\begin{equation} \label{eq:epsilonnu}
     \epsilon_\nu = \frac{\nu_i}{\Omega_i} \sim \epsilon^2 .
\end{equation}
\\
From \cref{eq:epsilonnu}, the electron-electron collision frequency $\nu_e \equiv \nu_{ee}$ is ordered as 

\begin{equation} \label{eq:nue}
\frac{\nu_e}{\Omega_i} \sim \sqrt{\frac{m_i}{m_e}} \left( \frac{T_i}{T_e}\right)^{3/2} \epsilon_\nu,
\end{equation}
\\
being $\nu_e \sim \sqrt{m_i/m_e}(T_i/T_e)^{3/2} \nu_i$. In order to estimate the level of collisionality in the plasma periphery, we look at the ratio of the electron mean-free path, $ \lambda_{mfp} = v_{the}/\nu_e$, to the parallel scale length $L_\parallel \sim 1/k_\parallel$. \Cref{eq:kparallel} and \cref{eq:nue} yield the estimate

\begin{equation} \label{eq:meanfreepath}
    k_\parallel \lambda_{mfp} \sim \frac{\epsilon \epsilon_\perp}{\epsilon_\nu}.
\end{equation}
\\
From \cref{eq:meanfreepath}, one infers that our model is compatible with high ($k_\parallel \lambda_{mfp} < 1$) and low ($k_\parallel \lambda_{mfp} > 1$) collisionality regime.

The ordering given in \cref{eq:epsilonnu} is compatible with experimental conditions. Indeed, with typical TCV SOL parameters, estimating $B = 1.4 $ T, $n_e \simeq 1 \times 10^{18}$ m$^{-3}$, $T_e \sim T_i \sim 25$ eV, one finds $\epsilon_\nu \simeq 0.0003$ \citep{Garcia2007,Nespoli2017}, and for JET discharges, estimating the SOL parameters by $B = 2.4$ T, $n_e \simeq 10^{18}$ m$^{-3}$ and $T_e \sim T_i \sim 40$ eV, one obtains $\epsilon_\nu \simeq 8.3 \times 10^{-5}$ \citep{Erents2000,Xu2009}. Lower values of $\epsilon_\nu$ are typical of the edge due the presence of ELMs, when temperatures are of the order of $T_e \sim T_i \sim 100$ eV, and $\epsilon_\nu \sim 10^{-6}$ \citep{Pitts2003,Kirk2005}.

\section{Single Particle Dynamics in the Plasma Periphery}
\label{Singleparticledynamicsinthetokamakperiphery}
Following previous gyrokinetic models \citep{Dimits1992,Qin2006,Hahm2009,Dimits2012,Madsen2013}, we derive a model for the single particle dynamics in the plasma periphery by performing two successive changes of phase-space coordinates. Taking advantage of the low-frequency ordering in \cref{eq:lowfrequency}, these two transformations allow us to pass from the particle phase-space, described by the coordinates $(\x,\vi,t)$ where $\x$ the particle position, $\vi$ the particle velocity measured in the laboratory reference frame and $t$ the time coordinate, to a set of new coordinates where the fast gyromotion, associated with the gyrophase dependent part of the particle motion, is decoupled and removed from the gyrophase independent low-frequency (compared to $\Omega_a$) drifts. In addition, the transformations ensure the adiabatic invariance of one phase-space coordinate. The transformations we perform are obtained by using Lie-transform perturbation theory, a class of continuous change of phase-space coordinates that preserve the Hamiltonian structure of the original system. We remark that, since the coordinate systems are connected by an invertible transformation, the physics is not affected by the use of the new coordinates \citep{Qin2004}.

The present section is organized as follows. In \cref{GeneralMethodology}, we describe the general methodology we follow to obtain the proper description of the single particle dynamics in the plasma periphery. The mathematical tools associated with Lie-transform perturbation theory are introduced in \cref{LieTransformPerturbationTheory}. Then, the first and second phase-space coordinate transformations are presented in \cref{GuidingCentertransformation,GyrocenterTransformation}, respectively.

\subsection{General Methodology}

\label{GeneralMethodology}

Within a coordinate independent formulation, the dynamics of a particle of species $a$ is prescribed by its action $\mathcal{A}_a = \int L_a dt = \int \gamma_a$, where $L_a$ is the particle Lagrangian and $\gamma_a = L_a dt$ is the Lagrangian one-form. In this section, we drop the particle subscript $a$ for simplicity. The Lagrangian one-form $\gamma$ is given, in the $(6+1)$-dimensional particle phase-space described by the coordinates $\z  = (\x,\vi,t)$, by \citep{Jackson2012}

\begin{equation} \label{eq:particlegamma}
\begin{aligned}
\gamma  & = \left[ q \A + m \vi \right] \cdot d \x - \left[ q \phi + \frac{m v^2}{2} \right] d t  \\
& \equiv \bm{\gamma} \cdot d \z = \gamma_\nu d \z^\nu.
\end{aligned}
\end{equation}
\\
where we use the Einstein's notation and we denote $\A = \A(\x,t)$ and $\phi = \phi(\x,t)$ the magnetic vector potential and electrostatic potential evaluated at the particle position $\x$, and $\z^{\nu}$ is the $\nu$ component of $\z$. In general, the Greek index $\nu$ runs from $1$ to $6+1$ (since it includes the time coordinate $t$), whereas the Latin index $i$ runs from $1$ to $6$. Thus, $\z = \left( \z^i,t\right)$ and $\bm{\gamma} = (\Lambda_i,- \mathcal{H})$ with $\Lambda_i$ and $\mathcal{H}$ the symplectic and Hamiltonian components of the one-form $\gamma$, respectively. In particular, in \cref{eq:particlegamma}, $\Lambda_i = q A_i  + m v_i$ for $i=1,2,3$ with $A_i$ the $i$th component of the magnetic vector potential $\A$, while $\Lambda_i =0$ for $i=4,5,6$ and $\mathcal{H} = q \phi + m v^2/2 $.

To decouple the fast gyromotion from to the low-frequency drifts in the particle dynamics, both contained in $\gamma$, two successive changes of coordinates are performed from the particle phase-space, $ \z$. In the first coordinate transformation, we introduce the lowest order guiding-center coordinates, $\Z_0 = (\x_0, v_{\parallel 0},\mu_{0},  \theta_0,t)$. In this coordinate system, $\x_0$ is the zeroth-order guiding-center position, i.e. $\x_0 \equiv \x$ corresponds to the particle position, and the coordinates $(v_{\parallel 0}, \mu_0, \theta_0)$ are introduced as a first step to decouple the gyrophase dependent and independent parts of the particle motion. In particular, $v_{\parallel 0}= \vi \cdot \b$ is the parallel velocity ($\b = \B / B$), $\theta_0$ is the lowest order gyrophase angle, and $\mu_0$ is the lowest order magnetic moment. More precisely, the coordinates $\theta_0$ and $\mu_0$ are defined by introducing the right-handed Frenet-Serret orthonormal vector basis $(\e_1,\e_2, \b)$ with $(\bm e_1,\bm e_2)$ spanning the plane perpendicular to $\b$ \citep{Frenet1852,Littlejohn1988}. We also introduce the associated cylindrical vector basis $(\a,\b,\c)$, where  $\a \equiv \a(\x_0,\theta_0,t)$ and $\c \equiv \c(\x_0,\theta_0,t)$ are defined with respect the coordinate angle $\theta_0$, i.e.

\begin{align} \label{eq:avector}
\a(\x_0,\theta_0,t) &= \cos \theta_0 \e_1(\x_0,t)  - \sin \theta_0 \e_2(\x_0,t),  \\
\c(\x_0,\theta_0,t) & = - \sin \theta_0 \e_1(\x_0,t) - \cos \theta_0 \e_2(\x_0,t). \label{eq:cvector}
\end{align}
\\
 From \cref{eq:avector,eq:cvector}, we remark that $\c=  \partial_{\theta_0} \a$. The $\theta_0$ coordinate is then explicitly introduced by writing the particle velocity as

\begin{equation} \label{eq:velocitydecomposition}
\vi = \U + \cperp,
\end{equation}
\\
with $\U$  and $\cperp \equiv v_\perp \c$ being the gyrophase independent and dependent parts of the particle velocity, respectively (discussed more in the details in \cref{GuidingCentertransformation}). Then, $v_\perp$, the perpendicular velocity measured in the frame moving with the velocity $\U$, is used to define the lowest order adiabatic magnetic moment $\mu_0 = m v_\perp^2 /(2 B)$. Given the velocity decomposition in \cref{eq:velocitydecomposition}, the gyrophase dependent part of the velocity, $\cperp$, can be isolated in the particle one-form $\gamma$ in \cref{eq:particlegamma}. This allows us to perform a perturbative change of coordinates from the $\Z_0$ coordinates to the guiding-center phase-space coordinates denoted by $\Z = \left(  \R, v_\parallel,\mu,  \theta ,t\right)$. The guiding-center coordinates, $\Z$, are constructed such that the dynamics is described by a gyrophase independent guiding-center one-form $\Gamma = \Gamma(\R,v_\parallel,\mu,t)$, removing the gyrophase terms associated with background gradients. As a consequence, the guiding-center magnetic moment $\mu$ is dynamically conserved. We remark that the time coordinate $t$ does not change under the transformations. \corr{We perform this perturbative transformation up to second order in the small parameter $\epsilon$. We note that this is equivalent to carry out an expansion in the $ m/q \sim \Omega^{-1}$ parameter, considered in previous litterature \citep{Northrop1963,kruskal1965,Littlejohn1979,Brizard1995,Hazeltine2003}. Then, we order the terms with respect to the parameter $\epsilon_E$ [see \cref{eq:defepsilon}] up to second order to obtain a suitable guiding-center model for the plasma periphery, considering potentially large amplitude electromagnetic fluctuations present on the $L_P$ scale length \citep{Dimits2012}. We neglect the $\epsilon_\delta$ electromagnetic fluctuations in the guiding-center transformation. The second order accuracy is motivated by the importance of retaining the lowest order FLR corrections and the effects associated with a time-dependent $\E \times \B$ drift.} 

The second perturbative change of coordinates is from the guiding-center to the gyrocenter coordinates, denoted by $\overline{\Z}= \left( \gyR,  \gyvparallel, \gymu,\gytheta,t \right)$. Within this change of coordinates, we consider the presence of small scale electromagnetic fluctuations, $ \phi_1 (\x)$ and $A_{\parallel 1}(\x)$, on the $\epsilon_\perp\sim 1$ scales, and perturb the guiding-center one-form. The conservation of the guiding-center adiabatic magnetic moment $\mu$, violated by the presence of the small scale fluctuations, is retrieved in the gyrocenter magnetic moment $\gymu$, up to second order in the small amplitude parameter $\epsilon_\delta$, by the perturbative construction of $\overline{ \Z}$. As a result, we obtain the gyrophase independent gyrocenter one-form $\overline{\Gamma} = \overline{\Gamma}(\gyR, \gyvparallel, \gymu,t)$ that we use to derive the gyrokinetic second order accurate electromagnetic equations of motion in the plasma periphery of tokamak devices.

\subsection{Lie-Transform Perturbation Theory}
\label{LieTransformPerturbationTheory}

We present the formalism we use to perform the perturbative coordinate transformations, i.e. a perturbation approach known as Lie-transform perturbation theory \citep{Cary1981,Littlejohn1982,Cary1983,Brizard2009}. This formalism let us pass from the coordinates $\z$ and the associated one-form $\gamma = \gamma_\nu d \z^\nu$ to a new set of coordinates $\Z$ with the associated new one-form $\Gamma = \Gamma_\nu d  \Z^\nu$, the two one-forms, $\gamma$ and $\Gamma$, being linked by the $\z$ to $\Z$ transformation. We consider a near-identical coordinate transformation around the small parameter $\epsilon \ll 1$ in the form

\begin{equation} \label{eq:nearidtransformation}
\Z^{\nu} = \phi_+^\nu\left( \z, \epsilon\right) = \sum_{n=0}^{\infty}  \frac{\epsilon^n}{n!} \frac{\partial^n  \phi_+^\nu\left( \z, \epsilon\right)   }{\partial \epsilon^n} \bigg\rvert_{\epsilon =0} ,
\end{equation}
\\
where $\phi_{+}^\nu = \phi_+^{\nu} \left( \z, \epsilon\right) $ is the mapping function that specifies the coordinate transformation, such that $\phi_+^\nu(\z,0) = \z^\nu$. In \cref{eq:nearidtransformation}, the function $\phi_+^\nu$ transforms the coordinates $\z$ to the new coordinates $\Z$, given $\epsilon$. Indeed, the coordinates $ \Z$ are the values of the function $\phi_+^\nu$ evaluated at $(\z,\epsilon)$. Symmetrically, we can define the inverse transformation of \cref{eq:nearidtransformation} by introducing the mapping function $\phi_{-}^\nu( \Z, \epsilon) $, such that

\begin{equation} \label{eq:inversenearidtransformation}
\z^\nu = \phi_{-}^{\nu}( \Z, \epsilon).
\end{equation}
\\
Our perturbation theory is built on the framework given by \cref{eq:nearidtransformation,eq:inversenearidtransformation}, and the Lie-transform is a special case of \cref{eq:nearidtransformation} where the function $\phi_+^\nu$ is specified by introducing a generating function, $g^\nu$, and asking that $\phi_+^\nu$ is solution of

\begin{equation} \label{eq:defgeneratingfunctions}
\frac{\partial \phi_+^\nu}{\partial \epsilon}(\z, \epsilon) = g^\nu(\phi_+^\nu(\z, \epsilon)).
\end{equation}
\\
We remark that \cref{eq:defgeneratingfunctions} is a functional relation since both sides are evaluated at $(\z, \epsilon)$. An equation for $\phi_-^\nu$ can be obtained by taking the derivative with respect to $\epsilon$ on both sides of \cref{eq:inversenearidtransformation} with $\Z^\nu = \phi_+^\nu(\z,\epsilon)$ and using \cref{eq:defgeneratingfunctions} yielding

\begin{equation} \label{eq:defgeneratingfunctionsinverse}
\frac{\partial \phi_-^\nu}{\partial \epsilon} = - g^\lambda \frac{\partial \phi^\nu_-}{\partial  \Z^\lambda},
\end{equation}
\\
being $d  \z^\nu / d \epsilon =0$. We now deduce the transformation rule of scalar functions induced by a Lie-transform specified by \cref{eq:defgeneratingfunctions}. Let $f$ be a scalar function of the coordinates $ \z$ and $F$ a scalar function of the new coordinates $\Z$. The scalar invariance requires that $f( \z) = F( \Z)$. More precisely, since the coordinate transformation in \cref{eq:nearidtransformation} depends explicitly on $\epsilon$, the function $F$ should also depend explicitly on $\epsilon$. Thus, we write 

\begin{equation} \label{eq:scalarinvariance}
F( \Z,\epsilon) = f( \z).
\end{equation}
\\
Taking the derivative with respect to $\epsilon$ of \cref{eq:scalarinvariance}, while noticing that $d f / d \epsilon = 0$, and using \cref{eq:defgeneratingfunctions}, we derive

\begin{equation} \label{eq:partialFpartialepsilon}
\frac{\partial F}{\partial \epsilon} = - g^\nu \partial_\nu F \equiv - \L_{g} F,
\end{equation}
\\
where we introduce the Lie-derivative $\L_{g} \equiv g^\nu  \partial_\nu$ applied to scalar functions. Since \cref{eq:partialFpartialepsilon} is a functional relation, it can be evaluated both at $\z$ and $\Z$. Indeed, the differential operator $\partial_\nu$ acting on $F$ is defined by 

\begin{equation}
\partial_\nu F =  \begin{cases}
\dfrac{\partial F(\Z)}{\partial \Z^\nu} \\	
\\
\dfrac{\partial F(\z)}{\partial \z^\nu}.\end{cases}
\end{equation}
\\
Expanding $F( \Z, \epsilon)$ around $\epsilon$, using \cref{eq:partialFpartialepsilon} recursively and the fact that $F(\z,0)  = f(\z)$, the functional relation between $F$ and $f$ under the transformation in \cref{eq:nearidtransformation} can be found, i.e.

\begin{equation} \label{eq:pushforward}
F = e^{ - \epsilon\L_{\g} } f.
\end{equation}
\\
The inverse relation follows directly from \cref{eq:defgeneratingfunctionsinverse}, i.e. 

\begin{equation}\label{eq:pullback}
f= e^{\epsilon \L_{\g} } F.
\end{equation}
\\
We emphasize again that \cref{eq:pushforward,eq:pullback} are relations between functions, and, therefore, can be evaluated at both $ \z$ and $\Z$. We also remark that $e^{\epsilon  \L_{g}}$ indicates a linear differential operator. We refer to \cref{eq:pushforward} as push-forward transformation, and to \cref{eq:pullback} as pull-back transformation \citep{Brizard2007,Brizard2009}. 

\Cref{eq:pushforward,eq:pullback} allow us to derive the functional form of the coordinate transformation in \cref{eq:nearidtransformation}, which is specified by \cref{eq:defgeneratingfunctions}. With the particular choice of scalar functions $F =  \phi_-^\nu$ and $f = I^{\nu}$ [$I^{\nu}$ is the coordinate function, such that $I^\nu(\z) = \z^\nu =  \phi_-^\nu( \Z, \epsilon )$], the push-forward transformation in \cref{eq:pushforward} evaluated at $ \Z$ yields

\begin{equation} \label{eq:inversecoordinatetransformation}
\z^{\nu} = e^{-\epsilon\L_{g}}  \Z^\nu.
\end{equation}
\\
The inverse coordinate transformation of \cref{eq:inversecoordinatetransformation} follows directly from the pull-back transformation in \cref{eq:pullback} with, in particular, $f = \phi_+^\nu$ and $F = I^\nu$ evaluated at $\z$, i.e.

\begin{equation} \label{eq:Znuexpznu}
     \Z^\nu = e^{\epsilon\L_{g}} \z^\nu.
\end{equation}

We now derive the transformation rule of a one-form [e.g., $\gamma$ in \cref{eq:particlegamma}] under the coordinate transformation in \cref{eq:nearidtransformation}. From the invariance $\Gamma_\nu d  \Z^\nu = \gamma_\nu d \z^\nu$, the components of $\Gamma$ transform as components of a covariant vector, 

\begin{equation} \label{eq:invarianceGammamu}
\Gamma_\nu( \Z,\epsilon) = \frac{\partial \phi_-^\lambda }{\partial  \Z^\nu} ( \Z,\epsilon)\gamma_\lambda\left( \phi^\nu_- ( \Z , \epsilon)\right)\corr{ + \partial_\nu S(\Z)},
\end{equation}
\\
with $S$ a gauge function. The gauge function $S$ reflects the invariance of the action $\mathcal{A} = \int  \Gamma$ under the addition of a total derivative. Evaluating the derivative with respect to $\epsilon$ on both sides of \cref{eq:invarianceGammamu}, using \cref{eq:defgeneratingfunctionsinverse}, and finally expanding $\Gamma_\nu( \Z,\epsilon)$ around $\epsilon$, we find the following functional relation

\begin{equation} \label{eq:pushforwardoneform}
\Gamma_\nu = e^{- \epsilon\L_{g}} \gamma_\nu  + \partial_\nu S,
\end{equation}
\\
with $\L_{g}$ the Lie-derivative acting on a one-form $\gamma$ being defined by

\begin{equation} \label{eq:Lieoneform}
 \left( \L_{g} \gamma \right)_\nu  = g^\lambda \left(\partial_\lambda \gamma_\nu  - \partial_\nu \gamma_\lambda \right).
\end{equation}
\\
Emphasis is made here on the fact that the Lie-derivative in \cref{eq:Lieoneform} is not equivalent to the one in \cref{eq:pushforward}, since they act on different mathematical objects.

When treating the single particle dynamics, a series of change of coordinates are performed in order to remove the gyrophase dependent parts of the particle dynamics at each $n$th-order in the expansion in the form of \cref{eq:pullback} \citep{Dragt1976,Cary1981}. In this case, the pull-back transformation in \cref{eq:pullback} can be carried out by using a perturbation approach and can be written as a composition of individual Lie-transforms, i.e.

\begin{equation} \label{eq:fTepsilonF}
f  \equiv T_{\epsilon} F = \prod_{n=1}^{\infty} e^{ \epsilon^n  \mathcal{L}_{n} } F,
\end{equation}
\\
where we introduce $\L_{n} \equiv \L_{g_n}$ as a shorthand notation for the Lie-derivative associated with the generating function $g_n^\nu$. Similarly to \cref{eq:pushforward}, the inverse transformation of \cref{eq:fTepsilonF} is

\begin{equation} \label{eq:FTepsilonf}
F = T_{\epsilon}^{-} f = \prod_{n=1}^{\infty} e^{- \epsilon^n \mathcal{L}_n} F.
\end{equation}
\\
From \cref{eq:fTepsilonF} with $T_\epsilon = e^{\epsilon \mathcal{L}_{1}} e^{\epsilon^2 \mathcal{L}_{2}} \dots$, the second order accurate pull-back transformation becomes

\begin{equation} \label{eq:pullbacksecondorder}
f   = F + \epsilon g_1^\nu \partial_\nu F + \epsilon^2 \left(  \frac{1}{2} g_1^\lambda \partial_\lambda \left(  g_1^\nu \partial_\nu F \right)+    g_2^\nu \partial_\nu F  \right) + O(\epsilon^3).
\end{equation}
\\
In particular, with $f = \phi_+^\nu$ and $F = I^\nu$ evaluated at $\z$ in \cref{eq:pullbacksecondorder}, the second order accurate coordinate transformation is 

\begin{equation} \label{eq:Znu}
  \Z^{\nu}  = \z^{\nu} + \epsilon g_1^\nu(\z) + \epsilon^2 \left( \frac{1}{2}g_1^\lambda(\z) \partial_\lambda g_1^\nu(\z)  + g_2^\nu(\z) \right) + O(\epsilon^3).
\end{equation}
\\
At the same time, the hierarchy giving the functional relation between the one-form $\Gamma = \sum_{n} \Gamma_n$ and the one-form $\gamma = \sum_n \gamma_n$ is

\begin{subequations}
\label{eq:systemorderbyorder}
\begin{align}
\Gamma_0 &= \gamma_0 + d S_0, \label{eq:systemorderbyorder0}\\
\Gamma_1 &= \gamma_1 - \L_1 \gamma_0 + d S_1 , \label{eq:systemorderbyorder1} \\
\Gamma_2 &= \gamma_2 - \L_1 \gamma_1  + \left( \frac{1}{2} \L_1^2 - \L_2\right) \gamma_0 + d S_2, \label{eq:systemorderbyorder2} \\
\Gamma_3 & = \gamma_3 - \L_1 \gamma_2 - \L_3 \gamma_0 - \L_2 \Gamma_1 + \frac{1}{3} \L_1^2\left( \gamma_1 + \frac{1}{2} \Gamma_1 \right) + d S_3,  \label{eq:systemorderbyorder3}\\
\vdots \nonumber 
\end{align}
\end{subequations}
\\
 In \cref{eq:systemorderbyorder}, the Lie-derivatives act on one-forms and are, therefore, defined by the relation in \cref{eq:Lieoneform}.
 
 In the following, we use the Lie-transform perturbation theory and solve the hierarchy in \cref{eq:systemorderbyorder} to obtain successively the gyrophase independent guiding-center one-form $\Gamma(\R,v_\parallel,\mu,t)$ and the gyrocenter one-form $\overline{\Gamma}(\gyR,  \gyvparallel,\gymu,t)$ starting from the fundamental particle one-form $\gamma(\x,\vi,t)$ in \cref{eq:particlegamma}. The two transformations are systematically obtained up to second order in $\epsilon$ and $\epsilon_\delta$, respectively. The inherent degrees of freedom in choosing the generating functions $g^\nu_n$ allow for different expressions of $\Gamma$ and $\overline{\Gamma}$ depending on the desired properties of the one-forms \citep[see, e.g.,][]{Miyato2009,Madsen2010}.

\subsection{Guiding-Center Transformation}
\label{GuidingCentertransformation}
In the first coordinate transformation $\Z = T_{\epsilon} \Z_0$ with $\bm{\Z}_0 = (\x_0, v_{\parallel 0},\mu_{0},  \theta_0,t) $ and $\Z = \left(  \R, v_\parallel,\mu,  \theta ,t\right)$, we use the Lie-transform method up to second order in \corr{$\epsilon$} aiming to obtain the gyrophase independent guiding-center one-form $\Gamma = \Gamma(\R,v_\parallel,\mu,t)$. Within this transformation, we consider the presence of \corr{potentially large amplitude drift-kinetic} electromagnetic fluctuations, i.e. $\phi_0$ and $A_{\parallel 0}$, while neglecting \corr{the gyrokinetic components $\phi_1$ and $A_{\parallel 1}$ following a two-step derivation \citep[see, e.g,][]{Brizard1995,Hahm1996,Hahm2009}. We remark that the consistency of the ordering is maintained up to second order as discussed in \cref{GyrocenterTransformation}. The present derivation follows closely the ones in \citet{Brizard1995,Madsen2010}, but differs in the functional forms of the symplectic components of $\Gamma$ at second order [see \cref{appendixGC}].}

For the equations of motion to be gyrophase independent, we require that

\begin{equation} \label{eq:gctransformationrule1}
    \frac{\partial \mathcal{H}_0}{\partial  \theta} = \frac{\partial \Gamma_{\R} }{\partial \theta } =  \frac{\partial \Gamma_\parallel }{\partial \theta }  = 0.
\end{equation}
\\
being $\mathcal{H}_0$ the guiding-center Hamiltonian, $\Gamma_{\R}$ and $\Gamma_{\parallel}$ the $ \R$ and $ v_\parallel$ symplectic components of the one-form $\Gamma$, respectively. Besides removing the gyrophase dependence from the equations of motion, we require the guiding-center magnetic moment, $\mu$, to be dynamically conserved, i.e. $\dot{\mu} =0$. The Euler-Lagrange equations of motion show that this can be obtained by imposing the following sufficient conditions in addition to \cref{eq:gctransformationrule1}, 

\begin{equation} \label{eq:gctransformationrule2}
    \frac{\partial \Gamma_\theta}{\partial \R }  = \frac{\partial \Gamma_\theta}{\partial v_\parallel}  = \frac{\partial \Gamma_\theta}{\partial t} = 0,
\end{equation}
\\
while $\partial_\mu \Gamma_\theta \neq 0$ in general. In the following, we choose the guiding-center generating functions such that the conditions in \cref{eq:gctransformationrule1,eq:gctransformationrule2} are satisfied at all the considered orders. We note that the physical time $t$ is not transformed, such that \cref{eq:Znu} imposes that  $g_{1}^t =g_2^t=0$. We also note that the same conditions in \cref{eq:gctransformationrule1,eq:gctransformationrule2} are used in \citet{Miyato2009} and in \citet{Madsen2010}.   

We first express $\gamma$ in \cref{eq:particlegamma} in the preliminary coordinates $  \Z_0 = (\x_0,v_{\parallel 0},\mu_{0},  \theta_0,t)$ separating the gyrophase dependent and independent parts of the particle velocity, i.e. $\U$ and $\cperp$, respectively. Thus, we write $\vi = \U + \cperp$ [see \cref{eq:velocitydecomposition}], imposing \citep{Littlejohn1981,Brizard1995,Hahm1996,Qin2006,Hahm2009,Madsen2010,Jorge2017},

\begin{equation} \label{eq:U}
\U = \u_E + v_{\parallel 0} \b.
\end{equation}
\\
With respect to standard gyrokinetic formalisms valid in the core, the inclusion of the $\E \times \B$ drift in $\U$ is due to the fact that, in the plasma periphery, strong time-dependent sheared radial electric fields can be present with a time-averaged perpendicular scale length $L_\phi \sim L_P$. The careful analysis of the role of these sheared electric fields is of special interest for stability and transport studies in the plasma periphery, especially when the L-H mode transition occurs. The Lagrangian one-form $\gamma$ in \cref{eq:particlegamma} can, therefore, be written as $\gamma = \gamma_0 + \gamma_1$, where $\gamma_1 \sim O(\epsilon \gamma_0)$ \citep{Northrop1963,kruskal1965,Littlejohn1979,Hazeltine2003}, with

\begin{subequations} \label{eq:gamma0gamma1}
	\begin{align}
    \gamma_{0} &=  q \A \cdot d \x_0 - q \phi_0 d t, \label{eq:gamma0}\\
    \gamma_{1} & = \left[ m \U + m \cperp \right] \cdot d \x_0 - \left[\frac{m}{2} v_{\parallel_0}^2 + \corr{\frac{m }{2 }u_E^2}+ \mu_0 B + m \U \cdot \cperp \right] d t\label{eq:gamma1}.
	\end{align}
\end{subequations}
\\
being $\A = \hat{\bm A} + \hb A_{\parallel 0}$. \corrs{We remark that, in writing \cref{eq:gamma0gamma1}, no ordering in the $\E \times \B$ drift is considered. This assumption is made at this point of the derivation in order to include the polarization drift along with the magnetic drifts \citep{Wimmel1984} and the important FLR corrections in the background fields. We show then that the second order accurate guiding-center model is appropriate to describe situations when, in particular, $|\u_E|/c_s \sim \epsilon_E$, relevant for the plasma periphery.} We also note that the terms in \cref{eq:gamma0,eq:gamma1} proportional to $m \cperp$ and $m  \U \cdot \cperp$ are gyrophase dependent. 

We now solve the hierarchy in \cref{eq:systemorderbyorder} and \cref{eq:Znu} to obtain the guiding-center one-from $\Gamma$ and the guiding-center coordinates $\Z$, respectively. Using the fact that \cref{eq:systemorderbyorder} are functional relations, we evaluate them at the guiding-center coordinates $\Z$, and specify later the analytical expressions of $\Z$ using \cref{eq:Znu}. Choosing $S_0 =0$, the zeroth-order guiding-centre one-form $\Gamma_{0}$ is 

\begin{equation} \label{eq:Gamma0}
\Gamma_{0} = q \A  \cdot d \R - q \phi_0 d t.
\end{equation}
\\
Using the vectorial identity $\grad \A - (\grad \A)^T \equiv \bm \epsilon \cdot \B$ with $\bm \epsilon$ the Levi-Cevita tensor, $(\bm \epsilon \cdot \B)_{ij} = \epsilon_{ijk} B_k$, $\B = \grad \times \A$ [see \cref{eq:totalB}], and $\grad  = \grad_\perp + \b \grad_\parallel \equiv \partial / \partial \R \equiv \grad_{\R}$, we derive the equations of motion $\dot \R =   \E \times \B  / B^2$ with $\E = - \grad \phi_0 - \partial_t \A$ from the Euler-Lagrange equations. This corresponds to the zeroth-order $\E \times \B$ drift of the guiding-center $\R$. \corr{Here, all the quantities are evaluated at $\R$, i.e. $\phi_0 = \phi_0(\R,t)$, $\hat{\A} = \hat{\A}(\R,t)$, and $A_{\parallel 0}= A_{\parallel 0}(\R,t)$}, and notice that $\phi_0$ and $A_{\parallel 0}$ are velocity-independent. Using the Lie-derivative in \cref{eq:Lieoneform} and \cref{eq:systemorderbyorder1}, we obtain the first order guiding-center correction,

\begin{align} \label{eq:Gamma1}
\Gamma_{1} & = \left[ m \bm{U}  + m \cperp + q
\g_1^{\R} \times \B + \grad S_1\right] \cdot d \R \nonumber \\  
& - \left[ \frac{m}{2}v_{\parallel }^2 + \frac{m}{2} u_E^2
  +    \mu B + m \U  \cdot \cperp + q \g_{ 1 }^{\R} \cdot \E  +
  \frac{\partial}{\partial t } S_1  \right] dt.
\end{align}
\\
Choosing $S_1 =0$ to remove one degree of freedom, we cancel the gyrophase dependent terms in \cref{eq:Gamma1}, as required by \cref{eq:gctransformationrule1}, by choosing the first order generating function $\gR_{1}$ to be 
 
\begin{equation} \label{eq:G1Rrhoa}
     \gR_{1  } =  \frac{v_\perp}{\Omega} \b \times \c \equiv- \bm \rho,
\end{equation}
\\
with $\bm \rho \equiv  \bm \rho(\R,\mu,\theta)$ the rotating $\vperp$-dependent gyroradius vector [see \cref{eq:avector}]. By inserting \cref{eq:G1Rrhoa} into \cref{eq:Gamma1}, $\Gamma_1$ reduces to 

\begin{equation}  \label{Gamma1final}
\Gamma_{1} =  m \U  \cdot d \R- \left[ \frac{m}{2} v_{\parallel}^2 + \frac{m}{2} u_E^2  + \mu B \right] d t.
\end{equation}
\\
The calculation of the  Lie-transform up to $O(\epsilon)$ is detailed in \cref{appendixGC}. From \cref{eq:systemorderbyorder2}, one deduces that

\begin{equation} \label{eq:Gamma2}
    \Gamma_2 = - \frac{ \mu B }{\Omega } \bm T \cdot d \R  + \frac{\mu B}{\Omega} d \theta + \left[  \frac{B \mu}{\Omega} S - \frac{B \mu}{2 \Omega} \b \cdot \grad \times \U    \right] dt,
\end{equation}
\\
where $S = \a \cdot \partial_t \c = \e_1 \cdot \partial_t \e_2  $ and $\bm{T} = (\grad \c) \cdot \a = \grad \e_2 \cdot \e_1$ are the gyrogauge field vectors introduced by \citet{Littlejohn1988}. \corr{While these two quantities are gyrophase independent, they dependent on the relative orientation of the vector basis $(\e_1,\e_2,\b)$. Indeed, performing a rotation of $(\e_1, \e_2, \b)$ in a plane perpendicular to $\b$ by a shift $\alpha(\R,t)$, such that $(\e_1, \e_2, \b) \mapsto (\e'_1, \e'_2, \b)  $ linked by $\theta \mapsto \theta' = \theta + \alpha(\R,t)$, the $\R$ and $t$ components of an arbitrary one-from, $\gamma$, transform according to $\gamma_{\R}' = \gamma_{\R} - \gamma_\theta \grad \alpha$ and $\gamma_t' = \gamma_t - \gamma_\theta \partial_t \alpha$, with $\grad \alpha = \bm T' - \bm T$ (and $\partial_t \alpha = S' -S$). Here, $\gamma'$ is the one-from defined with respect to $(\e'_1, \e'_2, \b)$. From the invariance of the one-form, we deduce that $\gamma_{\R} = \overline{\gamma}_{\R} - \gamma_\theta \bm T$ (and $\gamma_t = \overline{\gamma}_t - \gamma_\theta S $) to ensure that $\gamma_{\R}' = \overline{\gamma}_{\R}- \gamma_\theta \bm T'$ (and $\gamma_t' = \overline{\gamma}_t - \gamma_\theta S'$) where $\overline{\gamma}_{\R}$ and $\overline{\gamma}_t$ are the gyrogauge invariant parts of $\gamma_{\R}$ and $\gamma_t$, respectively. While \citet{Littlejohn1988} shows that one can make $\bm T$ vanish along a specific trajectory, it cannot, in general, be chosen to be zero over a finite region. However, Since the $\bm T$ and $S$ terms enters in the equations of motion as higher order terms, we neglect them and only keep the gyrogauge invariant parts of $\Gamma_2$ at second order that includes the lowest order FLR corrections in the fields, i.e.}

\begin{equation}
\Gamma_2 = \frac{\mu B}{\Omega} d \theta - \frac{B \mu}{2 \Omega} \b \cdot \grad \times \U dt.
\end{equation}
\\
 We notice that the last term in \cref{eq:Gamma2} contains the Ba$\tilde{\text{n}}$os drift, proportional to $ \mu v_\parallel B \b \cdot \grad \times \b / 2 \Omega$ \citep{Banos1967}. This term appears in the symplectic component of the guiding-center one-form derived by \citet{Madsen2010}, while it is neglected in the guiding-center one form in \citet{Hahm2009}. Indeed, the Ba\~nos can be transferred from the symplectic to the Hamiltonian components by a proper choice of gauge function [see \cref{eq:GammaBrizard}]. Adding the guiding-center corrections $\Gamma_1$ and $\Gamma_2$, given in \cref{eq:Gamma1} and \cref{eq:Gamma2} respectively, to $\Gamma_0$ in \cref{eq:Gamma0},
we obtain the $O(\epsilon^2)$ accurate gyrophase independent guiding-center one-form, $\Gamma = \Gamma_0 + \Gamma_1 + \Gamma_2$, \corr{with $\Gamma_1 \sim O(\epsilon \Gamma_0)$ and $\Gamma_2 \sim O(\epsilon \Gamma_1)$}, which can be written as

\begin{equation} \label{eq:GCGamma}
\Gamma \left( \R, v_\parallel, \mu ,t\right) = q \A^*  \cdot d \R +  \frac{\mu  B}{\Omega} d \theta - \mathcal{H}_0 d t.
\end{equation}
\\
Here, the velocity dependent effective vector potential, $\A^*$, is

\begin{equation} \label{eq:Astar}
    \A^* =    \A  +   \frac{m}{q} \U ,
\end{equation}
\\
and the guiding-center Hamiltonian $\mathcal{H}_0$,

\begin{equation} \label{eq:H0}
\mathcal{H}_0 = q \phi_0 +  \frac{m}{2} \vparallel^2 +   \frac{m }{2} u_E^2 +  \mu B +  \frac{\mu B}{2 \Omega} \b \cdot   \grad \times \U. 
\end{equation}
\\
We notice that $\Gamma$ in \cref{eq:GCGamma} is gyrophase independent thanks to guiding-center transformation. The first order accurate guiding-center coordinates $\Z =\left(  \R, v_\parallel,\mu, \theta,t\right)$, which we derive from \cref{eq:Znu}, are

\begin{equation} \label{eq:gccoordinates}
    \begin{aligned}
\R & = \x_0 - \bm \rho + O(\epsilon^2)  ,\\
v_{\parallel} & = v_{\parallel 0}+ O(\epsilon), \\
\mu& = \mu_0 +  O(\epsilon) ,\\ 
\theta& = \theta_0 + O(\epsilon).
\end{aligned}
\end{equation}
\\
 From the definition of the effective vector potential $\A^*$ in \cref{eq:Astar}, we introduce an effective magnetic field $\B^* =\grad \times \A^*$. The Jacobian of the guiding-center coordinate transformation $\mathcal{J}$, such that $d \x d \vi = \mathcal{J} d \R d \vparallel d \mu d \theta$ \citep{Cary2009}, is given by $ \mathcal{J} = B_\parallel^*/ m$  with

\begin{equation} \label{eq:Bparallelstar}
B^*_\parallel = \b \cdot \B^* = B \left( 1 + \frac{\b \cdot \grad \times \u_E}{\Omega} + \frac{v_\parallel \b \cdot \grad \times \b}{\Omega}\right).
\end{equation}
\\
 From \cref{eq:Bparallelstar}, we derive the approximation $B_\parallel^*  \simeq B  + O(\epsilon^2)$. \corr{We now order that the second order guiding-center one-form, $\Gamma$, is appropriate the describe situations where $|\u_E|/c_s \sim \epsilon_E$, that sets the proper amplitude and spatial scale of the fluctuating potential $\phi_0$ in the plasma periphery. In \cref{eq:H0}, we remark the presence of second order terms in $\epsilon_E$, $u_E^2 \sim \epsilon_E^2 c_s^2$ and $\mu B \b \cdot \grad \times \U /(2 \Omega) \sim \epsilon_\perp \epsilon_E q \phi_0 $, which is a second order term with $\epsilon_E \sim \epsilon_\perp \sim \epsilon$. We remark that the $\vparallel$ and $\u_E$ contributions to $\U$ [see \cref{eq:U}] should be retained. Indeed, the $\vparallel$ contribution, that is the Ba\~nos drift, is a second order term. This can be shown by using \cref{eq:totalB}, i.e. $\B \simeq \hB + \grad_\perp A_{\parallel 0} \times \hb$, and developing to find $\mu v_\parallel B \b \cdot \grad \times \b /(2 \Omega) \simeq \mu v_\parallel \hb \cdot [(\hb \cdot \grad) \grad_\perp  A_{\parallel 0} - \hb \grad_\perp^2 A_{\parallel 0}] / (2 \Omega) \sim \epsilon \epsilon_E q \phi_0$ with $\phi_0 \sim c_s A_{\parallel 0}$. The $\E \times \B$ contribution is proportional to $(\mu B / 2 \Omega)\b \cdot \grad \times \u_E \simeq \mu \grad_\perp^2 \phi_0/(2 \Omega)$, and is also the order of $\epsilon \epsilon_E q \phi_0$.} The latter term has a simple physical interpretation, being an FLR correction. Indeed, expanding the potential $\phi_0(\x)$ at the particle position around the guiding-center position $\R$ (according the coordinate transformation $\bm x = \R + \bm \rho$), the averaged potential $\gyaver{\phi_0(\x)} \equiv \int_0^{2 \pi} d \theta \phi_0(\R + \bm \rho)/(2 \pi)$ acting on the particle around its gyro-orbit is $\gyaver{\phi_0(\x)} = \phi_0(\R) + (\mu /2 q \Omega)  \grad_\perp^2 \phi_0(\R) + O(\epsilon^3)$. A similar development can be made with the Ba\~nos term as being FLR correction in $A_{\parallel 0}(\x)$. We also remark that the guiding-center one-form $\Gamma$ in \cref{eq:GCGamma} simplifies to the one used in \citet{Hahm2009} if the Ba\~nos term is neglected, and reduces to the guiding-center model derived in \citet{Jorge2017} if the FLR corrections in the fields are ignored. Finally, we note that the $\Gamma_\theta$ component of $\Gamma$, proportional to $\mu B /\Omega$, ensures $\dot{\mu} = 0$, as shown by direct application of the Noether's theorem \citep{Cary2009}.

\subsection{Gyrocenter Transformation}
\label{GyrocenterTransformation}

In the second coordinate transformation $\overline{\Z} =T_{\epsilon_\delta}\Z $, which maps the guiding-center coordinates $\Z = \left(  \R, v_\parallel,\mu,  \theta,t \right)$ to the gyrocenter coordinates $\overline{\Z }= \left(  \gyR, \gyvparallel,\gymu,  \gytheta ,t \right)$, we perturb the guiding-center one-form $\Gamma$ in \cref{eq:GCGamma} by introducing \corr{small amplitude (of the order of $\epsilon_\delta$) gyrokinetic electromagnetic fluctuations, $ \phi_1= \phi_1 (\x)$ and $  A_{\parallel 1} =A_{\parallel 1}(\x)$, that can vary at all scales, i.e. $\epsilon_\perp \sim 1$}. Since such fluctuations act at the particle position $\x = \R + \bm \rho$, the conservation of the guiding-center magnetic moment, $\mu$, is broken by the $\theta$-dependence reintroduced by $\bm \rho$ contained in the spatial argument of $\phi_1$ and $A_{\parallel 1}$. To retrieve the dynamical conservation of $\mu$, we construct the gyrocenter coordinates, $\overline{ \Z}$, that allows us to obtain the gyrophase independent gyrocenter one-form $\overline{\Gamma} = \overline{\Gamma}_i d \overline{ \Z}^i - \overline{\mathcal{H}}dt$ such that the conditions in \cref{eq:gctransformationrule2} are satisfied. The gyrocenter one-form $\overline{\Gamma}$ is derived by using Lie-transform perturbation theory around the small parameter $\epsilon_\delta$ up to second order. This allows us to describe the single particle dynamics in the plasma periphery in the presence of electromagnetic fluctuations at the particle gyroscale. 
We start by writing the guiding-center one-form $\Gamma$ in \cref{eq:GCGamma} in the gyrocenter coordinates $\gyZ$ using the fact that \cref{eq:systemorderbyorder} are functional relations. Then, we introduce the small-scale electromagnetic fluctuations in the guiding-center description obtaining the one-form $\Gamma + \delta \Gamma$ with $\delta \Gamma \sim O(\epsilon_\delta \Gamma)$ where $\delta \Gamma$ contains the contributions related to $\phi_1$ and $A_{\parallel 1 }$, i.e.  

\begin{align} \delta \Gamma & = q \A_1 \cdot d \x - q \phi_1 d t \nonumber \\
 &  = q \A_1 \cdot \left[ d \gyR + \frac{\partial \overline{\bm \rho}}{\partial \gymu}  d \gymu + \frac{\partial \overline{\bm \rho}}{\partial \gytheta} d \gytheta  \right] - q \phi_1 d t, \label{eq:GCperturbation}
\end{align}
\\
where we use that $\x \simeq \gyR + \overline{\bm{\rho}}$ with $\overline{\bm \rho} = \bm \rho(\gyR,\gymu,\gytheta)$ and neglect $\gygrad \overline{\bm \rho} \cdot d \gyR$ since it is proportional to $\lvert \bm \rho \cdot \gygrad \ln \rho \rvert  \sim |\overline{\bm \rho} \cdot \gygrad \ln \hat{B}| $ being of higher order. We take advantage of the degrees of freedom in the choice of the gyrocenter generating functions, denoted by $\overline{g}^\nu_{1}$ and $\overline{g}_2^\nu$, to impose that only the functional form of the guiding-center Hamiltonian, $\mathcal{H}_0$, is modified by the gyrocenter transformation. In this framework, the gyrocenter symplectic components $\overline{\Gamma}_i$ have the same functional form as $\Gamma_i$ in \cref{eq:GCGamma}, but are evaluated at the gyrocenter coordinates instead, i.e. $ \overline{\Gamma}_i \left( \gyR, \gyvparallel,\gymu ,t\right) =  \Gamma_i \left(  \gyR, \gyvparallel,\gymu  ,t\right)$. This formulation, known as the Hamiltonian representation \citep{Brizard2007,Miyato2011}, imposes that the symplectic components of $\overline{\Gamma}$ vanish at all $\epsilon_\delta^n$ orders with $n\geq 1$, while the gyrophase independence of $\overline{\Gamma}$ requires that $\partial_{\gytheta} \overline{\mathcal{H}} =0$. The Hamiltonian formulation has the advantage that the guiding-center Jacobian, $ B_{\parallel}^* / m$, is free from $\epsilon_\delta$ electromagnetic fluctuations, such that it preserves its guiding-center functional form, i.e. $
B_{\parallel}^*  d \R d \vparallel d \mu d \theta/m  =B_{\parallel}^*  d \gyR d \gyvparallel d \gymu d \gytheta /m$. \corr{We remark that, compared with previous Hamiltonian formulation where the independent velocity variable is the parallel canonical momentum \citep{Hahm1988,Brizard2007}, we use $\gyvparallel$ instead.}

We now solve the hierarchy in \cref{eq:systemorderbyorder} up to second order in $\epsilon_\delta$. From the zeroth-order transformation in \cref{eq:systemorderbyorder0}, we find $\overline{\Gamma}_0 = \Gamma$ with $S_0 =0$ and retrieve the guiding-center dynamics at the lowest order in $\epsilon_\delta$. The first order gyrocenter correction $\overline{\Gamma}_1$, given by \cref{eq:systemorderbyorder1} that is $\overline{\Gamma}_1 = \overline{\Gamma}_0 + \delta \Gamma$, is obtained by computing the Lie-derivative of $\overline{\Gamma}_0$, according to \cref{eq:Lieoneform}. This yields

\begin{equation} \label{eq:overlineGamma1full}
\begin{aligned}
\overline{\Gamma}_{1}& = \left[ q \gbR_1 \times \B^*- m \gbparallel_1 \b + q \A_1 + \gygrad S_1 \right] \cdot d \gyR +
 \left[ m \gbR_1 \cdot \b  + \frac{\partial S_1}{\partial \gyvparallel} \right] d \gyvparallel  \\
& +  \left[
 q \A_1  \cdot   \frac{\partial \bm \rho}{\partial \gytheta} - \frac{m}{q } \gbmu_1 + \frac{\partial S_1}{\partial \gytheta}  \right] d \gytheta +
\left[   q \A_1 \cdot \frac{\partial \bm \rho}{\partial \gymu} + \frac{m  }{q }\gbtheta_1 + \frac{\partial S_1}{\partial \gymu}    \right] d \gymu  \\ 
&+ \left[ - q \phi_1  + \gbmu_1 \gypmu  \overline{\mathcal{H}}_0 + \gbparallel_1 \gypvparallel \overline{\mathcal{H}}_0  + \gbR_1 \cdot  \left( \gygrad \: \overline{\mathcal{H}}_0 + \frac{\partial }{\partial t} \overline{\bm A}^* \right) + \frac{\partial S_1}{\partial t} \right]dt,
\end{aligned}
\end{equation}
\\
where $\gygrad  \equiv \partial / \partial \gyR$. Here, the overline notations $\overline{\bm A^*}$ and $\overline{\mathcal{H}}_0$ indicate that the guiding-center quantities are evaluated at $\left( \gyR, \gyvparallel , \gymu ,t\right)$, i.e. $\overline{\bm A^*} = \bm A^*(\gyR, \gyvparallel,\gymu,t)$ with $ \overline{\mathcal{H}}_0 = \mathcal{H}_0(\gyR, \gyvparallel,\gymu,t)$ defined in \cref{eq:Astar} and in \cref{eq:H0}, respectively. The gyrophase dependent parts of the fluctuations can be isolated by introducing the gyroaverage operator, $\gyaver{\chi}_{\gyR} = \gyaver{\chi}_{\gyR}(\gyR, \gymu, \gyvparallel,t)$ acting on a function \corr{$\chi = \chi(\x, \gymu, \gyvparallel,t)$, defined by 

\begin{align} \label{eq:gyaveroperator}
\gyaver{\chi}_{\gyR} & = \frac{1}{2 \pi} \int_{0}^{2 \pi} d \gytheta \int  d \x \delta(\gyR + \overline{\bm \rho}- \x) \chi(\x,\gymu, \gyvparallel,t)  \nonumber  \\
& = \sum_{i\geq 0} \frac{(\overline{\bm \rho} \cdot \overline{\bm \rho})^i \gygrad_\perp^{2i}}{2^{2i} i! i!}  \chi(\gyR, \gymu, \gyvparallel,t),
\end{align}}
\\
with $\gygrad_\perp^2 \equiv \gygrad \cdot \gygrad_\perp$. In \cref{eq:gyaveroperator}, we performed the Taylor expansion around $\gyR$ and evaluated the $\gytheta$-integral at constant $\gyR$. The gyroaverage operator allows us to separate the gyrophase dependent and independent parts of $\chi$, such that $\chi = \gyaver{\chi}_{\gyR} + \widetilde{\chi}$ where $\widetilde{\chi}$ is defined as the gyrophase dependent part of $\chi$, while $\gyaver{\widetilde{\chi}}_{\gyR} = 0$ by construction. We can therefore write 

\begin{equation} \label{eq:gyaverandwidetilde}
\quad \phi_1 = \gyaver{\phi_1 }_{\gyR} + \widetilde{\phi_1},\quad \A_1 = \gyaver{\A_1}_{\gyR}+ \widetilde{\A_1}.
\end{equation}
\\
Imposing that the symplectic components in \cref{eq:overlineGamma1full} vanish and that the Hamiltonian component remains gyrophase independent, one finds the first order gyrocenter generating functions with $\A_1 = \hb A_{\parallel 1}$,

\begin{equation} \label{eq:GYG1}
\begin{aligned}
\gbR_1 &= - \frac{1}{q B_{\parallel}^*}\b \times  \gygrad S_1  - \frac{\B^*}{m B_{\parallel}^*} \frac{\partial S_1}{\partial \gyvparallel} ,\\
\gbparallel_1 &= \frac{q}{m } A_{\parallel 1} + \frac{\B^*}{m B_\parallel^*}\cdot \gygrad S_1,\\
\gbmu_1 &=  \frac{q}{m} \frac{\partial S_1}{\partial \gytheta}, \\ \\
\gbtheta_1 &= - \frac{q}{m} \frac{\partial S_1}{\partial \gymu},
\end{aligned}
\end{equation}
\\
\corr{where the first order gauge function, $S_1$, is solved iteratively. Indeed, by expanding $S_1 = S_{10} +  S_{11} + \dots$ with $S_{11} \sim O(\epsilon S_{10})$, we derive $S_{10} = q \overline{\widetilde{\Phi_1}} / \Omega$ (with $\overline{ \chi} \equiv \int^{\gytheta} d \gytheta \chi)$ and $S_{11} = - d_{tgc}\overline{S_{10}}$. Here, we introduce the guiding-center convective derivative $d_{tgc} \equiv \partial_t + (\gyvparallel \b + \bm D_\perp) \cdot \gygrad$ with $\bm D_\perp = \b \times \gygrad \: \overline{\mathcal{H}}_0  / (qB)$, and the first order gyrokinetic potential $\Phi_1$, being defined as 

\begin{equation} \label{eq:Phi1}
\Phi_1 = \phi_1 -\gyvparallel A_{\parallel 1}.
\end{equation}
\\
The details of the evaluation of $S_1$ are reported in \cref{appendixGY}.} With \cref{eq:GYG1}, the first order gyrocenter correction, $\overline{\Gamma}_1$ in \cref{eq:overlineGamma1full}, reduces to

\begin{equation} \label{eq:overlineGamma1t}
\overline{\Gamma}_{1} = - q \gyaver{\Phi_1}_{\gyR} dt.
\end{equation}
\\
\corr{The second order perturbation analysis in $\epsilon_\delta$ is carried out by solving \cref{eq:systemorderbyorder2}. We report the details in \cref{appendixGY}. The second order gyrocenter correction, $\overline{\Gamma} = - \overline{\mathcal{H}}_2 dt$, is given by the second order gyrocenter Hamiltonian, 

\begin{align} \label{eq:fullH2}
 \gyaver{\overline{\mathcal{H}}_{2}}_{\R}  &=  \frac{1}{2B} \gyaver{\b \times \gygrad S_1 \cdot \gygrad \widetilde{\Phi_1}}_{\gyR} + \frac{q}{2m} \gyaver{\frac{\partial S_1}{\partial \gyvparallel} \gygrad_\parallel \widetilde{\Phi_1}}_{\gyR} \nonumber \\
& + \frac{q}{2m } \gyaver{ \frac{\partial S_1}{\partial \gyvparallel}  \left(  \frac{d_{gc}}{dt} \widetilde{A_{\parallel 1}} + \Omega \frac{\partial \widetilde{A_{\parallel 1}}}{\partial \gytheta} \right) }_{\gyR}  + \frac{q^2}{2m} \gyaver{A_{\parallel 1}}^2_{\gyR}\nonumber \\
&- \frac{q^2}{2 m } \left( \gyaver{\frac{\partial S_1}{\partial \gytheta}\frac{\partial \Phi_1}{\partial \gymu}}_{\gyR} - \gyaver{\frac{\partial S_1}{\partial \gymu} \frac{\partial \Phi_1 }{\partial \gytheta}}_{\gyR} \right),
\end{align}
\\
where $S_1 \simeq S_{10} +  S_{11}$. In \cref{eq:fullH2}, we notice the presence of mixed terms, i.e., terms that are proportional to products between the large amplitude and small amplitude fluctuations. While it has been recognized that these terms are an important elements in the elaboration of a gyrokinetic models for the plasma periphery \citep{Dimits2010}, we notice that they are absent in \citet{Brizard1995,Hahm1988,Qin2007,Hahm2009}. In these theories, since a large and time-independent potential yielding $\epsilon_E \sim 1$ is considered, such terms should be retained. However, within our ordering, these terms can be shown to be effectively smaller by a factor $\epsilon$ than the leading order one, i.e. the last term in \cref{eq:fullH2} having imposed $S_1 \simeq S_{10}$. Indeed, ordering the guiding-center convective derivative $d_{tgc} \sim \epsilon \Omega$ for $\epsilon_\perp \sim 1$ (since $\omega \sim k_\perp \lvert \u_E \rvert \sim c_s k_\parallel$) and $k_\parallel / k_\perp \sim \epsilon$ implies that the terms proportional to $d_{tgc}$, and $S_{11}$, are smaller by at least a factor $\epsilon$ compared to the leading order terms in \cref{eq:fullH2}. Thus, the leading order second order gyrocenter correction to $\overline{\Gamma}$ is given by}
\corrs{
\begin{align} \label{eq:overlineGamma2}
\overline{\Gamma}_2  =-\gyaver{\overline{\mathcal{H}}_{2}}_{\R} dt& =  \left[ \frac{q^3}{2  m \Omega } \gypmu  \left( \gyaver{\Phi_1^2}_{\gyR} - \gyaver{\Phi_1}_{\gyR}^2 \right) - \frac{q^2}{2  m} \gyaver{A_{\parallel 1}^2}_{\gyR} \right. \nonumber \\ & \left.  - \frac{q^2}{2 m \Omega^2} \gyaver{ \left(\b \times \gygrad\overline{ \widetilde{\Phi_1}} \right) \cdot \gygrad \widetilde{\Phi_1}}  \right] d t,
\end{align}}
\\
where the second order gauge function, $S_2$, is solution of \cref{eq:S2}. 
The corresponding second order gyrocenter generating functions are given by

\begin{equation} \label{eq:GYG2}
\begin{aligned}
\gbR_2 &=    \frac{ \B_{ 1}}{2 B^2 } \overline{\widetilde{A_{\parallel 1}}}  - \frac{1}{m \Omega} \b \times \gygrad S_2 - \frac{\b}{m} \frac{\partial S_2}{\partial \gyvparallel}, \\
\gbparallel_2 &=  \frac{q}{ 2 m \Omega B } \B_{  1} \cdot \gygrad_{\perp} \overline{ \widetilde{ \Phi_1 } } - \frac{q^2}{2 B}  \widetilde{ \Phi_1 }  \gypmu A_{\parallel 1} + \frac{q^3}{2 m  \Omega} \gypmu \overline{ \widetilde{ \Phi_1 } } \gyptheta A_{\parallel 1},  \\
\gbmu_2 &=    \frac{q^3}{2 m^2 \Omega } \overline{\widetilde{A_{\parallel 1}} }\gyptheta \widetilde{A_{\parallel 1}} +\frac{q}{m } \frac{ \partial S_2}{\partial \gytheta},\\
\gbtheta_2 &=  - \frac{q^3}{2 m^2  \Omega } \overline{\widetilde{A_{\parallel 1}}}\gypmu A_{\parallel 1} -\frac{q}{ m } \frac{\partial S_2}{\partial \gymu},
\end{aligned}
\end{equation}
\\
where we have neglected the higher order terms in \cref{eq:GY2}. Evaluating the functional expressions in \cref{eq:overlineGamma1t,eq:overlineGamma2} at the gyrocenter coordinates $\overline{\Z}$, we obtain the gyrocenter one-form $\overline{\Gamma}$, accurate up to $O(\epsilon^2, \epsilon_\delta^2)$, 

\begin{equation} \label{eq:GYGamma}
\overline{\Gamma}\left( \gyR, \gyvparallel , \gymu,t\right)  = q \overline{\bm A^*} \cdot d \overline{\R} + \frac{\overline{\mu} B}{\Omega } d \overline{ \theta} - \overline{\mathcal{H}}  d t,
\end{equation}
\\
where the gyrokinetic Hamiltonian is

\begin{equation} \label{eq:gyH}
\overline{\mathcal{H}} = \overline{\mathcal{H}}_0 + q \gyaver{\Psi_1}_{\gyR}
\end{equation}
\\
with the second order gyrokinetic potential, $\gyaver{\Psi_1}$, given by 

\begin{align} \label{eq:GYpotential}
\gyaver{\Psi_1}_{\gyR} & =  \gyaver{\Phi_1}_{\gyR} + \frac{q}{2  m} \gyaver{A_{\parallel 1}^2}_{\gyR} - \frac{q^2}{2 m \Omega } \gypmu \left( \gyaver{\Phi_1^2}_{\gyR} - \gyaver{\Phi_1}_{\gyR}^2 \right)\nonumber \\
& +\frac{q}{2 m \Omega^2} \gyaver{ \left(\b \times \gygrad  \overline{\widetilde{\Phi_1}} \right) \cdot \gygrad \widetilde{\Phi_1}}.
\end{align}
\\
By using \cref{eq:Znu} with \cref{eq:GYG1}, the $O(\epsilon_\delta)$ accurate gyrocenter coordinates $\overline{\Z}$ are obtained,

\begin{equation} \label{eq:GYcoordinates}
\begin{aligned}
\overline{\R } &= \R  - \frac{1}{\Omega B^*_{\parallel}}\b \times  \gygrad \overline{ \widetilde{ \Phi_1 } } + \frac{\B^*}{B B_{\parallel}^*} \overline{\widetilde{A_{\parallel 1}}} + O(\epsilon_\delta^2,\corr{\epsilon \epsilon_\delta}),\\
 \overline{v}_{\parallel} &= v_{\parallel} + \frac{q }{  m} A_{ \parallel 1}  + O(\epsilon_\delta^2,\corr{\epsilon \epsilon_\delta}) ,\\
\overline{\mu} & = \mu +  \frac{q}{B} \widetilde{ \Phi_1 } + O(\epsilon_\delta^2,\corr{\epsilon \epsilon_\delta}),  \\
   \overline{\theta} &= \theta  -\frac{q^2}{m \Omega} \pmu  \overline{ \widetilde{\Phi_1 } } + O(\epsilon_\delta^2,\corr{\epsilon \epsilon_\delta}).
   \\
\end{aligned}
\end{equation}
\\
The gyrokinetic potential $\Psi_1$ in \cref{eq:GYpotential} is evaluated at the particle position $\x$ expressed as a function of the gyrocenter coordinates $\gyZ$, i.e. $\bm x(\gyZ)$. To express $\x$ in terms of the $\overline{\Z}$ coordinates, we proceed as follows. As a first step, we write the particle position $\x$ as a function of the $\z = (\x,\vi,t)$ coordinates by introducing the coordinate function $I^{\x}(\z)$ such that $I^{\x}(\z) = \x$. To find the functional form of $I^{\x}$ in the gyrocenter phase-space, \cref{eq:FTepsilonf} is used with $f = I^{\x}$. We derive the guiding-center functional form of $I^{\x}$ that is $T_\epsilon^{-} I^{\x}(\Z) =  \R + \bm{\rho} + O(\epsilon^2)$ [see \cref{eq:gccoordinates}]. As a second step, we consider again \cref{eq:FTepsilonf} using now the first order gyrocenter generating functions given in \cref{eq:GYG1}. We obtain the function $ T_{\epsilon_\delta}^- T_{\epsilon}^- I^{\x}$, which gives the particle coordinate evaluated at the gyrocenter coordinates $\gyZ$, i.e.

\begin{equation} 
 T_{\epsilon_\delta}^- T_{\epsilon}^- I^{\x}(\overline{\Z}) = \gyR+ \overline{\bm \rho} - \gbR_1 - \left(\gbmu_1 \frac{\partial \overline{\bm \rho}  }{\partial \gymu} + \gbtheta_1 \frac{\partial \overline{\bm \rho}}{\partial \gytheta} \right) + O(\epsilon^2,\epsilon_\delta^2),
\end{equation}
\\
with $\overline{\bm \rho} = \bm \rho(\gyR,\gymu,\gytheta)$, being the function $\bm \rho$ defined in \cref{eq:G1Rrhoa}, and where we use the fact that $|\gbR_1 \cdot \gygrad \overline{\bm \rho} / \overline{ \rho} | \sim |\overline{\bm \rho} \cdot \gygrad \ln \hat{B}| \sim \epsilon_B$. At the leading order, it is sufficient to approximate 

\begin{equation} \label{eq:overlinex}
\overline{\bm x} \equiv  T_{\epsilon_\delta}^- T_{\epsilon}^- I^{\x}(\overline{\Z})  \simeq \gyR + \overline{\bm \rho} + O(\epsilon_\delta,\epsilon \epsilon_\delta,\epsilon^2,\epsilon_\delta^2),
\end{equation}
\\
in the argument of $\Psi_1$ since the other terms in \cref{eq:overlinex} are higher order corrections. This leads to evaluate $\gyaver{\Psi_1}_{\gyR} \equiv \gyaver{\Psi_1(\gyR + \overline{\bm \rho})}_{\gyR}$ consistently with \cref{eq:GCperturbation} \citep{Brizard1989,Sugama2000}.\\
From the variation of the gyrocenter action $\mathcal{A} = \int \overline{\Gamma} $, we obtain the second order accurate electromagnetic gyrokinetic equations of motion,

\begin{align} \label{eq:EulerLagrange}
  - q \B^* \times \dot{\gyR} - m \b \dot{\gyvparallel}  & = \gygrad \: \overline{\mathcal{H}} + q \frac{\partial }{\partial t} \overline{\A^*},
\end{align}
\\
where \corrs{$\gyvparallel = \b \cdot \dot{\gyR} + \gymu_{\parallel}$ (where we introduce $\gymu_{\parallel} = \gymu B (\b \cdot \gygrad \times \b) /(2 m \Omega)$)}, $\dot{\gymu}=0$. The $\dot{\gyR}$ and $\dot{\gyvparallel}$ equations of motion can be obtained by taking the vector and scalar products of \cref{eq:EulerLagrange} with $\b$ and $\B^*$, respectively. Using the definition of $B_\parallel^*$ in \cref{eq:Bparallelstar} and the fact that $\gyvparallel = \b \cdot \dot \gyR + \gymu_\parallel$, we derive the following gyrocenter equations of motion, 

\begin{align} 
\dot{ \gyR} & = \overline{\U} + \frac{B}{ B^*_{\parallel}\Omega } \b \times \left( \frac{d \overline{\U} }{d t}  + \frac{\gymu}{m} \gygrad B + \frac{\gymu B}{2  m} \frac{\gygrad \left( \b \cdot \gygrad \times \overline{\U} \right)}{\Omega}\right) +  \frac{\b}{B^*_{\parallel}}  \times \gygrad \gyaver{ \Psi_1}_{\gyR} \nonumber \\
&+ \gymu_\parallel \left( \b + \frac{B}{B_\parallel^*} \left[ \gygrad \times \U\right]_\perp \right)
\label{eq:GYdotR} \\ 
 m \dot{\gyvparallel} & = q E_{\parallel} - q \frac{\B^*}{B_{\parallel}^*} \cdot \gygrad \gyaver{ \Psi_1}_{\gyR} + m \u_E \cdot \frac{d \b}{d t}  - \gymu \b \cdot \gygrad B \nonumber  \\ &
 - \frac{m B}{ B_{\parallel}^*} \frac{\left(\gygrad \times \overline{\U} \right)}{\Omega} \cdot \left(  \frac{d \overline{\U} }{d t}   \bigg\rvert_{\perp} + \gymu \gygrad_{\perp} B \right)
  -  \frac{\gymu B\B^*}{2 B^*_{\parallel}} \cdot \frac{\gygrad \left( \b \cdot \gygrad \times \overline{\U} \right)}{\Omega}, \label{eq:GYdotvparallel}   \\  
 \dot{\gymu} & =0, \label{eq:GYdotmu}  \\
\dot \gytheta & =   \Omega + \frac{q^2}{m } \gypmu \gyaver{ \Psi_1}_{\gyR}  +  \frac{1}{2} \b \cdot \gygrad \times \overline{\U}, \label{eq:GYdottheta}
\end{align}
\\
where \corrs{the convective derivative 

\begin{equation}
\frac{d}{d t} = \partial_t  + \overline{\U} \cdot \gygrad,
\end{equation}}
\\
is evaluated with the gyrophase independent particle velocity $\overline{\U} = \u_E + \gyvparallel \b$. We remark that the parallel electric field, $E_\parallel \equiv \E \cdot \b = - \gygrad_\parallel \phi_0 - ( \partial_t \A )\cdot \b$, contains both its electrostatic and inductive parts. In particular, the parallel inductive part is $(\partial_t \A) \cdot \b = (\partial_t \hat{\A}) \cdot \b+ [\partial_t (A_{\parallel 0} \hb )]\cdot \b$, where $\partial_t \hat{\A}$ obeys \cref{eq:dAdt} while $\partial_t A_{\parallel 0}$ follows \cref{eq:lowfrequency}. \corr{We notice also the presence of the ratio of the magnetic field strength $B$ and of the effective magnetic field $\B^*$ to the gyrocenter phase-space volume element $B_\parallel^*$ in the above equations of motion}

\Cref{eq:GYdotR}, which describes the motion of a single gyrocenter in the plasma periphery, includes the polarization drift $ \b \times d_t \overline{\U}/\Omega_a$, the magnetic gradient drifts, e.g. $\gymu\b \times \gygrad B/\Omega_a$, and, finally, a number of transport terms driven by the fluctuations at the particle Larmor radius scale contained in the $\b \times \gygrad \gyaver{\Psi_1}_{\gyR}/B$ term [see \cref{eq:GYpotential}]. In particular, these terms are the perturbed electrostatic $\E \times \B$ drift, proportional to $  \B \times  \gygrad_\perp \gyaver{\phi_1}_{\gyR} /B^2$, the shear-Alfv\'en transport term, proportional to $ \gyvparallel \b \times \gygrad_\perp \gyaver{A_{\parallel 1}}_{\gyR}  /B$, also referred to as the magnetic-flutter velocity \citep{Brizard2007,Hahm2009}, and, finally, a nonlinear electromagnetic term proportional to $\b \times \gygrad_\perp \partial_{\gymu} \gyaver{ \widetilde{\Phi_1}^2}/B$ that drives ponderomotive effects \citep{Brizard2007,Hahm2009,Krommes2012}. \Cref{eq:GYdotvparallel} is the parallel momentum equation that includes the parallel forces associated with the parallel electric field $q E_\parallel$, mirror force $\gymu \gygrad_\parallel B$, and a FLR induced parallel force driven by the gyrokinetic potential proportional to $q \gygrad_\parallel \gyaver{\Psi_1}_{\gyR}$. The dynamical conservation of $\gymu$ is given by \cref{eq:GYdotmu}, whereas \cref{eq:GYdottheta} represents the evolution in time of the gyrocenter gyrophase $\gytheta$, which differs from the guiding-center gyroangle $\theta$ due to torsional effects driven by the small-scale electromagnetic perturbations and the FLR field corrections.
\corr{
We remark that the equations of motion in \cref{eq:GYdotR,eq:GYdotvparallel} contain terms proportional to the gradient of the magnetic field strength, being $B = \lvert \B\rvert$ with $\B \simeq \hB + \gygrad_\perp A_{\parallel 0} \times \b $. This yields a polarization drift, $\b \times d_t \U / \Omega \sim \epsilon^2 \epsilon_E c_s$, that can be of the same order as the $\gygrad B$ drift on the $L_B$ scale, while the $\gygrad B$ drift driven by perpendicular magnetic perturbations, $\delta \B_\perp \simeq \gygrad_\perp A_{\parallel 0} \times \b$, is of the order of $\epsilon^2 c_s$.}

Neglecting the field FLR corrections, proportional to $\b \cdot \gygrad \times \U / \Omega$, the gyrokinetic potential $\gyaver{\Psi_1}$ and the inductive contribution $(\partial_t \A) \cdot \b$, \cref{eq:GYdotR,eq:GYdotvparallel} reduce to the guiding-center equations of motion presented in \citet{Jorge2017} and obtained from a direct gyroaveraging of the particle Lagrangian, with the fields formally expanded around the particle position $\x$. In addition, the equations of motion in \cref{eq:GYdotR,eq:GYdotvparallel} constitute an improvement over previous gyrokinetic theories for the edge region. In particular, they are the generalization of the equations developed by \citet{Hahm2009,Dimits2012,Madsen2013} to consider time dependent large amplitude and scale electromagnetic fields. They also generalize the models derived by \citet{Dimits1992} and \citet{Qin2006} by describing electromagnetic fluctuations at second order.

\section{Gyrokinetic Boltzmann Equation}
\label{GyrokineticBoltzmannEquation}
Having derived the equations of motion of a single gyrocenter in the presence fluctuating electromagnetic fields in the plasma periphery, we can now address their collective dynamics. In the present section, we therefore express the gyrokinetic Boltzmann equation in the gyrocenter phase-space coordinates.

The distribution function $f_a(\x,\vi,t)$ of particles species $a$ (hereafter, we reintroduce the species subscript $a$) obeys the Boltzmann kinetic equation,

\begin{equation} \label{eq:Boltzmannparticlephasespace}
    \frac{\partial }{\partial t} f_a + \dot \x \cdot \grad f_a + \dot \vi \cdot \frac{\partial }{\partial \vi} f_a = C_a(f_a),
\end{equation}
\\
with $C_a(f_a)$ being the collision operator (see \cref{GyrokineticCollisionOperator}). To write \cref{eq:Boltzmannparticlephasespace} in the gyrocenter phase-space coordinates, $\overline{\Z}$, derived in \cref{GyrocenterTransformation}, we introduce the full gyrocenter distribution function $\gyFa(\overline{\Z}) =f_a(\x(\overline{\Z}), \vi(\overline{\Z}) )$ (the time coordinate $t$ is omitted for simplicity), and use the chain rule to express $(\x,\vi)$ in terms of the $\overline{\Z}$ coordinates, that is

\begin{equation} \label{eq:Boltzmann}
\frac{\partial  }{\partial t}  \gyFa(\overline{\Z}) +  \dot \gyR \cdot \gygrad \:   \gyFa (\overline{\Z}) +\dot \gyvparallel \frac{\partial  }{\partial \gyvparallel}  \gyFa(\overline{\Z}) + \dot \gytheta \frac{\partial  }{\partial \gytheta} \gyFa(\overline{\Z})   = C_a(\gyFa(\overline{\Z}) ),
\end{equation}
\\
where the dynamical conservation of the magnetic moment, $\dot{\gymu} =0$, is used. The gyrocenter equations of motion $\dot{\gyR}$, $\dot{\gyvparallel}$ and $\dot \gytheta$ are given by \cref{eq:GYdotR,eq:GYdotvparallel,eq:GYdottheta}. 

In \cref{eq:Boltzmann}, the full gyrocenter distribution function $\gyFa = \gyFa(\overline{\Z}) $ can be written as $\gyFa = \gyaver{\gyFa}_{\gyR}+ \widetilde{\gyFa} $, where $\gyaver{\gyFa}_{\gyR} \equiv \gyaver{\gyFa(\overline{\Z})}_{\gyR} = \gyaver{\gyFa}_{\gyR}(\gyR,\gymu,\gyvparallel)$ and $\widetilde{\gyFa} = \widetilde{\gyFa}(\overline{\Z})$ are the gyrophase independent and dependent parts of $\gyFa$, respectively. The operator $\gyaver{\cdot}_{\gyR}$ is the gyroaverage operator defined in \cref{eq:gyaveroperator}. We remark that no assumption is made of the spatial variation of the gyroaveraged distribution function, i.e.  $\gyaver{\gyFa}_{\gyR}$ is allowed to vary on both $\epsilon_\perp \sim 1$ and $\epsilon_\perp \sim \epsilon$ scales. The evolution equation of $\gyaver{\gyFa}_{\gyR}$ can be obtained by applying the gyroaverage operator to \cref{eq:Boltzmann} yielding

\begin{equation} \label{eq:GyaverBoltzmann}
\frac{\partial }{\partial t} \gyaver{\gyFa }_{\gyR} + \dot \gyR  \cdot \gygrad \gyaver{\gyFa }_{\gyR}+ \
\dot \gyvparallel  \frac{\partial  }{\partial \gyvparallel}  \gyaver{\gyFa }_{\gyR}   = \gyaver{ C_a(\gyFa)}_{\gyR},
\end{equation}
\\
where we use the gyrophase independence of $\dot \gyR$ and $\dot \gyvparallel$. Using the fact that the gyrocenter phase-space volume element, $B_\parallel^* / m_a$ in \cref{eq:Bparallelstar}, is conserved along the gyrocenter trajectories \citep{Brizard2007}, i.e. 
 
 \begin{equation} \label{eq:conservationBstar}
 \frac{\partial }{\partial t} B_\parallel^* + \gygrad \cdot \left( \dot{\gyR} B_\parallel^*     \right) + \frac{\partial }{\partial \gyvparallel} \left(  \dot{\gyvparallel} B_\parallel^* \right) =0,
 \end{equation}
\\
\cref{eq:GyaverBoltzmann} can be written in a conservative form,

\begin{equation} \label{eq:GYconservativeform}
\frac{\partial }{\partial t} \left( B_\parallel^* \gyaver{\gyFa }_{\gyR} \right)+ \gygrad \cdot \left(  B_\parallel^*  \dot \gyR  \gyaver{\gyFa}_{\gyR}   \right)+  \frac{\partial }{\partial \gyvparallel}\left(  B_\parallel^* \dot \gyvparallel \gyaver{\gyFa }_{\gyR}   \right)  =  B_\parallel^*  \gyaver{ C_a(\gyFa)}_{\gyR}.
\end{equation} 
\\
\Cref{eq:GYconservativeform} is a formulation more convenient than \cref{eq:GyaverBoltzmann} to derive the gyro-moment hierarchy equation (see \cref{GyroMomentHierarchy}). Subtracting \cref{eq:GyaverBoltzmann} to the gyrokinetic Boltzmann equation \cref{eq:Boltzmann}, one finds the evolution equation of $\widetilde{\gyFa}$,

\begin{equation} \label{eq:tildeBoltzmann}
\frac{\partial }{\partial t}\widetilde{\gyFa } + \dot \gyR  \cdot \gygrad \: \widetilde{\gyFa}+ \
\dot \gyvparallel  \frac{\partial   }{\partial \gyvparallel}  \widetilde{\gyFa }  + \dot \gytheta \frac{\partial }{\partial \gytheta}  \widetilde{\gyFa } = C_a(\gyFa) - \gyaver{ C_a(\gyFa)}_{\gyR}.
\end{equation}
\\
We notice that $\gyaver{\gyFa}_{\gyR}$ and $\widetilde{\gyFa}$ are coupled through the collision operator $C_a$ in \cref{eq:GyaverBoltzmann,eq:tildeBoltzmann}, since $C_a$ acts on the full gyrocenter distribution function. Aiming to obtain a closed gyrophase independent evolution equation for $\gyaver{\gyFa}_{\gyR}$, we estimate the magnitude of $\widetilde{\gyFa}$ with respect to $\gyaver{\gyFa}_{\gyR}$. This can be done by comparing the leading order terms of \cref{eq:tildeBoltzmann} at the left- and right-hand sides. At the left-hand side, the leading term is $\dot \gytheta \partial  \widetilde{\gyFa} / \partial \gytheta$, since $\dot \gytheta \partial   / \partial \gytheta  \sim  \Omega_a $ while $\partial / \partial t \sim \dot{ \gyvparallel} \partial / \partial \gyvparallel \sim \dot{\gyR} \cdot \gygrad  \sim \epsilon \epsilon_\perp \Omega_i$. At the right-hand side, expanding the full distribution function $\gyFa$ in the collision operator $C_a$ as $\gyFa = \overline{F_{a0}} + \epsilon_\nu \overline{F_{a1}} + \epsilon_\nu^2 \overline{F_{a2}} + \dots $ with $\overline{F_{a0}} =  \gyaver{\gyFa}_{\gyR}$ and using the ordering of the collision frequencies \cref{eq:epsilonnu,eq:nue}, such that $C_{e}(\gyFe) \sim \sqrt{m_i/m_e} (T_i/T_e)^{3/2} \epsilon_\nu \Omega_i \gyFe$ while $C_{i}(\gyFi) \sim \epsilon_\nu \Omega_i \gyFi$, we obtain the following estimate for the electrons, 

\begin{equation} \label{eq:tildeFe}
\frac{\widetilde{\overline{F_{e}}}}{\gyaver{\overline{F_e}}_{\gyR}} \sim \sqrt{\frac{m_e}{m_i}} \left( \frac{T_i}{T_e}\right)^{3/2} \epsilon_\nu \sim  \sqrt{\frac{m_e}{m_i}}   \left( \frac{T_i}{T_e}\right)^{3/2} \epsilon^2,  
\end{equation}
\\
and for the ions,

\begin{equation} \label{eq:tildeFi} 
\frac{\widetilde{\overline{F}_{i}}}{\gyaver{\overline{F}_i}_{\gyR}} \sim  \epsilon_\nu  \sim \epsilon^2.
\end{equation}
\\
As a consequence, up to second order in $\epsilon$ (or $\epsilon_\delta$), the gyrophase dependent part $\widetilde{\gyFa}$ can be neglected in the gyrokinetic Boltzmann equation in \cref{eq:GYconservativeform} for both electrons and ions. Further approximations can be done by noticing that the collision operator $C_a$, associated with the gyrocenter transformation [see \cref{GyrocenterTransformation}], can be expanded in powers of $\epsilon_\delta$, such that $C_{a}(F_a) = C_{a0}(F_a) + \epsilon_\delta C_{a1}(F_a) + \dots$ with $\epsilon_\delta \sim \epsilon$. Because of the ordering of the collision frequencies in \cref{eq:epsilonnu,eq:nue} and up to second order in $\epsilon$ (or $\epsilon_\delta$), we can neglect the gyrokinetic corrections $C_{a1}(F_a)$ and higher, since they are $O(\epsilon_\nu \epsilon_\delta)$. We remark that, while neglecting the $\epsilon_\delta$ corrections in $C_a$, the FLR effects contained in $\gyaver{\gyFa}_{\gyR}$ are retained at arbitrary order in $\epsilon_\perp$ in the collision operator. To conclude, the gyrokinetic Boltzmann equation in \cref{eq:GYconservativeform}, second order accurate in $\epsilon$ and $\epsilon_\delta$, is 

\begin{equation} \label{eq:GYBoltzmannfinal}
\frac{\partial }{\partial t} \left( B_\parallel^* \gyaver{\gyFa }_{\gyR} \right)+ \gygrad \cdot \left(  B_\parallel^*  \dot \gyR  \gyaver{\gyFa}_{\gyR}   \right)+  \frac{\partial }{\partial \gyvparallel}\left(  B_\parallel^* \dot \gyvparallel \gyaver{\gyFa }_{\gyR}   \right)  =  B_\parallel^*  \gyaver{ C_{a0}( \gyaver{\gyFa}_{\gyR})}_{\gyR},
\end{equation} 
\\
which is a closed equation for $\gyaver{\gyFa}_{\gyR}$.

\section{Gyro-Moment Hierarchy}
\label{GyroMomentHierarchy}
In this section, we address the development of a gyrokinetic moment hierarchy, referred to as gyro-moment hierarchy, that we propose as a technique to evolve the gyrokinetic Boltzmann equation, \cref{eq:GYBoltzmannfinal}, that is derived in \cref{GyrokineticBoltzmannEquation}. In the present section, we focus on the collisionless part of the gyrokinetic Boltzmann equation, while the collisional part is the subject of \cref{GyrokineticCollisionOperator}. In \cref{GyroMomentExpansion}, we introduce a velocity-space Hermite-Laguerre decomposition of the gyroaveraged distribution function and relate the coefficients of this expansion to fluid-like quantities. \Cref{GyroMomentHierarchyEquation} describes the gyro-moment hierarchy equation that sets their evolution. The hierarchy is obtained by projecting the gyrokinetic Boltzmann equation onto the Hermite-Laguerre basis. In the process, we retain the parallel and perpendicular phase-mixing terms, i.e. the coupling terms between gyro-moments arising from the parallel and perpendicular drifts and FLR corrections. \corr{As an example of the evaluation of the gyro-moment hierarchy equation, we derive the lowest order equations that describe the evolution of the fluid moments in \cref{AReducedGyroMomentHierarchy}.} Finally, the gyro-moment hierarchy equation is completed in \cref{HermiteLaguerreRepresentationOfGyroaverageOperator} by providing a gyro-moment expansion of the terms that contain the gyroaveraged gyrokinetic potential.
\subsection{Gyro-Moment Expansion}
\label{GyroMomentExpansion}
As a technique to solve the gyrokinetic Boltzmann equation, \cref{eq:GYBoltzmannfinal}, we derive a gyro-moment hierarchy based on decomposing the full gyroaveraged gyrocenter distribution function $\gyaver{\gyFa}_{\gyR}$ onto a complete velocity-space basis provided by the Hermite-Laguerre polynomials. Therefore, we write

\begin{equation} \label{eq:GYFadecomposition}
\gyaver{\gyFa}_{\gyR} = \overline{F}_{Ma}   \sum_{l =0}^{\infty} \sum_{k =0}^{\infty}  \gyN^{lk}  H^{lk}_a,
\end{equation}
\\
where $\gyN^{lk} = \gyN^{lk}(\gyR,t)$ are the expansion coefficients of $\gyaver{\gyFa}_{\gyR}$, hereafter referred to as gyro-moments, and $H_a^{lk}$ are the Hermite-Laguerre basis elements defined by

\begin{equation} \label{eq:Halk}
 \quad H^{lk}_a = \frac{H_l(\sparallel) L_k(\sperp^2)}{\sqrt{2^l l!}}.
\end{equation}
\\
Here, $H_l$ denotes the physicits' Hermite polynomials of order $l$ defined by the Rodrigues' formula

\begin{equation}
    H_l(x) = (-1)^l e^{x^2 } \frac{d^l}{d x^l} e^{-x^2},
\end{equation}
\\
while $L_k$ are the Laguerre polynomials given by the Rodrigues' formula \citep{Abramowitz1974} 

\begin{equation} 
    L_k(x) = \frac{e^{x}}{k!} \frac{d^k}{d x^k} x^k e^{-x}.
\end{equation}
\\
The Hermite polynomials $H_l$ are orthogonal over the interval $] -\infty, \infty[$ weighted by $e^{-x^2}$ such that

\begin{equation} \label{eq:Hermiteorthogonality}
    \int_{-\infty}^\infty d x H_l(x) H_{l'}(x) e^{-x^2} = 2^l l! \sqrt{\pi} \delta^l_{l'},
\end{equation}
\\
with $\delta_{l'}^l$ the Kronecker delta, whereas the Laguerre polynomials $L_k$ are orthogonal over the interval $[0,+ \infty[$ weighted by $e^{-x}$ via the relation

\begin{equation} \label{eq:Laguerreorthogonality}
    \int_0^\infty d x L_k(x) L_{k'}(x) e^{-x} = \delta_{k'}^k.
\end{equation}
\\
In \cref{eq:GYFadecomposition}, we also introduce the gyrocenter shifted Maxwellian distribution function $\overline{F}_{Ma}$ by

\begin{equation} \label{eq:shiftedMaxwellian}
\overline{F}_{Ma} =\gyN \frac{ e^{- \sparallel^2 - \sperp^2}}{ \pi^{3/2} \vthparallel \vthperp^2},
\end{equation}
\\
with $\gyN = \gyN(\gyR,t)$ the gyrocenter density defined as $\gyN = \int d \gymu d \gyvparallel d \gytheta B \gyaver{\gyFa}_{\gyR}/m_a$. The overline notation indicates that the fluid quantities are those associated with the gyrocenters, i.e. defined as moments of gyroaveraged gyrocenter distribution $\gyaver{\gyFa}_{\gyR}$. The velocity variables $\sparallel = (\gyvparallel- \uparallel)/\vthparallel$ and $\sperp^2 = \gyvperp^2 /\vthperp^2  = \gymu B /  \Tperp$ represent the normalized shifted parallel and perpendicular gyrocenter velocity. Indeed, $\uparallel$ is the gyrocenter parallel fluid velocity, while $\vthparallel^2= 2 \Tparallel\
 / m_a$ and $\vthperp^2 = 2 \Tperp/ m_a $ are, respectively, the parallel and perpendicular thermal velocities associated with the parallel and perpendicular temperatures $\Tparallel$ and $\Tperp$. These are defined by $\gyN \Tperp = \int  d \gymu d \gyvparallel d \gytheta B \gyaver{\gyFa}_{\gyR} \gymu B /m_a$ and $\gyN \Tparallel =  \int d \gymu d \gyvparallel d \gytheta B \gyaver{\gyFa}_{\gyR} ( v_\parallel - \uparallel )^2 $, respectively, with the parallel fluid velocity $ \gyN \uparallel = \int d \gymu d \gyvparallel d \gytheta B \gyvparallel \gyaver{\gyFa}_{\gyR} /  m_a $. Our definitions of the perpendicular and parallel thermal speeds motivate the choice of the physicists' Hermite polynomials as basis, being orthogonal with respect to a Maxwellian distribution, instead of the probabilists' Hermite polynomials, used, e.g., in \citet{Mandell2018}, that are orthogonal to a Gaussian function of the form $e^{-x^2/2}$. 
 
 \corr{Since the Hermite-Laguerre polynomials define an orthogonal basis of the space of functions $f$ such that \citep{Wong1998}
 
 \begin{equation} \label{eq:L2cond}
 \int d \gymu d \gyvparallel d \gytheta \frac{B}{m_a} |f|^2 e^{-\sparallel^2 - \sperp^2} < + \infty,
 \end{equation}
 \\
then, any distribution function $\gyaver{\gyFa}_{\gyR}$ that fulfills the requirement given in \cref{eq:L2cond} can be decomposed as in \cref{eq:GYFadecomposition}. We remark that the number gyro-moments that must be kept in the Hermite-Laguerre decomposition in order to provide a good pseudo-spectral representation is directly related to the deviation of the distribution function from Maxwellian. In general, a large number of coefficients is needed if fine velocity structures are present \citep{Schekochihin2016}.} We note that the choice of $\sparallel$ and $\sperp$ as arguments of the basis functions [see \cref{eq:Halk}], provides an efficient representation in both the strong ($\uparallel/\vthparallel \sim 1$) and the weak ($\uparallel / \vthparallel \ll 1$) flow regimes, \corr{ensuring that the Hermite-Laguerre decomposition, in \cref{eq:GYFadecomposition}, is also efficient in the case of distribution functions that do not deviate significantly from a Maxwellian with finite $\gyuparallel$ \citep{Jorge2017}.} 

Using the orthogonality relations in \cref{eq:Hermiteorthogonality,eq:Laguerreorthogonality}, the gyro-moments $\gyN^{lk}$ are evaluated as 

\begin{equation} \label{gyNalk}
\gyN^{lk} = \frac{1}{\gyN}\int  d \gymu d \gyvparallel   d \gytheta \frac{B}{m_a}  \gyaver{\gyFa}_{\gyR}  H^{lk}_a,
\end{equation}
\\
and correspond, indeed, to generalized moments of the full gyroaveraged gyrocenter distribution function $\gyaver{\gyFa}_{\gyR}$, i.e. to fluid-like quantities. To conveniently derive the gyro-moment hierarchy, which describes the spatial and time evolution of $\gyN^{lk}$, we introduce the Hermite-Laguerre projector operator of order $(l,k)$, $\moment{lk}{\cdot}$  applied to a phase-space function $\chi = \chi(\overline{\Z})$, 

\begin{equation} \label{eq:projectorlk}
 \moment{lk}{\chi} \equiv \int  d \gyvparallel  d \gymu d \gytheta \frac{B}{m_a}  \chi \gyaver{\gyFa}_{\gyR}  H^{lk}_a.
\end{equation}
\\
To simplify the notation, we denote the $(l,k) = (0,0)$ Hermite-Laguerre projector operator simply by $\moment{}{\cdot}$ (i.e. $\moment{}{\cdot} \equiv \moment{00}{\cdot}$). The definition in \cref{eq:projectorlk} allows us to define the gyrocenter density $\gyN = \momenta{1}$ and, more in general, $\gyN^{lk} = \moment{lk}{1}/\gyN$ in terms of the Hermite-Laguerre projector operator. Analogously, the gyrocenter perpendicular and parallel pressures can be defined as $\Pperp = \gyN \gyTperp = \moment{}{\gymu B}$ and $\Pparallel = \gyN \Tparallel = m_a  \momenta{ ( \gyvparallel - \uparallel)^2}$ with $\gyN \uparallel = \momenta{\gyvparallel}$, and the parallel and perpendicular heat fluxes as $\overline{Q}_{\parallel a} = m_a \momenta{(\gyvparallel - \uparallel)^3}$ and $\overline{Q}_{\perp a} =  \momenta{\gymu B (\gyvparallel - \uparallel) }$, respectively, yielding the following lowest order expansion coefficients, $\gyN^{01} = \gyN^{10} = \gyN^{20} = 0$, $\gyN^{00}=1$, and, finally, 

\begin{equation}
\gyN^{30} =  \frac{\Qparallel }{\sqrt{3} \hspace{1pt} \Pparallel \vthparallel}, \quad \gyN^{11} = - \frac{\sqrt{2} \:\Qperp}{\Pperp \vthparallel}.
\end{equation}
\\
In the present model, contrary to previous gyro-moment hierarchies \citep[see, e.g.,][]{Beer1996,Snyder2001,Madsen2013} that approximate $B_\parallel^*$ by neglecting the $O(\epsilon)$ terms, we retain the velocity-dependence of $B_\parallel^*/m_a$ [see \cref{eq:Bparallelstar}] in the gyrokinetic Boltzmann equation \cref{eq:GYconservativeform}. For this purpose, we introduce the star Hermite-Laguerre projector operator of order $(l,k)$, defined as

\begin{equation} \label{eq:projectorstarlk}
 \momentstar{lk}{\chi} \equiv  \frac{1}{ \gyN B } \moment{lk}{ B_\parallel^* \chi} =  \frac{1}{\gyN}\int d \gyvparallel  d \gymu d \gytheta   \frac{B^*_\parallel}{ m_a}  \chi \gyaver{\gyFa}_{\gyR}  H^{lk}_a.
\end{equation}
\\
 To simplify the notation also in this case, the $(l,k) =(0,0)$ star Hermite-Laguerre projector operator is denoted by $\momentastar{\cdot} \equiv \momentstar{00}{\cdot}$. Similarly to $\gyN^{lk}$, we define the star gyro-moments $\gyN^{*lk} = \momentstar{lk}{1}$ with, in particular, $\gyN^* \equiv \momentstar{}{1}$. The star gyro-moments $\gyN^{*lk}$ can be expressed in terms of the gyro-moments $\gyN^{lk}$ using the recursive property of the Hermite polynomials $H_{l+1}(x) = 2 x H_l(x)  - 2 l H_{l-1} (x)$ in \cref{eq:projectorstarlk}, obtaining
\corr{
\begin{equation}
 \label{eq:star2nostar}
\gyN^{*lk}  = \frac{\overline{B}_{\parallel a}^*}{B} \gyN^{lk}  + \frac{ \vthparallel \b \cdot \gygrad \times \b }{\sqrt{2}\Omega_a } \left( \sqrt{l+1} \; \gyN^{l+1k} + \sqrt{l} \;\gyN^{l-1k} \right),
\end{equation}
\\
where $\overline{B}_{\parallel a}^* = \overline{\B}_a^* \cdot \b$ with

\begin{equation} \label{eq:Bstarfluid}
\frac{\bB^*_a}{B} = \b + \frac{ \gygrad \times \u_E}{\Omega_a} + \frac{\uparallel   \gygrad \times \b}{\Omega_a},
\end{equation}}
\\
being the normalized effective magnetic field evaluated at $ \gyvparallel = \uparallel$ [see \cref{eq:Bparallelstar}]. By using \cref{eq:star2nostar}, we derive

\begin{subequations} \label{LowestgyNstar1}
    \begin{align}
     \gyN^{*} & = \frac{\overline{B}_{\parallel a}^* }{B},  \\
     \gyN^{*10} & = \frac{\vthparallel \b \cdot \gygrad \times \b}{\sqrt{2} \Omega_a}, \\
     \quad \gyN^{*01}  & = - \frac{\Qperp}{\Pperp} \frac{\b \cdot \gygrad \times \b}{\Omega_a }, \\
     \gyN^{*20} &= \frac{\Qparallel}{\Pparallel} \frac{\b \cdot \gygrad \times \b}{\sqrt{2} \Omega_a}.
    \end{align}
\end{subequations}
As a final remark, we note that Hermite polynomials are a well-known velocity-space basis in plasma physics \citep[see, e.g.,][]{Grant1967,Dorland1993,Hammett1993,Scott2010,Zocco2011,Zocco2015,Schekochihin2016,Adkins2018,Pezzi2019}. On the other hand, the use of Laguerre polynomials is more recent \citep[see, e.g.,][]{Sugama2008,Belli2012,Omotani2015,Zocco2015,Jorge2017,Mandell2018,Jorge2018,Jorge2019}. \corr{In particular, it is motivated for the perpendicular velocity dynamics by their relation with Bessel functions [see \cref{HermiteLaguerreRepresentationOfGyroaverageOperator}].}

\corr{The Hermite-Laguerre decomposition derived in this work presents several advantages over velocity-space grid methods for a numerical resolution of the gyrokinetic Boltzmann equation given in \cref{eq:GYBoltzmannfinal}. For example, velocity-spectral representations allow the evaluation of the $\gymu$- and $\gyvparallel$-derivatives present in the gyrokinetic Boltzmann equation given in \cref{eq:GYBoltzmannfinal} with no truncation error, and ensure a spectral convergence property such that the error decreases exponentially with the number of gyro-moments retained in the case of distribution functions that are not discontinuous, while velocity-space grid methods follow a power law convergence rate \citep{Boyd2001}. A possible equivalence between a velocity-space grid method with resolution $(\Delta \vparallel, \Delta v_\perp)$ and a Hermite-Laguerre decomposition with order $(L,K)$ can be obtained by estimating the smallest velocity-space structures $\delta v_\parallel $ and $\delta v_\perp$ that both methods need to resolve. By balancing the collisional diffusion term, $C \sim \nu / \delta v_\parallel^2$, and the time variation, $\partial_t \sim \omega$, one finds $\Delta v_\parallel \lesssim \delta v_\parallel \sim \sqrt{\nu / \omega} v_{th \parallel}$, while the Hermite order $L$, at which the gyro-moments become negligible because of collisional effects, can be estimated by following the procedure described in \citet{Zocco2011,Loureiro2016,Jorge2018}. From nonlinear perpendicular phase-mixing \citep{Schekochihin2008}, uncorrelated $\delta v_\perp$ structures of the order of $v_{th \perp} / k_\perp \rho$ can develop. Thus, one has $ \Delta v_\perp \lesssim \delta v_\perp \sim v_{th \perp} / k_\perp \rho$, while a rough estimate for the Laguerre order $K$ is $K \gtrsim (k_\perp \rho)^2$ since $k_\perp \rho \simeq \sqrt{2 K}$ is the position of the maximum of the kernel function $\kernel{K}$ [see \cref{HermiteLaguerreRepresentationOfGyroaverageOperator}].}

\subsection{Gyro-Moment Hierarchy Equation}
\label{GyroMomentHierarchyEquation}

The Hermite-Laguerre decomposition  of $\gyaver{\gyFa}_{\gyR}$ in \cref{eq:GYFadecomposition} is the key step to provide an efficient technique to approach the solution of the five-dimensional (including time) gyrokinetic Boltzmann equation \cref{eq:GYBoltzmannfinal}. In fact, the evolution of $\gyaver{\gyFa}_{\gyR}$ can be obtained by solving an infinite set of coupled three-dimensional (and time dependent) equations for the gyro-moment $\gyN^{lk}(\gyR,t)$ and the accuracy of the solution, i.e. the degree of fidelity, is directly related to the number of retained gyro-moments. The gyro-moment equation hierarchy is obtained by multiplying the gyrokinetic Boltzmann equation in \cref{eq:GYBoltzmannfinal} by the Hermite-Laguerre basis element $H_a^{lk}$ and performing the integral over the velocity space. Before proceeding, we highlight the velocity dependence of the gyrokinetic equations of motion in \cref{eq:GYdotR,eq:GYdotvparallel} in terms of $\sparallel$ and $\sperp^2$. Thus, we write

\begin{align} 
\dot \gyR &= \U_{0a} + \frac{\b}{B^*_{\parallel}} \times \gygrad  \gyaver{\Psi_1}_{\gyR}  + \U_{pa}^*   + \sparallel^2 \U_{\curvature a}^* + \sperp^2 \left( \U_{\grad a}^* + \U_{\omega a}^* + \U_{\mu a}^\parallel + \U_{\mu  a}^{\perp *}   \right) \nonumber  \\ 
&  + \sparallel \sperp^2 \left(  \U_{B a}^* +\U_{\mu  a}^{\parallel \perp *} \right) + \sparallel \left( \U_{pa}^{th*} +  \vthparallel \b  \right) , \label{eq:GYdotRvelocityspace} \\
m_a \dot{\gyvparallel} & = F_{\parallel a} - \sperp^2 F_{Ma} + \sparallel F_{pa}^{th} - q_a \frac{\B^*}{B_{\parallel}^*} \cdot \gygrad \gyaver{ \Psi_1}_{\gyR} - \sperp^2 \left( F^*_{\omega a} + \sparallel F_{Ba}^*\right) - m_a \mathcal{A}_a^*,\label{eq:GYdotvparallelvelocityspace}
\end{align}
\\
with the gyrocenter drifts
\begin{equation} \label{eq:GYdrifts}
\begin{aligned}
&\U_{pa}^* = \frac{1}{\Omega_a^*} \b \times \frac{d_0}{d t } \U_{0a} , \\
&\U_{pa}^{th*} = \vthparallel \frac{1}{\Omega_a^*} \b \times \left( \b \cdot \gygrad \u_E + \u_E \cdot \gygrad \b + 2 \uparallel \curvature  \right) ,\\
&\U_{\curvature a}^* = \frac{2 \Tparallel }{m_a} \frac{\b \times \curvature}{\Omega_a^*},\\
&\U_{\grad a}^* = \frac{\Tperp}{m_a B } \frac{\b \times \gygrad B}{\Omega_a^*},  \\
&\U_{\omega a}^* = \frac{\Tperp}{2 m_a \Omega_a} \frac{\b \times \gygrad \left(  \b \cdot \gygrad \times \u_E \right)}{\Omega_a^*},\\
& \U_{Ba }^* = \frac{\Tperp \vthparallel}{2 m_a \Omega_a} \frac{\b \times \gygrad \left( \b \cdot \gygrad \times \b \right)}{\Omega_a^*}, \\
& \U_{\mu  a  }^\parallel = \frac{\Tperp}{2 m_a } \frac{\b \b \cdot \gygrad \times \b}{\Omega_a}, \\
& \U_{\mu  a  }^{\perp *} = \frac{\Tperp}{2 m_a } \frac{ \b \cdot \gygrad \times \b}{\Omega_a^*} \left[ \gygrad \times \u_E \right]_\perp, \\
& \U_{\mu   a }^{\parallel \perp *} =  \frac{\Tperp \vthparallel}{2 m_a } \frac{\b \cdot \gygrad \times \b}{\Omega_a^*} \left[ \gygrad \times \b  \right]_\perp
\end{aligned}
\end{equation}
\\
 and with the parallel forces

\begin{equation} \label{eq:GYparallelforces}
\begin{aligned}
&F_{\parallel a} = q_a E_{\parallel} + m_a \u_E \cdot \frac{d_0\b}{d t }  ,\\
&F_{M a} = \frac{\Tperp}{B} \b \cdot \gygrad B, \\
&F_{pa}^{th} = \vthparallel m_a \b \cdot \left( \frac{\curvature \times \E}{B}  \right) ,\\
&F^*_{\omega a} = \frac{\Tperp }{2 B} \frac{\B^* \cdot \gygrad \left ( \b \cdot \gygrad \times \u_E \right)}{\Omega_a^*}, \\
& F^*_{B a} =  \frac{\Tperp \vthparallel}{2 B} \frac{\B^* \cdot \grad \left( \b \cdot \grad \times \b\right)}{\Omega_a^*},\\
&\mathcal{A}^*_a = \frac{\left( \gygrad \times \U \right)}{\Omega_a^*} \left(  \frac{d}{d t }\U \bigg \vert_{\perp} + \Tperp \sperp^2  \gygrad_{\perp} \ln B \right).
\end{aligned}
\end{equation}
\\
We note that the lowest order convective fluid derivative $d_{0t} = \partial_t + \U_{0a} \cdot \gygrad$, in \cref{eq:GYdrifts,eq:GYparallelforces}, is associated with the gyrocenter fluid velocity $\U_{0a} = \u_E + \uparallel \b$ and that $\Omega_a^*= q_a B_{\parallel}^*/ m_a$ with $B_\parallel^*$ defined in \cref{eq:Bparallelstar}.  

Performing the integral over the velocity space of \cref{eq:GYBoltzmannfinal}, we obtain the gyro-moment equation hierarchy of particle species $a$, describing the spatial and temporal evolution of the gyro-moments $\gyN^{*lk}$,

\begin{equation} \label{eq:GyromomentHierarchyEquation}  
\frac{\partial \gyN^{*lk} }{\partial t }  + \gygrad \cdot \momentstar{lk}{\dot \gyR}  - \frac{\sqrt{2l}}{ \vthparallel}\momentstar{l-1k}{\dot \gyvparallel} + \mathcal{F}_a^{lk} = C_a^{lk},
\end{equation}
\\
with $C_a^{lk}$ the Hermite-Laguerre projection of the collision operator (see \cref{GyrokineticCollisionOperator}), and $\mathcal{F}_{a}^{lk}$ the fluid operator given by 

\begin{align} \label{eq:fluidoperatoralk}
\mathcal{F}_{a}^{lk} & = \frac{d_a^{*lk}}{d t} \ln \left(\gyN \Tparallel^{l/2}  \Tperp^k B^{-k}\right)  + \frac{\sqrt{l(l-1)}}{2} \frac{d_a^{*l-2k}}{d t} \ln \Tparallel \nonumber \\
 &  - k \frac{d_a^{*lk-1}}{d t} \ln \left(   \frac{\Tperp}{B} \right) + \frac{\sqrt{2l}}{\vthparallel} \frac{d_a^{*l-1k}}{d t} \uparallel,
\end{align}
\\
having introduced the gyro-moment convective fluid derivative

\begin{equation} \label{eq:convectivederivative}
\frac{d_a^{*lk}}{d t} = \gyN^{*lk} \frac{\partial }{\partial t} + \momentstar{lk}{\dot \gyR} \cdot \gygrad.
\end{equation}
\\
The fluid operator in \cref{eq:fluidoperatoralk}, $\mathcal{F}_a^{lk}$, defines the evolution of the fluid quantities $\gyN$, $\uparallel$, $\Pperp$ and $\Pparallel$, similarly as for the drift-kinetic model in \citet{Jorge2017}. The Hermite-Laguerre projections of the electromagnetic gyrokinetic equations of motion, appearing in \cref{eq:GyromomentHierarchyEquation}, are

\begin{align}
\momentstar{lk}{\dot \gyR} &= \sum_{p,j} \left[ \left( \U_{pa}\delta_p^l \delta_j^k    + \U_{\curvature a} \phaseV_{pj}^{2lk}  + \left( \U_{\grad a} + \U_{\omega a} + \U_{\mu  a}^\perp \right) \phaseM_{pj}^{lk}  +   \U_{pa}^{th}  \phaseV_{pj}^{lk} \right. \right. \nonumber \\
& \left. \left. +\left( \U_{Ba} + \U_{\mu  a }^{\parallel \perp }  \right) \mathcal{M}_{lj}^{lk} \mathcal{V}_{pj}^{lj}\right) \gyN^{pj}  + \left( \U_{0a} \delta_p^l \delta_j^k   + \vthparallel \b  \phaseV_{pj}^{lk}   +  \U_{\mu a}^\parallel \mathcal{M}_{pj}^{lk}\right) \gyN^{*pj}  \right] \nonumber \\
& + \momentstar{lk}{\frac{\b}{B_\parallel^*} \times \gygrad  \gyaver{\Psi_1 }_{\gyR} } ,\label{eq:momentdotR}  \\
 m_a \momentstar{lk}{\dot{\gyvparallel}} &= \sum_{p,j} \left[  \left( F_{\parallel a} \delta_p^l \delta_j^k  - F_{M a} \phaseM_{pj}^{lk}  + F_{pa}^{th} \phaseV_{pj}^{lk}  \right) \gyN^{*pj}\right. \nonumber \\
&  \left.  -  \phaseM_{pj}^{lk} \momentstar{pj}{F^*_{\omega a}} - \mathcal{M}_{lj}^{lk} \mathcal{V}_{pj}^{lj} \momentstar{pj}{F_{Ba}^*} \right]- m_a \momentstar{lk}{\mathcal{A}_a^*}  - q_a \momentstar{lk}{\frac{\B^*}{B_{\parallel}^*} \cdot \gygrad \gyaver{\Psi_1}_{\gyR}}  , \label{eq:momentdotvparallel} 
\end{align}
\\
where the drifts and forces have equivalent definitions to the ones in \cref{eq:GYdrifts,eq:GYparallelforces}, having replaced all $\Omega_a^*$ with $\Omega_a$. The gyro-moment expansion of $\mathcal{A^*}_a$, $F_{\omega a}^*$ and $F_{B a}^*$ in \cref{eq:momentdotvparallel} are given, respectively, by 

\begin{align}
  \momentstar{lk}{\mathcal{A}_a^*} & = \sum_{p,j} \left[ \left( \frac{\gygrad \times \u_E }{\Omega_a}\delta_p^l \delta_j^k  +  \frac{\gygrad \times \b }{\Omega_a} \mathcal{V}_{apj}^{lk}  \right) \cdot  \moment{pj}{ \frac{d}{d t }\gyU  \bigg \vert_{\perp}}  \right. \nonumber \\
                                 & \left. + 
 \sum_{r,s} \Tperp \curvature \cdot \phaseM_{pj}^{lk} \left(\frac{\gygrad \times \u_E}{\Omega_a}  \delta_r^p \delta_s^j  +\frac{ \gygrad \times \b }{\Omega_a} \mathcal{V}_{ars}^{lk}   \right) \gyN^{rs} \right],
\end{align}
\\
and
\begin{align}
\momentstar{pj}{F^*_{\omega a}} &= \frac{\Tperp }{2 }  \frac{\gygrad \left ( \b \cdot \gygrad \times \u_E \right)}{\Omega_a} \cdot \moment{pj}{\B^*}, \\
      \momentstar{pj}{F_{Ba}}& =  \frac{\Tperp \vthparallel}{2 B} \frac{ \gygrad \left( \b \cdot \gygrad \times \b\right)}{\Omega_a} \cdot \moment{pj}{\B^*},
\end{align}
\\
where

\begin{align}
\moment{pj}{ \frac{d}{d t }\gyU  \bigg \vert_{\perp}} & = \left[ \frac{\partial \u_E}{\partial t  } + \uparallel \frac{\partial \b }{\partial t  }  \right]_\perp \gyN^{pj} +   \sum_{r,s} \vthparallel  \left[  \frac{\partial \b }{\partial t  } +  \b \cdot \grad \u_E + \u_E \cdot \grad \b  \right]_\perp  \phaseV_{rs}^{pj}  \gyN^{rs}  \nonumber \\
 & +\curvature  \sum_{r,s}\mathcal{V}_{ars}^{2lk} \gyN^{rs},
\end{align}
\\
and

\begin{equation}
\moment{pj}{\B^*}  = \overline{\B}_a^* \gyN^{pj}  + \frac{\vthparallel B \gygrad \times \b }{\sqrt{2} \Omega_a} \left( \sqrt{p+1} \: \gyN^{p+1j} + \sqrt{p} \: \gyN^{p-1j}  \right),
\end{equation}
\\
having used \cref{eq:Bstarfluid}. In \cref{eq:momentdotR,eq:momentdotvparallel}, the parallel and perpendicular phase-mixing operators, $\mathcal{V}^{lk}_{pj}$, $\mathcal{V}_{apj}^{lk}$, $\mathcal{V}^{2lk}_{pj}$ and $\mathcal{M}^{lk}_{pj}$, are defined, respectively, by

\begin{subequations} \label{eq:phasemixingops}
\begin{align} 
\mathcal{V}_{pj}^{lk} & = \left( \sqrt{\frac{l+1}{2 }} \delta_{p}^{l+1} + \sqrt{\frac{l}{2}} \delta_{p}^{l-1}\right) \delta_{j}^{k} , \label{eq:phaseV1} \\ 
    \mathcal{V}_{apj}^{lk} &= \gyuparallel \delta_p^l \delta_j^k + \gyvthparallel \mathcal{V}_{pj}^{lk}, \label{eq:phaseVa}\\
\mathcal{V}_{pj}^{2lk} &  = \left( \frac{\sqrt{(l+1)(l+2)}}{2} \delta^{l+2}_p     + (l + \frac{1}{2}) \delta^l_p  + \frac{\sqrt{l(l-1)}}{2} \delta_p^{l-2} \right) \delta_j^k, \label{eq:phaseV2}\\
\mathcal{M}_{pj}^{lk} & = \left( \left( 2k+1  \right) \delta_j^k - k \delta_j^{k-1} - \left(  k+1 \right) \delta_j^{k+1}\right)\delta_{p}^{l}, \label{eq:phaseM}
\end{align}
\end{subequations}
\\
and $\mathcal{V}_{apj}^{2lk} =\sum_{r,s} \mathcal{V}_{ars}^{lk} \mathcal{V}_{apj}^{rs}$. These phase-mixing operators are derived from the recursive properties of the Hermite and Laguerre polynomials, that are $H_{l+1}(x) = 2 x H_l (x) - 2 l H_{l-1}(x)$ and $x L_k(x) =(2k+1) L_k(x) - k L_{k-1}(x) - (k+1) L_{k+1}(x)$. From the definitions in \cref{eq:phaseV1,eq:phaseM}, we see that the phase-mixing operator $\mathcal{V}_{pj}^{lk}$ couples the $(l-1)$ and $(l+1)$ Hermite gyro-moments due to parallel streaming, whereas the $\mathcal{M}_{pj}^{lk}$ couples the $(k-1)$, $k$ and $(k+1)$ Laguerre gyro-moments because of the presence of the curvature and gradient of the magnetic field [see \cref{eq:GYdrifts}]. We remark that, in previously developed gyrofluid models, closure approximations in $\mathcal{V}_{pj}^{lk}$ are used to model the associated linear response of Landau damping, a technique pioneered by \citet{Hammett1990} and \citet{Dorland1993}. Also, toroidal closures to model perpendicular phase-mixing in $\mathcal{M}_{pj}^{lk}$ are provided for these gyrofluid models \citep[see, e.g.,][]{Beer1996,Snyder2001,Madsen2013}. Instead, our model retains the full coupling between gyro-moments in both parallel and perpendicular directions, with the aim of evolving the number of moments necessary to obtain the desired accuracy.

We now turn to the evaluations of the Hermite-Laguerre projections of the FLR induced transport term in \cref{eq:momentdotR}, that is

\begin{equation} \label{eq:FLRtransport}
 \momentstar{lk}{\frac{\b }{B_\parallel^*}\times \gygrad \gyaver{\Psi_1}_{\gyR} } =\frac{1}{\gyN} \frac{\b}{B}\times  \moment{lk}{ \gygrad \gyaver{\Psi_1}_{\gyR}}, 
\end{equation}
\\
and of the FLR induced parallel force present in \cref{eq:GYdotvparallel},

\begin{align}
\label{eq:FLRforces}
   \momentstar{lk}{\frac{\B^*}{B_{\parallel}^*} \cdot \gygrad \gyaver{\Psi_1}_{\gyR} }  &  = \frac{\bB^*_{a}}{B \gyN} \cdot \moment{lk}{  \gygrad \gyaver{  \Psi_1}_{\gyR}} \nonumber \\
   & + \frac{ \vthparallel \gygrad \times \b}{\gyN\sqrt{2} \Omega_a}   \cdot \left( \sqrt{l+1}  \moment{l+1k}{  \gygrad \gyaver{  \Psi_1}_{\gyR}} + \sqrt{l}    \moment{l-1k}{  \gygrad \gyaver{ \Psi_1}_{\gyR}} \right).
\end{align}
\\
\Cref{eq:FLRtransport,eq:FLRforces} show that the Hermite-Laguerre projections of the FLR induced terms are reduced to the evaluation of $\moment{lk}{\gygrad \gyaver{\Psi_1}_{\gyR}}$. The definition of $\Psi_1$ in \cref{eq:GYpotential} and the phase-mixing operators in \cref{eq:phasemixingops} lead to

\begin{align} \label{eq:momentstarlkgygradPsi1}
\moment{lk}{\gygrad \gyaver{\Psi_1}_{\gyR}} &=\sum_{p,j}\left[ \delta_p^l \delta_j^k \moment{pj}{\gygrad \gyaver{\phi_1}_{\gyR}} - \mathcal{V}_{apj}^{lk} \moment{pj}{\gygrad \gyaver{A_{\parallel 1}}_{\gyR}} +\frac{q_a}{2m_a}\delta_p^l \delta_j^k \moment{pj}{\gygrad\gyaver{A_{\parallel 1}^2}_{\gyR}}  \right. \nonumber \\
&\left.- \frac{q_a^2}{ 2 m_a \Omega_a} \left( \delta_p^l\delta_j^k \moment{pj}{\gygrad\frac{\partial}{\partial \gymu} \left(\gyaver{\phi_1^2}_{\gyR}  - \gyaver{\phi_1}_{\gyR}^2 \right)} \right. \right. \nonumber \\
& \left. \left. +  \mathcal{V}_{apj}^{2lk} \moment{pj}{\gygrad \frac{\partial}{\partial \gymu} \left( \gyaver{A_{\parallel 1}^2}_{\gyR} - \gyaver{A_{\parallel 1}}_{\gyR}^2\right)}    \right.   \right.  \nonumber \\
    & \left. \left.  - 2 \mathcal{V}_{apj}^{lk} \moment{pj}{ \gygrad \frac{\partial}{\partial \gymu} \left( \gyaver{\phi_1 A_{\parallel 1}}_{\gyR} - \gyaver{\phi_1}_{\gyR} \gyaver{A_{\parallel 1}}_{\gyR}\right)   }  \right) \right. \nonumber \\
    & \left. + \frac{ q_a}{2 m_a \Omega^2_a} \delta_p^l \delta_j^k  \moment{pj}{\gygrad \gyaver{ \left(\b \times \gygrad \overline{\widetilde{\Phi_1}} \right) \cdot \gygrad \widetilde{\Phi_1}}_{\gyR}} \right].
\end{align}
\\
The first two terms on the right-hand side of \cref{eq:momentstarlkgygradPsi1} are the $O(\epsilon_\delta)$ contributions, whereas the following terms represent the $O(\epsilon_\delta^2)$ nonlinear contributions to the gyrocenter dynamics given by $\phi_1$ and $A_{\parallel 1}$. We note that the first $O(\epsilon_\delta)$ term provides the Hermite-Laguerre projection of the nonlinear $\E \times \B$ advection, proportional to $ \b \times \gygrad \gyaver{\phi_1}_{\gyR}/ B \cdot \gygrad_\perp \gyaver{\gyFa}_{\gyR}$ appearing in \cref{eq:FLRtransport}. As shown in previous gyrofluids models \citep[see, e.g., ][]{Brizard1992,Dorland1993,Beer1996,Madsen2013}, the gyroaverage operator present in this term couples the Laguerre gyro-moments, yielding FLR phase-mixing terms \citep{Schekochihin2008,Tatsuno2009,Schekochihin2009}. 
\corr{
\subsection{Lowest order Moment Equations}
\label{AReducedGyroMomentHierarchy}

The infinite set of coupled fluid equations that the gyro-moment hierarchy leads to is valid for arbitrarily far from equilibrium distribution functions. For practical and numerical application, a finite number of gyro-moments should be retained, still ensuring a proper description of the plasma dynamic and be explicitly evaluated. As an example, we derive here the equations for the six lowest order gyro-moments from the gyro-moment hierarchy equation given in \cref{eq:GyromomentHierarchyEquation}. These equations are obtained by a truncation of the Hermite-Laguerre expansion of $\gyaver{\gyFa}_{\gyR}$ in \cref{eq:GYFadecomposition}, i.e. retaining only the lowest order Hermite-Laguerre polynomials (i.e. $l \leq 3$ and $k \leq1$), yielding a closed set of fluid equations. This system states the evolution of the gyrocenter density $\gyN$, of the parallel gyrocenter fluid velocity, $\uparallel$, of the parallel and perpendicular gyrocenter temperature, $\Tparallel$ and $\Tperp$, and, finally, of the parallel and perpendicular heat fluxes, $\Qparallel$ and $\Qperp$. While the Hermite-Laguerre decomposition is truncated here and, in particular, the gyro-moments $\gyN^{40}$ and $\gyN^{02}$ are neglected, we note that a high-collisional closure to express $\Qparallel$ and $\Qperp$ in terms of $\gyN$, $\uparallel$, $\Tparallel$ and $\Tperp$ can be developed. This can be done by explicit evaluation of the collisional terms. The procedure can be extended to arbitrary collisionality by including a larger number of gyro-moments \citep{Zocco2011,Loureiro2016,Jorge2017}. We leave to a future work the development of closures of the gyro-moment hierarchy.

The fluid equation of the gyrocenter density, $\gyN$, is obtained by setting $(l,k) =(0,0)$ in \cref{eq:GyromomentHierarchyEquation}. This yields 

\begin{align} \label{eq:Na}
 \frac{1}{\gyN }\frac{d^{*0}_a}{d t } \gyN = C_a^{00} -  \frac{\partial}{\partial t} \left( \frac{\overline{B}_{\parallel a}^*}{B} \right) -  \gygrad \cdot \u_a^0,
\end{align}
\\
having defined the lowest order convective derivative, $d_{ta}^{*0} = \gyN^* \partial_t + \u_a^0 \cdot \gygrad $, and

\begin{align} \label{eq:ua0}
\u_a^0 & = \frac{\overline{B}_{\parallel a}^*}{B} \left(   \U_{a0}  + \U_{\mu  a}^\parallel \right) +  \frac{\Tparallel}{m_a} \frac{\tau}{ \Omega_a} \b  + \U_{pa} + \frac{1}{2} \U_{\kappa a} + \U_{\grad a} + \U_{\omega a} + \U_{\mu  a}^\perp  \nonumber \\
& +  \frac{\Qperp  }{\Pperp \vthparallel} \left( \U_{B a} + \U_{\mu a}^{\parallel \perp} + \frac{\vthparallel \tau}{\Omega_a} \U_{\mu a}^\parallel  \right)  +  \frac{1}{N_a B}\b \times \moment{}{\gygrad \gyaver{\Psi_1}_{\gyR}}.
\end{align}
\\
In \cref{eq:ua0} we introduce $\tau = \b \cdot \gygrad \times \b$, where $\b \simeq \hb + \gygrad A_{\parallel 0} \times \hb/\hB$, as a shorthand notation. \Cref{eq:ua0} is indeed the $(l,k) = (0,0)$ Hermite-Laguerre projection of $\dot \gyR$ [see \cref{eq:momentdotR}]. In \cref{eq:Na}, we note the presence of polarization effects due to the time-dependence of $\overline{B}_{\parallel a}^*/B$, which can be written, from \cref{eq:Bparallelstar}, at the leading order as 

\begin{equation} \label{eq:partialtBparallelstar}
\frac{\partial}{\partial t} \left(\frac{\overline{B}^*_{\parallel a}}{B} \right) \simeq\frac{\partial}{\partial t} \left(\frac{\lvert \gygrad_\perp A_{\parallel 0} \rvert}{B} \right) + \frac{\partial}{\partial t}\left[ \frac{\gygrad_\perp^2 \phi_0}{\Omega_a }  - \frac{\gyuparallel \gygrad_\perp^2 A_{\parallel 0}}{\Omega_a} \right].
\end{equation}
\\
In addition, a non-vanishing lowest-order collisional Hermite-Laguerre moment $C_a^{00}$, typically due to finite FLR effects [see \cref{GyrokineticCollisionOperator}], may be present, leading to classical diffusion of the gyrocenter density. Finally, we note that \cref{eq:Na} preserves the positivity of $\gyN$.

The parallel momentum equation describing the evolution of $\uparallel$ is derived by setting $(l,k) = (1,0)$ in \cref{eq:GyromomentHierarchyEquation}, i.e.

\begin{align} \label{eq:Uparallel}
\frac{d_a^{*0}}{d t} \uparallel & = \frac{ \vthparallel}{\sqrt{2}}C_{a}^{10} - \frac{1}{ \gyN \sqrt{2}}  \gygrad \cdot \left( \u_a^{\parallel 1} \gyN \vthparallel \right)  - \frac{1}{m_a \gyN} \frac{\partial}{\partial t} \left(  \frac{\Pparallel  \tau }{ \Omega_a} \right)     -  \momentstar{}{\dot \gyvparallel},
\end{align}
\\
with 

\begin{align} 
\u_a^{\parallel 1} & = \frac{1}{\sqrt{2}} \frac{\Qparallel \U_{\curvature a} }{ \Pparallel  \vthparallel} - \left( \U_{\grad a } + \U_{\omega a} + \U_{\mu a}^\perp \right) \frac{\sqrt{2} \:\Qperp}{\Pperp \vthparallel} + \frac{1}{\sqrt{2}}\U_{pa}^{th} + \frac{1}{\sqrt{2}} \U_{Ba}  + \frac{1}{\sqrt{2}} \U_{\mu  a}^{\parallel  \perp} \nonumber \\
 & + \frac{1}{\sqrt{2}}\U_{0a} \frac{\vthparallel \tau}{ \Omega_a} + \frac{\vthparallel}{\sqrt{2}} \left( \frac{\Qparallel}{\Pparallel} \frac{\tau}{\Omega_a} + \frac{\overline{B}_{\parallel a}^*}{B}\right) \b+ \left(\frac{\vthparallel \tau}{\sqrt{2} \Omega_a} + \frac{\overline{B}_{\parallel a}^*}{B} \frac{\sqrt{2} \:\Qperp}{\Pperp \vthparallel} \right)\U_{\mu a}^\parallel \nonumber \\
 & + \frac{1}{\gyN B}\b \times \moment{10}{\gygrad \gyaver{\Psi_1}_{\gyR}}.
\end{align}
\\
The parallel and perpendicular temperature equations for $\Tparallel$ and $\Tperp$ are obtained by setting $(l,k) = (2,0)$ and $(0,1)$ respectively in \cref{eq:GyromomentHierarchyEquation}. This yields for the parallel temperature $\Tparallel$,

\begin{align} \label{eq:Tparallel}
\frac{\gyN}{\sqrt{2}} \frac{d_a^{*0}}{d t} \Tparallel & = \Pparallel C_a^{20} -    \gygrad \cdot \left(  \Tparallel \gyN \u_{a}^{ \parallel 2}\right) - \frac{\sqrt{2}\tau}{\Omega_a} \frac{\partial}{\partial t}  \uparallel - \frac{2}{\vthparallel} \u_{ a}^{\parallel 1} \cdot \gygrad \uparallel \nonumber \\
& +  2 \sqrt{m_a \Tparallel} \momentstar{10}{\dot \gyvparallel} -  \frac{\partial}{\partial t} \left( \frac{\Qparallel \tau }{\sqrt{2} \Omega_a} \right),
\end{align}
\\
with

\begin{align}
\u_{a}^{\parallel  2} &= \frac{1}{\sqrt{2}}\U_{\curvature a} +  \left( \U_{pa}^{th} - \U_{Ba} - \U_{\mu  a}^{\parallel \perp} \right) \frac{\Qparallel}{\Pparallel \sqrt{2}\vthparallel} + \frac{\sqrt{2} \Qperp}{\Pperp \vthparallel}  (\U_{Ba} +   \U_{\mu a}^{\parallel \perp})  \nonumber \\
& + \U_{0a} \frac{\Qparallel}{\Pparallel} \frac{\tau}{\Omega_a}  + \left( \frac{\Tparallel \sqrt{2}\tau}{m_a \Omega_a} + \frac{\overline{B}_{\parallel a}^*}{B} \frac{\Qparallel}{\Pparallel \sqrt{2} }\right) \b \nonumber \\
&+  \U_{\mu \parallel a} \frac{\tau}{\Omega_a} \left( \frac{\Qparallel}{\Pparallel\sqrt{2}}  + \frac{\sqrt{2} \;\Qperp}{\Pperp}\right)+ \frac{1}{\gyN B} \b \times \moment{20}{\gygrad \gyaver{\Psi_1}},
\end{align}
\\
while for the perpendicular temperature, $\Tperp$, it yields

\begin{align} \label{eq:Tperp}
\gyN \frac{d_a^{*0}}{dt} \left( \frac{\Tperp}{B}\right) = \frac{\gyN \Tperp}{B} C_a^{01} -  \gygrad \cdot \left( \frac{\u_{a}^{\perp 1} \Pperp}{B}  \right)
 + \frac{\partial }{\partial t} \left( \frac{\Qperp}{B}\frac{\tau }{\Omega_a}  \right),
\end{align}
\\
with

\begin{align}
\u_{ a}^{\perp 1} &= - \U_{\grad a} - \U_{\omega a} - \U_{\mu a }^\perp -\left( \U_{pa}^{th} + 3 \U_{Ba} + 3 \U_{\mu  a}^{\parallel \perp} \right) \frac{\Qperp}{\Pperp \vthparallel}- \frac{ \U_{0a}  \Qperp}{\Pperp}\frac{\tau}{\Omega_a} \nonumber \\
&  -  \frac{\overline{B}_{\parallel a}^*}{B} \frac{\Qperp}{\Pperp }\b  - \U_{\mu  a}^\parallel \left(  3 \frac{\Qperp}{\Pperp} \frac{\tau}{\Omega_a } + \frac{\overline{B}_{\parallel a}^* }{B}  \right) +   \frac{1}{\gyN B} \b \times \moment{01}{\gygrad \gyaver{\Psi_1}}.
\end{align}
\\
Finally, the parallel heat flux equation for $\Qparallel$ is obtained by setting $(l,k) = (3,0)$  in \cref{eq:GyromomentHierarchyEquation} yielding

\begin{align} \label{eq:Qparallel}
  \frac{d_a^{*0}}{ dt } \Qparallel & = \gyN \Tparallel \sqrt{3} \vthparallel C_a^{30} + \Qparallel \frac{\partial}{\partial t} \left( \frac{\overline{B}_{\parallel a}^*}{B} \right) - \frac{\sqrt{3} \vthparallel \tau}{2 \Omega_a \Tparallel } \frac{\partial}{\partial t} \Tparallel  \nonumber \\
& +   \frac{3 \Qparallel \tau }{ \Omega_a }  \frac{\partial}{\partial t} \uparallel  -  \gygrad \cdot \left( \u_a^{\parallel 3} \gyN \Tparallel \sqrt{3} \vthparallel \right) -  \gyN  \vthparallel \frac{3}{\sqrt{2} }  \u_{a}^{\parallel 1} \cdot \gygrad \; \Tparallel
\nonumber \\
&  +  \gyN \Tparallel 3 \sqrt{2}  \u_{a}^{\parallel 2}  \cdot \gygrad  \uparallel  + \u_a^0 \cdot \gygrad \; \Qparallel-  \gyN \Tparallel 3 \sqrt{2}  \momentstar{20}{\dot \gyvparallel},
\end{align}
\\
while the perpendicular heat flux equation for $\Qperp$ follows from $(l,k) = (1,1)$,

\begin{align} \label{eq:Qperp}
\frac{d_a^{*0}}{d t } \Qperp & =  - \frac{\gyN \Tperp \vthparallel}{\sqrt{2}} C_a^{11} + \frac{B}{\sqrt{2} }\u^{ \parallel \perp 1}_a  \cdot \gygrad  \left( \frac{\gyN \Tperp \vthparallel}{B}\right) - \frac{\gyN \Tparallel  \tau  }{ q_a } \frac{\partial}{\partial t} \left( \frac{\Tperp}{B} \right)\nonumber \\
& -\frac{\gyN  \vthparallel B}{ \sqrt{2} }\u_a^{\parallel 1} \cdot \gygrad \left( \frac{\Tperp}{B} \right)- \frac{\Qperp \tau}{\Omega_a} \frac{\partial}{\partial t} \uparallel + \gyN \Tperp  \u_a^{\perp 1} \cdot \gygrad \uparallel \nonumber \\
& - \gyN \Tperp \momentstar{01}{\dot \gyvparallel} + \gygrad \cdot \left( \u_a^0 \Qperp \right) -  \Qperp B  \frac{\partial}{\partial t} \left( \frac{\overline{B}_{\parallel a}^*}{B^2} \right)-\frac{\gyN \Tperp  \vthparallel}{\sqrt{2}} \gygrad \cdot \u_a^{\parallel \perp 1},
\end{align}
\\
 with

\begin{align}
\u_{ a}^{ \parallel 3}   & = \left( \U_{pa}^{th}  + \frac{7}{2} \U_{\curvature a} + \U_{\grad a} + \U_{\omega a } + \U_{\mu  a}^\perp + \frac{\overline{B}_{\parallel a}^*}{B} \U_{0a} \right) \frac{\Qparallel}{\sqrt{3} \; \Pparallel \vthparallel }   \nonumber \\
& + \left(\frac{\sqrt{3} \tau \vthparallel}{2 \Omega_a}    + \frac{\sqrt{2} \;\overline{B}_{\parallel a}^* }{\sqrt{3} B  }\right) \frac{\Qparallel}{\Pparallel} \b  +\U_{\mu  a}^\parallel \frac{\overline{B}_{\parallel a}^*}{B}  \frac{\Qparallel }{\sqrt{3} \hspace{1pt} \Pparallel \vthparallel}+  \frac{1}{\gyN B} \b \times \moment{30}{\gygrad \gyaver{\Psi_1}},
\end{align}
\\
and 

\begin{align}
\u_{ a}^{\parallel \perp 1}  & = -\left( \U_{pa}  + \frac{3}{2} \U_{\curvature a} + 3 \U_{\grad a} + 3 \U_{\omega a}  + 3 \U_{\mu  a}^\perp\right) \frac{\sqrt{2} \;\Qperp}{\Pperp \vthparallel} \nonumber \\
& - \frac{1}{\sqrt{2}} \U_{B a}  - \frac{1}{\sqrt{2}} \U_{\mu  a}^{\parallel \perp} +  \frac{\sqrt{2} \; \Qperp \U_{0 a}}{\Pperp \vthparallel}  - \vthparallel  \left( 1 + \sqrt{2} \right) \frac{\tau \vthparallel}{\Omega_a}\frac{\Qperp}{\Pperp}\b \nonumber \\
& - \U_{\mu a}^\parallel \left( 3\frac{\overline{B}_{\parallel a}^*}{B} \frac{\sqrt{2} \:\Qperp}{\Pperp \vthparallel} + \frac{\vthparallel \tau}{\sqrt{2} \Omega_a} \right) +\frac{1}{\gyN B} \b \times \moment{11}{\gygrad \gyaver{\Psi_1}}.
\end{align}
\\
In the above fluid equations, we note the presence of multiple partial time derivatives that emerge from the fluid operator $\mathcal{F}_a^{lk}$, given in \cref{eq:fluidoperatoralk}, and from the gyro-moment convective fluid derivative $d_{at}^{*lk}$ defined in \cref{eq:convectivederivative}. In a numerical implementation, these partial time derivatives can be handled by introducing generalized fluid quantities that include also field dependencies \citep[see, e.g, ][]{Halpern2016}. 

This system of \cref{eq:Na,eq:Uparallel,eq:Tparallel,eq:Tperp,eq:Qparallel,eq:Qperp} constitutes the gyrokinetic electromagnetic extension of the collisional drift-reduced fluid equations obtained in \citet{Jorge2017}. Indeed, the present equations contain several new terms, proportional to $\tau$, and terms related to the small scale and amplitude electromagnetic fluctuations through the gradient of the gyroaveraged gyrokinetic potential, $\gygrad \gyaver{\Psi_1}$. Also, they retain exact polarization effects associated with $B_\parallel^*$ [see \cref{eq:partialtBparallelstar}]. The evaluation of these terms was approximated in previous reduced gyro-moment hierarchy \citep{Madsen2013}.}

\subsection{Hermite-Laguerre Representation of Gyroaverage Operator}
\label{HermiteLaguerreRepresentationOfGyroaverageOperator}
In this section, we overcome the limitations of previous gyrofluid models \citep[see, e.g.,][]{Beer1996,Snyder2001,Scott2010,Madsen2013} deriving closed expressions for the FLR induced transports and forces appearing in \cref{eq:FLRtransport,eq:FLRforces}. More precisely, we aim to express the gyroaveraged small-scale electromagnetic fluctuations, evaluated at $\overline{\bm x} = \gyR + \gyrhoa$ [see \cref{eq:overlinex}], in terms of their values at $\gyR$, i.e. $\phi_1(\gyR)$ and $A_{\parallel 1}(\gyR)$ and of the gyro-moments $\gyN^{lk}$. This requires a suitable analytical form of the gyroaverage operator to perform the Hermite-Laguerre projection, while retaining the linear and nonlinear couplings accurately at all orders in $\epsilon_\perp$ and translating the spatial dependence of the fields from $\overline{\bm \x}$ to $\gyR$. In the following, we first evaluate the Hermite-Laguerre representation of the transport term associated with the small-scale electrostatic potential $\gyaver{\phi_1}_{\gyR}$, i.e. $\moment{pj}{\gygrad \gyaver{\phi_1}_{\gyR}}$. We then generalize the approach to the nonlinear terms, such as $\gygrad \partial_{\gymu} \gyaver{A_{\parallel 1}^2}_{\gyR}$, $\gygrad \partial_{\gymu} \gyaver{\phi_1^2}_{\gyR}$, $\gygrad \partial_{\gymu} \gyaver{\phi_1 A_{\parallel 1}}_{\gyR}$, and $\gygrad \partial_{\gymu} (\gyaver{\phi_1}_{\gyR} \gyaver{A_{\parallel 1}}_{\gyR})$ appearing in \cref{eq:momentstarlkgygradPsi1}, and report the Hermite-Laguerre projection of the last term, proportional to $\gygrad \gyaver{(\b\times \gygrad \overline{\widetilde{\Phi_1}})\cdot \gygrad \widetilde{\Phi_1}}_{\gyR}$ in \cref{eq:momentstarlkgygradPsi1}, in \cref{appendixHLnonlinearpot}.

In order to obtain $\moment{pj}{\gyaver{\phi_1}_{\gyR}}$, we first gyroaverage the potential $\phi_1$ in Fourier space. This is preferred to the Taylor expansion of $\phi_1(\overline{\bm x})$ around the gyrocenter position $\gyR$ since a large number of expansion terms would be needed for $\epsilon_\perp \sim 1$. By introducing the Fourier decomposition $\phi_1(\overline{\bm x} ) = \sum_{\bm k} \phi_1(\bm k) e^{ i \bm k \cdot \overline{\bm x}}$ and using the fact that $\overline{\bm x} = \gyR + \gyrhoa$ [see \cref{eq:overlinex}], one obtains

\begin{equation} \label{eq:gyaverFourier}
\gyaver{\phi_1}_{\gyR} \equiv \gyaver{\phi_1(\overline{\bm x})}_{\gyR} = \sum_{\bm k}  \gyaver{ e^{i  \bm k \cdot \gyrhoa }}_{\gyR} \phi_1 (\bm k) e^{ i \bm k \cdot \gyR} =  \sum_{\bm k}   J_0(b)  \phi_1(\bm k) e^{ i \bm k \cdot \gyR},
\end{equation}
\\
where we have oriented $(\e_1,\e_2,\b)$ such that \corr{$\bm k = k_\perp \e_1  + k_\parallel \b$}, and the Jacobi-Anger expansion \citep{Abramowitz1974} is used,

\begin{equation} \label{eq:jacobianger}
e^{ i \bm k \cdot \gyrhoa } = \sum_{m} i^m J_m(b) e^{i m \gytheta},
\end{equation}
\\
with $b = k_\perp v_\perp / \Omega_a$, and $J_m$ the $m$th-order Bessel function of the first kind, yielding $\gyaver{e^{i \bm k \cdot \gyrhoa}}_{\gyR} = J_0(b)$. We remark that, from \cref{eq:gyaverFourier} and the fact $J_0(b) = \sum_{l} (-1)^l b^{2 l}/(2^{2l} l!^2)$, perpendicular FLR couplings between gyro-moments arise with the gyroaverage operator.

\corr{To perform the Hermite-Laguerre projection of \cref{eq:gyaverFourier} and express $\gyaver{\phi_1}_{\gyR}$ in terms of gyro-moments $N_a^{lk}$, we expand $J_m$ in terms of generalized Laguerre polynomials $L_n^{m+1/2}$ defined by \citep{gradshteyn2007}

\begin{equation}
L_n^{m+1/2}(x) = \sum_{l=0}^n L_{nl}^{m} x^l,
\end{equation}
\\
where

\begin{equation}
L_{nl}^{m}  = \frac{(-1)^l (m + n + 1/2)!}{(n - l)!( m +l+ 1/2 )! l!},
\end{equation}
\\
that is 

\begin{equation} \label{eq:Jn2Laguerre}
    J_m(b) = \sigma_m \left( \frac{b_a \sperp}{2}\right)^{|m|} \sum_{n \geq 0} \frac{n!}{(n+|m|)!}\kernel{n}(b_a) L_n^{|m|} (\sperp^2),
\end{equation}
\\
with $\sigma_0 =1$ and $\sigma_m = \text{sgn}(m)^m$ for $m \neq 0$.} and where we introduced the thermal velocity based parameter $b_a = b/\sperp = k_\perp/ \Omega_a\sqrt{ 2 \gyTperp/ m_a } $, and the $n$th-order kernel functions $\kernel{n}$,

\begin{equation} \label{eq:kerneldef}
    \kernel{n}(b_a)  = \frac{1}{n!} \left( \frac{b_a}{2}\right)^{2n} e^{- (b_a/2)^2}  \corr{= \frac{(-1)^n}{n!} b_a^{2n} \frac{\partial^n}{\partial (b_a^2)^n} e^{ - (b_a/2)^2}},
\end{equation}
\\
From \cref{eq:kerneldef}, one can observe that the kernel functions satisfy the recursive formula $b_a^2 \kernel{n}(b_a) = 4 (n+1) \kernel{n+1}(b_a)$ and the normalization relation $\sum_{n \geq0 } \kernel{n}(b_a)=1$. In addition, the asymptotic behaviour of the kernel function is $\kernel{n}(b_a) \sim b_a^{2n}$ for $b_a \ll 1$, while $\kernel{n} \sim 1/n!$ at large $n$ for all values of $b_a$. Also, we notice that, for ions, $b_i \sim \epsilon_\perp$ for ions, while $b_e \sim \sqrt{m_e / m_i} \epsilon_\perp$ for electrons. 

Using \cref{eq:Jn2Laguerre} to expand the Bessel function $J_0$, the gyroaveraged electrostatic potential $\gyaver{\phi_1}_{\gyR}$ can be expressed as a Laguerre series,

\begin{equation} \label{eq:gyaverLaguerre}
    \gyaver{\phi_1(\overline{\bm x})}_{\gyR} = \sum_{\bm k }\sum_{n\geq0} \kernel{n}(b_a) \phi_1(\bm k )  L_n(\sperp^2) e^{i \bm k\cdot 
\gyR},
\end{equation}
\\
which decouples the spatial and velocity dependencies. In fact, the velocity dependence that involves the perpendicular FLR coupling between gyro-moments is embedded into the Laguerre polynomials $L_n$, whereas the spatial gyroaveraging is handled by the kernel $\kernel{n}$. As a consequence of \cref{eq:gyaverLaguerre}, the accuracy of FLR effects in our description is directly related to the Laguerre resolution \citep{Zocco2015,Mandell2018}. 

\corr{
We remark that a spatial (instead of Fourier) representation of gyroaveraged quantities, e.g., $\gyaver{\phi_1}$, can be obtained by performing the transformation $\bm k \to \gygrad/i$ such that $b_a^{2j} \to  (-1)^j 2^j(\Tperp/ m_a \Omega_a^2)^j  \gygrad_\perp^{2j}$ in \cref{eq:kerneldef} . Thus, at the lowest order, we can write

\begin{align} \label{eq:gyaverspatial}
\gyaver{\phi_1}_{\gyR} & = \sum_{n \geq 0} \sum_{m \geq0} \frac{(-1)^{n}}{n!m!} L_{n}(\sperp^2) \left( \frac{ \Tperp }{2 m_a \Omega_a^2} \gygrad_\perp^2    \right)^{n+m}  \phi_1(\gyR) \nonumber \\
& = \phi_1(\gyR) + \frac{\gyTperp}{2 m_a \Omega_a^2} \gygrad^2_\perp \phi_1(\gyR)  + \dots.
\end{align}
\\
 In \cref{eq:gyaverspatial}, $\gyTperp$ and $\Omega_a$ are spatially dependent, but are held constant when gyroaveraging $\phi_1$.}

We now apply the gradient operator $\gygrad$ to \cref{eq:gyaverLaguerre} and the Hermite-Laguerre projector. Since $b_a$ and $\sperp$ are spatially varying as they depend on $B$ and $\Tperp$, we use the property of the Laguerre polynomials, $x L_n'(x) = n [L_{n}(x) -L_{n-1}(x)]$, and the fact that $ 2 \kernel{n}'(x) = x[\kernel{n-1}(x) - \kernel{n}(x)]$ (here, the prime denotes the derivative with respect to the argument). In order to handle the products of Laguerre polynomials arising from the Hermite-Laguerre projector, such as, e.g., $H_a^{lk} L_n$ [see \cref{eq:Halk}], we use the following relationship \citep{Gillis1959}, 

\begin{equation} \label{eq:laguerrelaguerre2laguerre}
 L_k L_n = \sum_{s = \lvert k - n\rvert}^{ \lvert k + n\rvert} \alpha^{kn}_s L_s.
\end{equation}
The expansion coefficients $\alpha^{kn}_s$ are determined by the Laguerre polynomial orthogonality relation [see \cref{eq:Laguerreorthogonality}],

\begin{equation} \label{eq:alphaknsdef}
\alpha^{kn}_{s} = \int_0^{\infty} d x e^{-x} L_{k}(x) L_{n}(x) L_s(x),
\end{equation}
\\
and their closed formula is given by

\begin{equation} \label{eq:alphaknt}
\alpha^{kn}_s = \left( - 1 \right)^{k+n - s } \sum_{m} \frac{ 2^{2m - k -n +s }   \left( k + n - m \right)!}{(k-m)!(n-m)!(2m - k - n +s)!(k+n -s -m)!},
\end{equation}
\\
where the summation is performed over all possible values of $m$ such that the factorials are positive. From \cref{eq:alphaknsdef}, we note that the coefficients $\alpha_s^{kn}$ are symmetric in all pairs of indices and, in particular, that $\alpha^{kn}_0  = \delta_k^n$ [see \cref{eq:Laguerreorthogonality}]. We then derive 

\begin{equation}  \label{eq:momentlkgradgyaverphi1}
\frac{1}{\gyN}\moment{lk}{\gygrad \gyaver{\phi_1(\overline{\bm x})}_{\gyR}}  = \sum_{\bm k} \sum_{n \geq 0 } \bm{\mathcal{D}}^{lkn}_{an}(b_a,  \bm k) \phi_{1 }(\bm k) e^{i \bm k \cdot \gyR } ,
\end{equation}
\\
where the FLR gradient operator $\bm{\mathcal{D}}^{lkj}_{an}$   is introduced, that is

\begin{align} \label{eq:bmDanlkj}
\bm{\mathcal{D}}^{lkj}_{an} (b_a, \bm k) & =   \mathcal{D}_{an}^{lkj}(b_a)   i \bm k  + j (\gygrad \ln \eta) \left( \mathcal{D}_{an}^{lkj}(b_a)  -\mathcal{D}_{an}^{lkj-1}(b_a)  \right)   \nonumber \\
&   +     (\gygrad \ln \iota )  \left( \frac{ b_a^2}{2} \right) \left( \mathcal{D}_{an-1}^{lkj}(b_a)\ - \mathcal{D}_{an}^{lkj}(b_a)\right),
 \end{align}
\\
with $\eta = B/ \gyTperp$, $\iota = \sqrt{\gyTperp}/ B$, and

\begin{equation} \label{eq:Danlkj}
\mathcal{D}_{an}^{lkj}(b_a)= \sum_{r = |j-k|}^{|j+k|} \alpha_r^{jk} \gyN^{lr} \kernel{n}(b_a).
\end{equation}
\\
In absence of temperature and magnetic gradients, the gyroaverage and $\gygrad$ operators commute, i.e. $\gygrad \gyaver{\phi_1}_{\gyR} = \gyaver{\gygrad \phi_1}_{\gyR}$. In this case, $\bm{\mathcal{D}}^{lkj}_{an} (b_a, \bm k) = \mathcal{D}_{an}^{lkj}(b_a) i \bm k$. We note that \cref{eq:Danlkj} corresponds to the expression obtained within a similar Laguerre treatment of the gyroaverage operator by \citet{Mandell2018}. In addition, in the case of a Maxwellian equilibrium, such that $\gyN^{lk} =0$ for $l,k > 0$ [see \cref{eq:GYFadecomposition}], we notice that \cref{eq:Danlkj} reduces to $\mathcal{D}_{a0}^{000} =  \kernel{0}(b_a) = e^{-(b_a/2)^2} \equiv \int d \gymu d \gyvparallel  d \gytheta B J_0(b) F_{aM}/m_a $ \citep{Dorland1993}.The second term of the right-hand side of \cref{eq:momentstarlkgygradPsi1} can be evaluated similarly. When projected, this shear-Alfv\'en term has the following expression,

\begin{equation} \label{eq:momentlkgradgyaverA1}
    \frac{1}{\gyN}    \moment{lk}{\gygrad \gyaver{A_{\parallel 1}(\overline{\bm x})}_{\gyR}}   =   \sum_{\bm k} \sum_{n \geq 0}    \bm{\mathcal{D}}^{lkn}_{an}(b_a,\bm k) A_{ \parallel 1 }(\bm k) e^{i \bm k \cdot \gyR},
\end{equation}
\\
which is equivalent to \cref{eq:momentlkgradgyaverphi1}. 

We now evaluate the Hermite-Laguerre projections of the second order terms in \cref{eq:momentstarlkgygradPsi1}, such as $\gygrad \partial_{\gymu} \gyaver{\phi_1 A_{\parallel 1}}_{\gyR}$, $\gygrad \gyaver{A_{\parallel 1}^2}_{\gyR}$, and  $\gygrad \partial_{\gymu}( \gyaver{A_{\parallel 1}}_{\gyR} \gyaver{\phi_1}_{\gyR})$. We consider, as an example, the term $\gygrad \partial_{\gymu} \gyaver{\phi_1 A_{\parallel 1}}_{\gyR}$. As a first step, we write the product of $\phi_1$ and $A_{\parallel 1}$ in Fourier space,

\begin{align} \label{eq:phi1Aparallel1}
    \phi_1(\overline{\bm x}) A_{\parallel 1}(\overline{\bm x}) = \sum_{\bm k, \bm k'} \sum_{n} \sum_{m} i^{n +m} J_n(b) J_m(b') \phi_1(\bm k) A_{\parallel 1}(\bm k') e^{i(n+m) \gytheta} \corr{e^{-i m \alpha}} e^{i \bm K\cdot \gyR} .
\end{align}
\\
where $\bm K = \bm k + \bm k'$ \corr{with $\bm k' = k'_\parallel \b + k_\perp' (\cos \alpha \e_1 + \sin \alpha \e_2)$ (the angle $\alpha$ being the phase shift between $\bm k$ and $\bm k'$ in the plane perpendicular to $\b$, such that $i \bm k' \cdot \gyrhoa = i k_\perp' \gyvperp \cos (\theta - \alpha)/\Omega_a$)}, and $J_n$ and $J_m$ are the Bessel functions introduced via the Jacobi-Anger identity in \cref{eq:jacobianger}, with arguments $b = k_\perp v_\perp / \Omega_a$ and $b' = k_\perp' v_\perp / \Omega_a$, respectively. \Cref{eq:phi1Aparallel1} is in a suitable form to perform the gyroaverage operator. Indeed, using the fact that $\int_{0}^{2 \pi} d \gytheta e^{ i(n+m) \gytheta} = 2 \pi \delta_{-n}^{m} $ \citep{gradshteyn2007}, one obtains

\begin{equation} \label{eq:gyaverphiAparallel1}
      \gyaver{ \phi_1(\overline{\bm x}) A_{\parallel 1}(\overline{\bm x})}_{\gyR} = \sum_{\bm k, \bm k'} \corr{\sum_{n} (-1)^n e^{ in \alpha} J_n(b) J_n(b') \phi_1(\bm k) A_{\parallel 1}(\bm k') e^{i \bm K \cdot \gyR}}.
\end{equation}
\\
We now express the Bessel functions $J_n(b)$ and $J_n(b')$ in terms of generalized Laguerre polynomials according to \cref{eq:Jn2Laguerre},

\begin{align} \label{eq:phi1Aparallel1assLaguerre}
      \gyaver{ \phi_1(\overline{\bm x}) A_{\parallel 1}(\overline{\bm x})}_{\gyR} & = \sum_{\bm k, \bm k'} \corr{\sum_{n } \sum_{r,s \geq 0 } \frac{(-1)^{n} r!s!}{(|n|+r)!(|n|+s)!}e^{in\alpha}\left( \frac{b_a}{2}\right)^{|n|} \left( \frac{b_a'}{2}\right)^{|n|}  \kernel{r}(b_a) \kernel{s}(b_a') } \nonumber \\
& \corr{   \times     L_r^{|n|}(\sperp^2) L_s^{|n|}(\sperp^2) \sperp^{2|n|} \phi_1(\bm k) A_{\parallel 1}(\bm k') e^{i \bm K \cdot \gyR}.}
\end{align}
\\
\corr{The product between two generalized Laguerre polynomials weighted by $\sperp^{2|n|}$, i.e., $  L_r^{|n|}(\sperp^2) L_s^{|n|}(\sperp^2) \sperp^{2|n|}$ appearing in \cref{eq:phi1Aparallel1assLaguerre}, can be written in terms of single Laguerre polynomials using the two following identities,

\begin{equation} \label{eq:LrnLjsperp2Le}
L_r^n(\sperp^2)  \sperp^{2n} = \sum_{e  =0}^{r  + n} d_{re}^n L_e(\sperp^2),
\end{equation}
\\
and 

\begin{equation} \label{eq:LrnLj2Lf}
L_r^n(\sperp^2)  L_j(\sperp^2)  = \sum_{f =0}^{r+j} \overline{d}^n_{rjf} L_f(\sperp^2),
\end{equation}
\\
where 

\begin{equation}
d_{re}^n = \sum_{r_1 =0}^r  \sum_{e_1=0}^e L_{r r_1}^{n-1/2}  L_{e e_1}^{-1/2} (r_1  + e_1 + n) !,
\end{equation}
\\
and 

\begin{equation}
\overline{d}^n_{rjf} = \sum_{r_1 = 0}^r \sum_{j_1 =0}^j   \sum_{f_1=0}^f L_{j j_1}^{-1/2}  L_{rr_1}^{n-1/2}  L_{f f_1}^{-1/2} (r_1 + j_1 + f_1)!.
\end{equation}}
\\
Using consecutively \cref{eq:LrnLjsperp2Le} and \cref{eq:LrnLj2Lf}, we derive

\begin{align}
      \gyaver{ \phi_1(\overline{\bm x}) A_{\parallel 1}(\overline{\bm x})}_{\gyR} & = \sum_{\bm k, \bm k'} \corr{\sum_{n } \sum_{r,s \geq 0 } \sum_{e=0}^{|n|+r} \sum_{f =0}^{s+e} K_{rsef}^n 
           \left( \frac{b_a}{2}\right)^{|n|} \left( \frac{b_a'}{2}\right)^{|n|}  \kernel{r}(b_a) \kernel{s}(b_a') } \nonumber \\
& \corr{   \times    L_f(\sperp^2) \phi_1(\bm k) A_{\parallel 1}(\bm k') e^{i \bm K \cdot \gyR} .}
\end{align}
\\
\corr{where $K_{rsef}^n = (-1)^n e^{i n\alpha}d_{re}^{|n|} \overline{d}_{sef}^{|n|} r!s!/[(|n|+r)!(|n|+s)!]$}. By taking into account the spatial variation of the velocity variable $\sperp^2$ and using the fact that $\partial / \partial \gymu = \eta \partial / \partial \sperp^2$ (with $\eta = B / \Tperp$), one can write

\begin{equation} \label{eq:gradpmugyvaerphiA1}
\gygrad \gypmu \gyaver{ \phi_1 A_{\parallel 1}}_{\gyR}  = \eta \gygrad \left(  \frac{\partial}{\partial \sperp^2} \gyaver{ \phi_1 A_{\parallel 1}}_{\gyR} \right) + \left( \gygrad  \ln \eta \right) \frac{\partial}{\partial \gymu} \gyaver{ \phi_1 A_{\parallel 1}}_{\gyR}  .
\end{equation}
\\
We focus on the first term in \cref{eq:gradpmugyvaerphiA1}. Taking the $\sperp^2$-derivative of \cref{eq:gyaverphiAparallel1} with $L_n'(x) = (-1) L^1_{n-1}(x)$ for $n > 0$ and the property $L_m^1 = \sum_{l =0}^{m} L_l $ \citep{gradshteyn2007}, one derives

 \begin{align} \label{eq:psperpgyaverphi1A1}
 \frac{\partial}{\partial \sperp^2}  \gyaver{ \phi_1(\overline{\bm x}) A_{\parallel 1}(\overline{\bm x}) }_{\gyR}  & = - \sum_{\bm k, \bm k'} \corr{\sum_{n } \sum_{r,s \geq 0 } \sum_{e=0}^{|n|+r} \sum_{f =1}^{s+e} \sum_{g=0}^{f-1} K_{rsef}^n            \left( \frac{b_a}{2}\right)^{|n|} \left( \frac{b_a'}{2}\right)^{|n|}  \kernel{r}(b_a) \kernel{s}(b_a') } \nonumber \\
& \corr{   \times    L_g(\sperp^2) \phi_1(\bm k) A_{\parallel 1}(\bm k') e^{i \bm K \cdot \gyR}.}
 \end{align}
\\
\Cref{eq:psperpgyaverphi1A1} has the proper form to apply the Hermite-Laguerre projector in \cref{eq:projectorlk}. Indeed, applying the gradient operator $\gygrad$, we obtain  

 \begin{align} \label{eq:momentlkgradpmugyaverphi1A1}
 \frac{1}{\gyN}\moment{lk}{\gygrad  \frac{\partial}{\partial \sperp^2}  \gyaver{ \phi_1(\overline{\bm x}) A_{\parallel 1}(\overline{\bm x}) }_{\gyR}  } & =  -    \sum_{\bm k, \bm k'} \sum_{n } \sum_{r,s \geq 0 } \sum_{e=0}^{|n|+r} \sum_{f =1}^{s+e} \sum_{g=0}^{f-1} K_{rsef}^n \nonumber \\ 
 & \times\bm{\overline{\mathcal{D}}}_{anrs}^{lkg}(b_a,b_a',\bm K) \phi_1(\bm k) A_{\parallel 1} (\bm k') e^{i \bm K \cdot \gyR},
 \end{align}
\\
where the FLR gradient operator $\bm{\overline{\mathcal{D}}}_{anrs}^{lkg}$ is introduced, and defined by

\corr{
\begin{align} \label{eq:Danrslkg}
\bm{\overline{\mathcal{D}}}_{anrs}^{lkg}(b_a,b_a',\bm K) & = i \bm K \overline{\mathcal{D}}_{anrs}^{lkg}(b_a,b_a') + g (\gygrad \ln   \eta  )\left( \overline{\mathcal{D}}_{anrs}^{lkg}(b_a,b_a') - \overline{\mathcal{D}}_{anrs}^{lkg-1}(b_a,b_a')\right) \nonumber \\
& + (\gygrad \ln \iota )  \left[ \frac{b_a^2}{2} \left( \overline{\mathcal{D}}_{anr-1s}^{lkg}(b_a,b_a') - \overline{\mathcal{D}}_{anrs}^{lkg}(b_a,b_a')\right)  \right. \nonumber \\
& \left.  + \frac{b_a'^{2}}{2} \left( \overline{\mathcal{D}}_{anrs-1}^{lkg}(b_a,b_a')- \overline{\mathcal{D}}_{anrs}^{lkg}(b_a,b_a') \right)  + 2 n  \overline{\mathcal{D}}_{anrs}^{lkg}(b_a,b_a') \right],
\end{align}
\\
where 

 \begin{equation}
 \mathcal{\overline{D}}_{anrs}^{lkg}(b_a,b_a') = \sum_{h = |k-g|}^{|k+g| }  \alpha_{h}^{kg}  \gyN^{lh}          \left( \frac{b_a}{2}\right)^{|n|} \left( \frac{b_a'}{2}\right)^{|n|}  \kernel{r}(b_a) \kernel{s}(b_a') .
 \end{equation}}
 \\
We now consider the second term in \cref{eq:gradpmugyvaerphiA1}. Using the fact that $\partial / \partial \gymu = \eta \partial/ \partial \sperp^2$ with \cref{eq:psperpgyaverphi1A1} yields

\corr{
  \begin{align} \label{eq:momentlkpmugyaverphi1A1}
 \frac{1}{\gyN}\moment{lk}{ \frac{\partial}{\partial \gymu} \gyaver{ \phi_1(\overline{\bm x}) A_{\parallel 1}(\overline{\bm x}) }_{\gyR} }  & =   - \eta  \sum_{n } \sum_{r,s \geq 0 } \sum_{e=0}^{|n|+r} \sum_{f =1}^{s+e} \sum_{g=0}^{f-1} K_{rsef}^n \nonumber \\
&  \times\mathcal{\overline{D}}_{anrs}^{lkg}(b_a,b_a') \phi_1(\bm k) A_{\parallel 1}(\bm k') e^{i \bm K \cdot \gyR} .
 \end{align}}

In conclusion, we obtain, from \cref{eq:momentlkpmugyaverphi1A1,eq:momentlkgradpmugyaverphi1A1}, the Hermite-Laguerre representation of the second order nonlinear electromagnetic term $\gygrad \partial_{\gymu} \gyaver{\phi_1 A_{\parallel _1}}_{\gyR}$,
\corr{
\begin{align} \label{eq:momentlkgradpmugyaverphiA1}
\frac{1}{\gyN}\moment{lk}{\gygrad\frac{\partial}{\partial \gymu}\gyaver{\phi_1(\overline{ \x}) A_{\parallel 1}(\overline{ \x})}_{\gyR}  } & =- \eta  \sum_{\bm k, \bm k'}  \sum_{n } \sum_{r,s \geq 0 } \sum_{e=0}^{|n|+r} \sum_{f =1}^{s+e} \sum_{g=0}^{f-1} K_{rsef}^n \nonumber \\ 
& \times  \left[\bm{\overline{\mathcal{D}}}_{anrs}^{lkg}(b_a,b_a',\bm K) + ( \gygrad \ln \eta )\mathcal{\overline{D}}_{anrs}^{lkg}(b_a,b_a')\right] \nonumber \\
& \times \phi_{1}(\bm k) A_{\parallel 1 }(\bm k') e^{i \bm K\cdot \gyR}.
\end{align}


A similar procedure can be used to derive

\begin{align}
\frac{1}{\gyN} \moment{lk}{\gygrad \gyaver{A_{\parallel 1}(\overline{\bm x})^2}_{\gyR}} & = \sum_{\bm k, \bm k'}\sum_{n } \sum_{r,s \geq 0 } \sum_{e=0}^{|n|+r} \sum_{f =0}^{s+e} K_{rsef}^n \nonumber \\
& \times \bm{\overline{\mathcal{D}}}_{anrs}^{lkf}(b_a,b_a',\bm K)
            A_{\parallel 1}(\bm k) A_{\parallel 1}(\bm k') e^{i \bm K \cdot \gyR}, \label{eq:momentlkgradgyaverA12} \\
\frac{1}{\gyN}\moment{lk}{\gygrad\frac{\partial}{\partial \gymu}\gyaver{\phi_1(\overline{ \x})^2}_{\gyR}  } & =- \eta  \sum_{\bm k, \bm k'}  \sum_{n } \sum_{r,s \geq 0 } \sum_{e=0}^{|n|+r} \sum_{f =1}^{s+e} \sum_{g=0}^{f-1} K_{rsef}^n \nonumber \\ 
& \times  \left[\bm{\overline{\mathcal{D}}}_{anrs}^{lkg}(b_a,b_a',\bm K) + ( \gygrad \ln \eta )\mathcal{\overline{D}}_{anrs}^{lkg}(b_a,b_a')\right] \nonumber \\
& \times\phi_{1}(\bm k) \phi_{1 }(\bm k') e^{i \bm K\cdot \gyR} , \label{eq:momentlkgradpmugyaverphi12}\\
\frac{1}{\gyN}\moment{lk}{\gygrad\frac{\partial}{\partial \gymu}\gyaver{A_{\parallel 1}(\overline{ \x})^2}_{\gyR}  } & =- \eta  \sum_{\bm k, \bm k'}  \sum_{n} \sum_{r,s \geq 0 } \sum_{e=0}^{|n|+r} \sum_{f =1}^{s+e} \sum_{g=0}^{f-1} K_{rsef}^n \nonumber \\ 
& \times  \left[\bm{\overline{\mathcal{D}}}_{anrs}^{lkg}(b_a,b_a',\bm K) + ( \gygrad \ln \eta )\mathcal{\overline{D}}_{anrs}^{lkg}(b_a,b_a')\right] \nonumber \\
& \times A_{\parallel 1}(\bm k) A_{\parallel 1 }(\bm k') e^{i \bm K\cdot \gyR}.\label{eq:momentlkgradpmugyaverA12}
\end{align}
}


We finally consider the terms containing the product of gyroaveraged fluctuating fields, such as $\gygrad \partial_{\gymu} (\gyaver{\phi_1}_{\gyR} \gyaver{A_{\parallel 1}}_{\gyR})$ appearing in \cref{eq:momentstarlkgygradPsi1}. The product $\gyaver{\phi_1}_{\gyR} \gyaver{A_{\parallel 1}}_{\gyR}$ can be written in Fourier space and expanded in Laguerre polynomials as follows

\begin{align} \label{eq:gyaverphigyaverAparallel}
    \gyaver{\phi_1(\overline{\bm x})}_{\gyR}\gyaver{A_{\parallel 1}(\overline{\bm x})}_{\gyR} & = \sum_{\bm k, \bm k'}  \sum_{n,n' \geq 0}  \sum_{r= |n-n'|}^{|n +n'|} \alpha_r^{nn'}\kernel{n}(b_a) \kernel{n'}(b_a')  L_{r}(\sperp^2)   \nonumber \\
    & \times \phi_1(\bm k) A_{\parallel 1}(\bm k') e^{i \bm K \cdot \gyR},
\end{align}
\\
having made use of \cref{eq:laguerrelaguerre2laguerre}. Following the procedure used to evaluate $\gygrad \partial_{\gymu} \gyaver{\phi_1 A_{\parallel 1}}_{\gyR}$, we obtain the Hermite-Laguerre representation of $\gygrad \partial_{\gymu} \left( \gyaver{\phi_1}_{\gyR}\gyaver{ A_{\parallel 1}}_{\gyR}\right)$, i.e.

\begin{align}
\frac{1}{\gyN}\moment{lk}{\gygrad\frac{\partial}{\partial \gymu} \left( \gyaver{\phi_1(\overline{ \x})}_{\gyR} \gyaver{ A_{\parallel 1}(\overline{ \x})}_{\gyR} \right) } & =- \eta  \sum_{\bm k, \bm k'} \sum_{n,n' \geq 0}  \sum_{\substack{r= |n-n'|\\ r \neq 0}}^{|n +n'|} \alpha_r^{nn'} \sum_{s=0}^{r-1}\nonumber \\
& \times \left[\bm{\mathcal{D}}_{a nn'}^{lks}(b_a,b_a', \bm k,\bm k') +  \left( \gygrad \ln \eta \right) D_{ann'}^{lks}(b_a,b_a')  \right] \nonumber \\
& \times \phi_{\parallel 1}(\bm k) A_{\parallel 1 }(\bm k') e^{i \bm K\cdot \gyR},
\end{align}
\\
where the FLR gradient operator is defined by

 \begin{align} \label{eq:bmDannlkj}
\bm{D}_{ann'}^{lks}(b_a,b_a',\bm k,\bm k')  & =  D_{ann'}^{lks}(b_a,b_a') i \bm K +   (\gygrad \ln \iota )   \nonumber  \\
&\left.   \times  \left[  \frac{ b_a^2}{2} \left( D_{an-1n'}^{lks}(b_a,b_a') - D_{ann'}^{lks}(b_a,b_a')\right) \right. \right. \nonumber \\
& \left. \left. +  \frac{ b_a'^2}{2} \left( D_{ann'-1}^{lks}(b_a,b_a')   - D_{ann'}^{lkp}(b_a,b_a') \right)
\right] \right. \nonumber \\
&   +    s     (\gygrad \ln \eta) \left( D_{ann'}^{lks}(b_a,b_a') - D_{ann'}^{lks-1}(b_a,b_a') \right)  ,
 \end{align}
\\
with

\begin{equation} \label{eq:Dannlkj}
     D_{ann'}^{lks}(b_a,b_a') = \sum_{t= |s-k|}^{|s+k|} \alpha_t^{sk}N_a^{lt}  \kernel{n}(b_a) \kernel{n'}(b_a').
\end{equation}
\\
The remaining $O(\epsilon_\delta^2)$  Hermite-Laguerre projections in \cref{eq:momentstarlkgygradPsi1} are then

\begin{align}
\frac{1}{\gyN}\moment{lk}{\gygrad\frac{\partial}{\partial \gymu}\gyaver{\phi_1(\overline{ \x})}_{\gyR}^2  } & = - \eta  \sum_{\bm k, \bm k'} \sum_{n,n' \geq 0}  \sum_{\substack{r= |n-n'|\\ r \neq 0}}^{|n +n'|} \alpha_r^{nn'} \sum_{s=0}^{r-1}\nonumber \\
& \times \left[     \bm{\mathcal{D}}_{a n}^{lks}(b_a,b_a', \bm k,\bm k')  + \left( \gygrad \ln \eta\right) D_{ann'}^{lks}(b_a,b_a') \right] \nonumber \\
& \times \phi_{1}(\bm k) \phi_{1 }(\bm k') e^{i \bm K\cdot \gyR},\\
\frac{1}{\gyN}\moment{lk}{\gygrad\frac{\partial}{\partial \gymu}\gyaver{A_{\parallel 1}(\overline{ \x})}_{\gyR}^2  } & =- \eta  \sum_{\bm k, \bm k'} \sum_{n,n' \geq 0}  \sum_{\substack{r= |n-n'|\\ r \neq 0}}^{|n +n'|} \alpha_r^{nn'} \sum_{s=0}^{r-1}\nonumber \\
& \times \left[   \bm{\mathcal{D}}_{a n}^{lks}(b_a,b_a', \bm k,\bm k')  + \left( \gygrad \ln \eta \right)D_{ann'}^{lks}(b_a,b_a')  \right] \nonumber \\
& \times  A_{\parallel 1}(\bm k) A_{\parallel 1 }(\bm k') e^{i \bm K\cdot \gyR}.
\end{align}
\\
\Cref{eq:momentlkgradgyaverphi1,eq:momentlkgradgyaverA1} for the first order and \cref{eq:momentlkgradpmugyaverphiA1,eq:momentlkgradgyaverA12,eq:momentlkgradpmugyaverA12,eq:momentlkgradpmugyaverA12} for the second order terms provide the complete Hermite-Laguerre projection of the gradient of the gyroaveraged gyrokinetic potential $\Psi_1$ in \cref{eq:momentstarlkgygradPsi1}. By using these results, $\gygrad \gyaver{\Psi_1}_{\gyR}$ can be fully expressed in terms of the gyro-moments $\gyN^{lk}(\gyR)$ and of the Fourier components of the fluctuating fields, $\phi_1(\bm k)$ and $A_{\parallel 1}(\bm k')$. 

\corr{We remark that a spatial (instead of Fourier) representation of the different FLR terms can be adopted by performing the inverse Fourier transform in the above expressions. The FLR closure can be obtained by using a Pad\'e-based approximation closure at arbitrary wavelength for a numerical implementation, avoiding cumbersome convolutions in Fourier space \citep{Held2019}. Indeed, a Pad\'e-based approximation can be applied be noticing that the kernel function $\kernel{n}$, defined in \cref{eq:kerneldef}, can be approximated by 

\begin{equation}
\kernel{n}(b_a) \simeq  \frac{(-1)^n}{n!} b_a^{2n} \frac{\partial^n}{\partial (b_a^2)^n} \sqrt{\Gamma_0(b_a^2/2)},
\end{equation}
\\
where $\sqrt{\Gamma_0}$ is the FLR operator introduced by \citet{Dorland1993}, and $\Gamma_n(x) = I_n(x)e^{-x}$ (with $I_n(x) = i^{-n} J_n(i x)$ the modified Bessel function). This FLR operator stems from the fact that $\int d \gymu d \gyvparallel d \gytheta B J_0(b) \overline{F}_{aM}/m \simeq \sqrt{\Gamma_0}$. The approximated higher order kernels follow by using exact recursive properties.
}

With the Hermite-Laguerre representation of the FLR terms and the phase-mixing operators in \cref{eq:phasemixingops}, the full perpendicular and parallel FLR couplings between gyro-moments are exactly represented at arbitrary order in $\epsilon_\perp$ in the gyro-moment hierarchy, \cref{eq:GyromomentHierarchyEquation}. Our formulation provides an exact benchmark for FLR closures considered by previous models \citep{Dorland1993,Beer1996,Snyder2001,Madsen2013}, and allows for a possible dynamical refinement of the plasma description based on the Hermite-Laguerre resolution.

\section{Gyrokinetic Collision Operator}
\label{GyrokineticCollisionOperator}
The development of a proper gyrokinetic collision operator has been subject of large analytical \citep{Catto1977a,Sugama2009,Li2011,Madsen2013,Hirvijoki2017,Pezzi2019,Pan2019} and numerical \citep{Abel2008,Barnes2009,Esteve2015} efforts, since collisions provide a transport mechanism and influence turbulence and its associated transport. We provide herein the gyro-moment expansion of a relatively simple nonlinear inter-species collision operator, the Dougherty collision operator \citep{Dougherty1964}. The expansion is valid at arbitrary $\epsilon_\perp$ values. Despite its functional simplicity, the Dougherty collision operator has the necessary field-particle terms ensuring conservation of particles, momentum, and energy. It contains pitch-angle scattering, and satisfies the H-theorem \citep[see][]{Dougherty1964}. 

Starting from the Fokker-Planck form of the collision operator,

\begin{equation} \label{eq:FokkerPlanck}
C_a(f_a(\z)) = \nu_a  \frac{\partial}{\partial \vi} \bigg \rvert_{\x} \cdot \left[ \bm D(\z)\cdot \frac{\partial}{\partial \vi} \bigg \rvert_{\x}  f_a(\z) - \bm P(\z) f_a(\z) \right],
\end{equation}
\\
where $\nu_a = 4 \pi n_a q_a^4 \ln \Lambda /( m_a^2 v_{tha}^3)$ is the velocity-independent collision frequency, $\ln \Lambda$ is the Coulomb logarithm, $\bm D(\z)$ is the velocity diffusion tensor, and $\bm P(\z)$ is the friction drag from the background particles.The Dougherty collision operator can be expressed in a Fokker-Planck form, with the velocity diffusion tensor taken to be isotropic and velocity-independent, i.e. $\bm D(\z) \equiv T_a(\x) \bm 1$, and the friction drag to be proportional to the velocity difference with the mean particle fluid velocity $\u_a$, i.e. $\bm P(\z) \equiv \u_a(\x)- \vi$. More precisely, the Dougherty collision operator expressed in the $ \z = (\x,\vi)$ coordinates is defined by

\begin{equation} \label{eq:Dougherty}
C_a(f_a(\z))  = \nu_a \frac{\partial}{\partial \vi} \bigg \rvert_{\x}  \cdot \left[  \left(\frac{T_a(\x)}{m_a} \right) \frac{\partial }{\partial \vi } \bigg \rvert_{\x} f_a(\z)   + \left(   \vi- \u_a(\x)\right) f_a(\z)\right],
\end{equation}
\\
We remark that the temperature $T_a$ and the fluid velocity $\u_a$ appearing in \cref{eq:Dougherty} are moments of the particle distribution function $f_a$, and are spatially dependent, i.e. $T_a = T_a(\x)$ and $\u_a = \u_a(\x)$. They are defined in terms of the particle distribution function $f_a = f_a(\z)$ as $T_a = \int d \vi f_a m_a (\vi - \u_a)^2/3$ and as $n_a \bm{u}_a = \int d \vi f_a \vi$ with the particle density $n_a = \int d \vi f_a$.  \corr{We note that the collision operators expressed in a Fokker-Planck form - and therefore the Dougherty collision operator - conserve the positivity of the distribution function $f_a$, and of $\gyFa$ from the scalar invariance $f_a(\z) = \gyFa(\gyZ(\z))$. In fact, if $f_a \geq  0$ is twice differentiable, and if it exists $\vi^*$ such that $f_a(\x,\vi^*) =0$ and $f_a(\x,\vi) > 0 $ for all $\vi \neq \vi^*$ with $\partial_{\vi} f_a(\x,\vi^*) =0$, then $\partial_t f_a(\x, \vi^*) = \nu_a \bm D : {\partial^2_{\vi \vi}} f_a(\x, \vi^*) \geq 0$ provided that $\bm D : \partial^2_{\vi \vi} f_a(\x,\vi^*)$ is semi positive-definite, such as in the Dougherty collision operator.}

\corr{The gyrokinetic Dougherty collision operator is obtained as an expansion in the small parameter $\epsilon_\delta$ up to $O(\epsilon_\nu \epsilon_\delta)$ (the positivity of the distribution function $\gyFa$ is therefore ensured up to $O(\epsilon_\nu \epsilon_\delta)$).} We first express the velocity derivatives of $f_a(\z)$ appearing in \cref{eq:Dougherty} in terms of the gyrocenter coordinates $\overline{\Z}$ derivatives by using the chain rule,

\begin{equation} \label{eq:partialviatx}
\frac{\partial}{\partial \vi}  f_a(\z)\bigg \rvert_{\x}  =   \frac{\partial \gyZ}{\partial \vi }\bigg \rvert_{\x} \cdot \frac{\partial }{\partial \gyZ}\gyFa(\gyZ(\z))
\end{equation}
\\
where we use the fact that $f_a(\z) = \gyFa(\gyZ(\z))$. Consistently with the accuracy requirement for the collision operator (see \cref{GyrokineticBoltzmannEquation}), we approximate $C_a(\gyFa) \simeq C_{a0}(\gyFa)$ with $C_{a0}(\gyFa)$ the lowest order collision operator in $\epsilon_\delta$ and, therefore, we neglect the $O(\epsilon_\delta)$ terms in the derivatives in \cref{eq:partialviatx} [see \cref{eq:GYcoordinates}]. At the lowest-order in $\epsilon_\delta$, we have

\begin{equation} \label{eq:dvdZ}
    \frac{\partial \gyvparallel}{\partial \vi}\bigg \rvert_{\x} = \b + O(\epsilon_\delta), \quad \frac{\partial \gymu}{\partial \vi}\bigg \rvert_{\x} = \frac{m_a \c' }{B}  + O(\epsilon_\delta), \quad \frac{\partial \gyR}{\partial \vi} \bigg \rvert_{\x} \cdot \gygrad = -  \frac{\a}{\Omega_a}  \cdot \gygrad + O(\epsilon_\delta).
\end{equation}
\\
Additionally, for the same reason, we approximate $\gyFa\simeq \gyaver{\gyFa}_{\gyR}$ for both electrons and ions [see \cref{eq:tildeFe,eq:tildeFi}] and, in particular, $\partial_{\gytheta} \gyaver{\gyFa}_{\gyR} =0$ in \cref{eq:partialviatx}. In order to express the collision operator in \cref{eq:Dougherty} as a function of $\gyZ$, we first write the fluid quantities $T_a$ and $\u_a$ in Fourier harmonics, i.e. $T_a(\x) = \sum_{\bm k'} T_a(\bm k') e^{i \bm k' \cdot \x} $, $\u_a(\x) =  \sum_{\bm k'} \u_a(\bm k') e^{i \bm k' \cdot \x} $. We also note that $\x = \gyR + \gyrhoa$. Then, introducing the Fourier expansion $\gyaver{\gyFa}_{\gyR} = \sum_{\bm k}  \gyaver{\gyFa}_{\gyR}(\bm k) e^{i \bm k \cdot \gyR}$ (we use the shorthand notation $ \gyaver{\gyFa}_{\gyR}(\bm k) \equiv \gyaver{\gyFa}_{\gyR}(\bm k,\gymu,\gyvparallel)$) and noticing that $- \a / \Omega_a \cdot \gygrad  = - i k_\perp \cos \gytheta / \Omega_a$ in \cref{eq:dvdZ}, the Dougherty collision operator, expressed as a function of the gyrocenter coordinates $\overline{\Z}$, is

 \begin{align} \label{eq:CagyRFourier}
 & C_{a0}(\gyaver{\gyFa}_{\gyR})   = \nu_a \sum_{\bm k,\bm k'} \left[ \frac{T_a(\bm k')  e^{i \bm k' \cdot \gyrhoa}}{m_a} \left(  \frac{\partial^2 }{\partial \gyvparallel^2}  \gyaver{\gyFa}_{\gyR}(\bm k)    +   \frac{2 m_a}{B} \gypmu \left( \gymu \gypmu \gyaver{\gyFa}_{\gyR} (\bm k)  \right)  \right. \right. \nonumber \\
 & \left. \left.   -  \frac{2 m_a}{B}  \frac{i k_\perp \cos \gytheta}{\Omega_a}   \gyvperp \gypmu   \gyaver{\gyFa}_{\gyR}(\bm k)   -   \frac{ i k_\perp \cos \gytheta}{\Omega_a}    \frac{1}{\gyvperp} \gyaver{\gyFa}_{\gyR}(\bm k)  - \frac{  k_\perp^2 \cos^2 \gytheta}{\Omega_a^2} \gyaver{\gyFa}_{\gyR}(\bm k)  \right)    \right. \nonumber \\
 &\left. + 3 \gyaver{\gyFa}_{\gyR}(\bm k)  + \left(  \gyvparallel - u_{\parallel a} (\bm k')  e^{i \bm k' \cdot \gyrhoa} \right) \frac{\partial}{\partial \gyvparallel } \gyaver{\gyFa}_{\gyR}(\bm k)    \right. \nonumber \\ &  \left. + \left( \gyvperp -  \u_{ a }(\bm k')  \cdot \c  e^{i \bm k' \cdot \gyrhoa} \right)  \left(  \frac{m_a \gyvperp }{B}\gypmu \gyaver{\gyFa}_{\gyR}(\bm k)    - \frac{i k_\perp \cos \gytheta}{\Omega_a} \gyaver{\gyFa}_{\gyR}(\bm k)    \right) \right] e^{i \bm K \cdot \gyR},
 \end{align}
 \\
 where $\bm K = \bm k + \bm k'$ and $ u_{\parallel a} = \b \cdot \u_a$. The collision operator in \cref{eq:CagyRFourier} is in a suitable form to be gyroaveraged. Applying the gyroaverage operator to \cref{eq:CagyRFourier} by noticing that the gyrophase dependence is present in $e^{i \bm k' \cdot \gyrhoa}$ and in the terms proportional to $\cos \gytheta$ and $\cos^2 \gytheta$, and by using the Jacobi-Anger expansion, \cref{eq:jacobianger}, to evaluate  

\begin{align}
    \gyaver{e^{i \bm k' \cdot \gyrhoa} \cos \gytheta }_{\gyR} = i J_1(b')\corr{\frac{\Delta_1}{2}}, \quad \gyaver{e^{i \bm k' \cdot \gyrhoa} \cos^2 \gytheta }_{\gyR} = \frac{1}{2} \left( J_0(b') - \corr{\frac{\Delta_2}{2}}J_2(b')\right),
\end{align}
\\
where $\Delta_n = e^{-in \alpha} + e^{i n \alpha}$, we obtain 

 \begin{align} \label{eq:GyaverCagyRFourier}
 & \gyaver{ C_{a0}(\gyaver{\gyFa}_{\gyR})}   = \nu_a \sum_{\bm k,\bm k'} \left[ \frac{T_a(\bm k')  }{m_a} \left( J_0(b') \frac{\partial^2 }{\partial \gyvparallel^2}  \gyaver{\gyFa}_{\gyR}(\bm k)    +  J_0(b')  \frac{2 m_a}{B} \gypmu \left( \gymu \gypmu \gyaver{\gyFa}_{\gyR} (\bm k)  \right)  \right. \right. \nonumber \\
 & \left. \left.   +  \frac{2 m_a}{B}  \corr{\frac{\Delta_1}{2}} \frac{ J_1(b') k_\perp }{\Omega_a}   \gyvperp \gypmu   \gyaver{\gyFa}(\bm k)_{\gyR}   +   \frac{  k_\perp J_1(b')}{\Omega_a} \corr{\frac{\Delta_1}{2}}   \frac{1}{\gyvperp} \gyaver{\gyFa}_{\gyR}(\bm k)     - \frac{1}{2}\left( J_0(b') -\corr{\frac{\Delta_2}{2}} J_2(b')\right)  \right. \right. \nonumber \\
 & \left. \left. \times \frac{  k_\perp^2 }{\Omega_a^2} \gyaver{\gyFa}_{\gyR}(\bm k)  \right)  + 3  \gyaver{\gyFa}_{\gyR}(\bm k)  + \left(  \gyvparallel - u_{\parallel a} (\bm k')  J_0(b') \right) \frac{\partial}{\partial \gyvparallel } \gyaver{\gyFa}_{\gyR}(\bm k)   +
 2 \gymu \gypmu \gyaver{\gyFa}_{\gyR}(\bm k)  \right. \nonumber \\ &  \left. 
-  \u_{ a }(\bm k')  \cdot \e_2  \left( i J_1(b')\corr{\frac{\Delta_1}{2}} \left( \frac{m_a \gyvperp }{B}\right) \gypmu \gyaver{\gyFa}_{\gyR}(\bm k)     - \left( J_0(b') - \corr{\frac{\Delta_2}{2}}J_2(b') \right)\frac{i k_\perp}{2 \Omega_a} \gyaver{\gyFa}_{\gyR}(\bm k)    \right)
 \right] e^{i \bm K \cdot \gyR}.
 \end{align}
 \\
 To derive the Hermite-Laguerre representation of the gyroaveraged Dougherty collision operator, i.e.
 
 \begin{equation} \label{eq:defCalk}
C_a^{lk}=\frac{1}{\gyN}  \int d \gymu  d \gyvparallel d \gytheta  \frac{B_\parallel^*}{m_a} \gyaver{ C_{a0} (\gyaver{\gyFa}_{\gyR}) }_{\gyR} H_a^{lk},
\end{equation}
\\
appearing on the right-hand side of \cref{eq:GYBoltzmannfinal}, we relate the Fourier harmonics of particle fluid quantities, $T_a(\bm k')$ and $\u_a(\bm k')$ in \cref{eq:CagyRFourier} to gyrocenter fluid quantities, i.e. to the moments of $\gyaver{\gyFa}_{\gyR}(\bm k')$. First, we note that the velocity integrals can be expressed as phase-space integrals. In fact, by approximating $  \overline{\Z} \simeq \Z $, we have $f_a(\z) = \int d \gyR \: \gyFa(\gyZ) \delta(\gyR + \gyrhoa -\x)$ with  $\gyFa \simeq \gyaver{\gyFa}_{\gyR}$. Second, expanding in Fourier harmonics $f_a(\z) = \sum_{\bm k} f_{a}(\bm k) e^{i \bm k\cdot \x}$ with $f_{a}(\bm k) \equiv f_a(\bm k,\vi)$, and $\gyaver{\gyFa}_{\gyR}(\gyZ) = \sum_{\bm k} \gyaver{\gyFa}_{\gyR}(\bm k) e^{i \bm k \cdot \gyR}$, we derive that $f_{a}(\bm k)  = \gyaver{\gyFa}_{\gyR}(\bm k) e^{i \bm k \cdot \gyrhoa}$. Then, the particle density $n_a(\bm k') = \int d \vi f_a(\bm k')$ can be written as  

\begin{align} \label{eq:na2Na}
    n_a(\bm k') &=  \int d \gymu d \gyvparallel d \gytheta \frac{B^*_{\parallel}}{m_a} \gyaver{\overline{F}_a}_{\gyR}(\bm k') e^{-i \bm k' \cdot \gyrhoa}  \nonumber \\
    & =  \sum_{n \geq 0} \kernel{n}(b_a')  \gyN \gyN^{*0n}(\bm k').
\end{align}
\\
where the Jacobi-Anger expansion in \cref{eq:jacobianger}, the Laguerre expansion of Bessel function in \cref{eq:Jn2Laguerre}, and the definition of $N_a^{*lk}$ [see \cref{eq:projectorstarlk}] are used. Analogously, the particle temperature $T_a(\bm k') = [T_{\parallel a}(\bm k') + 2 T_{\perp a }(\bm k')]/3$, with  $T_{\parallel a}(\bm k') = \int d \vi m_a f_a(\bm k')  (v_\parallel - u_{\parallel a})^2/ n_a$ and $ T_{\perp a}(\bm k') = \int d\vi f_a(\bm k') \mu  B/ n_a$, is expressed using 

 \begin{equation}
T_{\perp a }(\bm k')  = \frac{1 }{n_a} \sum_{n \geq 0}\kernel{n}(b'_a) \sum_{j}  \mathcal{M}_{0j}^{0n}  \Pperp  \gyN^{*0j} (\bm k'), \label{eq:pTperp}
\end{equation}
\\
 and
 
\begin{equation} \label{eq:pTparallel}
T_{\parallel a}(\bm k') =  \frac{ m_a}{n_a }  \sum_{n \geq 0} \kernel{n}(b'_a)  \sum_{p} \left( \phaseV_{apn}^{20n} - 2 u_{\parallel a}(\bm k') \phaseV_{apn}^{0n} + u_{\parallel a}(\bm k')^2 \delta_{p}^0  \right) \gyN \gyN^{*pn} (\bm k').
\end{equation}
  \\
The parallel fluid velocity $u_{\parallel a}(\bm k')$ is given by 
  
 \begin{equation} \label{eq:puparallel}
u_{\parallel a}(\bm k')  = \frac{1}{n_a} \sum_{n \geq0} \kernel{n} (b'_a)\sum_{p} \phaseV_{apn}^{0n}  \gyN \gyN^{*pn}(\bm k').
 \end{equation}
 \\
Finally, the perpendicular fluid velocity $\u_{\perp a}(\bm k') = \u_a(\bm k') - \b \cdot \u_a(\bm k') $, expressed by

\begin{equation} \label{eq:puperpdef}
 \u_{\perp a}(\bm k') = \frac{\e_2}{n_a}  \int d \gymu d \gyvparallel d \gytheta \frac{B_\parallel^*}{m_a} \gyvperp i J_1(b') \gyaver{\gyFa}_{\gyR}(\bm k'),
\end{equation}
\\
can be written in terms of gyro-moments by projecting the first order Bessel function $J_1$ into Laguerre polynomials using \cref{eq:Jn2Laguerre}. Thus, we derive from \cref{eq:puperpdef} that

 \begin{equation} \label{eq:puperp}
 \u_{\perp a}(\bm k') =  \frac{\e_2}{n_a} \sum_{n\geq0} \sum_{m=0}^{n}  \sum_{j}   \frac{i b_a' \kernel{n}(b_a')}{(n+1)}\phaseM_{0j}^{0m}  \sqrt{ \frac{ \gyTperp}{2 m_a}}  \gyN \gyN^{*0j}  (\bm k').
 \end{equation}
 \\
We remark that, even if the direction of $\u_{\perp a}$ depends on the basis vector $\e_2$ in \cref{eq:puperp}, the collisional friction term in \cref{eq:CagyRFourier} does not. Indeed, the perpendicular component of the collisional friction term appearing in \cref{eq:GyaverCagyRFourier} is proportional to $(\vi_\perp - \u_{\perp a}) \cdot \c= \vperp - \cos \theta \u_{\perp a}\cdot \e_2$. Finally, we multiply \cref{eq:GyaverCagyRFourier} by the Hermite-Laguerre basis element, $H_a^{lk}$, and perform the velocity integral by expanding the first and second-order Bessel function, $J_1$ and $J_2$, in Laguerre polynomials using \cref{eq:Jn2Laguerre}. Thus, the Hermite-Laguerre representation of the nonlinear gyrokinetic Dougherty collision operator, accurate at arbitrary values of $\epsilon_\perp$, is given by 

\begin{align} \label{eq:Calk}
C_a^{lk} & = \nu_a \sum_{\bm k}\sum_{p,q} \left[\mathcal{C}_{apq}^{lk} + \sum_{n \geq 0}\left( D_{anpq}^{lk}[T_a,b_a] + P_{anpq}^{lk}[\u_a,b_a]\right) \right] \gyN^{*pq}(\bm k) e^{i \bm k\cdot \gyR}.
\end{align}
 \\
Here, the test-particle pitch-angle scattering term $C_{apq}^{lk}$ is, 

 \begin{equation} \label{eq:Capqlk}
     C_{apq}^{lk} = 2k \delta_p^l \delta_q^{k-1} - (l+2k) \delta_p^l \delta_q^k - \sqrt{l(l-1)} \delta_p^{l-2}\delta_q^k - \frac{\sqrt{2 l }\gyuparallel}{\gyvthparallel} \delta_p^{l-1} \delta_q^k,
 \end{equation}
 \\
and the field-particle collisional term $D_{anpq}^{lk}$, associated with velocity diffusion, is

 \begin{align} \label{eq:Danpqlk}
 D_{anpq}^{lk} [T_a,b_a] =  \sum_{\bm k'}\left(D_{anpq}^{1lk}\left[ T_a(\bm k'), b_a \right] + D_{anpq}^{2lk}\left[ T_a(\bm k'), b_a \right] + D_{anpq}^{3lk}\left[ T_a(\bm k'), b_a \right] \right)e^{i \bm k' \cdot \gyR } ,
 \end{align}
 \\
 with

 \begin{align} \label{eq:Danpqlk1}
      D_{anpq}^{1lk}\left[ T_a , b_a\right] & =  (\kernel{n} (b_a') T_a(\bm k'))  \sum_{r = |n-k|}^{|n+k|} \alpha_r^{nk} \left[ \frac{ \sqrt{l (l-1)}}{ \gyTparallel    } \delta_p^{l-2} \delta_q^r -  \frac{2r}{\gyTperp}\delta_p^l \delta_q^{r-1} \right. \nonumber \\
      & \left. - \frac{1}{4 \gyTperp} b_a^2 \delta_p^l \delta_q^r\right], \\\label{eq:Danpqlk2}
        D_{anpq}^{2lk}\left[ T_a , b_a\right] & =  \sum_{m =0}^n\sum_{r = |m-k|}^{|m+k|} \corr{\frac{\Delta_1}{2}} \frac{( b_a' \kernel{n}(b'_a) T_a(\bm k')) }{(n+1)}  \alpha_{r}^{mk}    \left[ \frac{1}{ \gyTperp}  b_a \left( r \delta_p^l \delta_q^{r-1} - (1+r)\delta_p^l \delta_q^r \right) \nonumber \right.\\
&        \left. + \frac{b_a}{4 \gyTperp}\delta_p^l \delta_q^{r}    \right],  \\\label{eq:Danpqlk3}
         D_{anpq}^{3lk}\left[ T_a, b_a \right] & = \sum_{m=0}^{n}\sum_{r = |m-k|}^{|m+k|}\corr{\frac{\Delta_2}{2}} \frac{(n-m +1)}{(n+2)}  \kernel{n+1}(b_a') T_a(\bm k') \alpha_r^{mk} \frac{b_a^2 }{4\gyTperp} \mathcal{M}_{pq}^{lr}.
 \end{align}
 \\
Finally, in \cref{eq:Capqlk}, we introduce the field-particle collision term $P_{anpq}^{lk}$, associated with the fluid friction force, defined by
 
 \begin{equation} \label{eq:Panpqlk}
     P_{anpq}^{lk}\left[ \u_a,b_a\right] = \sum_{\bm k'} \left( P_{anpq}^{1lk}\left[ \u_a(\bm k'),b_a\right] + P_{anpq}^{2lk}\left[ u_{\perp a}(\bm k'),b_a\right] + P_{anpq}^{3lk}\left[ u_{\perp a}(\bm k'),b_a\right] \right) e^{i \bm k' \cdot \gyR},
 \end{equation}
 \\
 where

 \begin{align}\label{eq:Papqlk1}
     P_{anpq}^{1lk}\left[ \u_a,b_a\right] & =   \sum_{r =|n-k|}^{|n+k|} \alpha_r^{nk} \left[ \left( \kernel{n}(b_a') u_{\parallel a}(\bm k') \right) \ \frac{\sqrt{2l}}{\gyvthparallel}  \delta_p^{l-1 } \delta_q^r \right. \nonumber \\
&     \left. + \frac{1}{2}(\kernel{n}(b_a') u_{\perp a}(\bm k') ) \sqrt{\frac{m_a}{2 \gyTperp}}(i b_a) \delta_p^l \delta_q^r \right] , \\\label{eq:Papqlk2}
      P_{anpq}^{2lk}\left[ u_{\perp a} ,b_a\right] & =  \sum_{m =0}^{n} \sum_{r = |m-k|}^{|m+k|} \sqrt{\frac{m_a}{2 \gyTperp}} \corr{\frac{\Delta_1}{2}}\frac{(i b_a' \kernel{n}(b_a') u_{\perp a}(\bm k'))}{(n+1)} \nonumber \\
      & \times \alpha_r^{km}( (1+r) \delta_p^l \delta_q^r -r \delta_p^l \delta_q^{r-1} ),  \\ \label{eq:Papqlk3}
       P_{anpq}^{3lk}\left[ u_{\perp a},b_a\right] & =  \sum_{m=0}^{n} \sum_{r = |m-k|}^{|m+k|}  \corr{\frac{\Delta_2}{2}} \frac{(n-m+1)}{4(n+2)}
       \kernel{n+1}(b_a') u_{\perp a}(\bm k') \alpha_{r}^{mk} \sqrt{\frac{m_a}{2 \gyTperp}}(-i b_a) \mathcal{M}_{pq}^{lr}, 
 \end{align}
 \\
 respectively. In \cref{eq:Capqlk}, the first three terms are associated with pitch-angle and energy scattering due to inter-species collisions, whereas the last term arises from the $\uparallel$ dependence of the velocity coordinate $\sparallel$ [see \cref{eq:GYFadecomposition}]. Also, we remark the presence of hyper collisional gyro-diffusion $\sim \nu_a b_a^{2n}$ in velocity-space in \cref{eq:Danpqlk,eq:Panpqlk}. These terms arise from the gyroaverage of the collision operator \citep{Catto1977a,Abel2008,Li2011}. We notice that this collisional gyro-diffusion yields a classical diffusion in the gyrocenter continuity equation, obtained by setting $(l,k)=(0,0)$ in the gyro-moment hierarchy equation in \cref{eq:GyromomentHierarchyEquation}, associated with FLR effects \citep{Sugama2015,Sugama2017,Mandell2018}.

Finally, it is instructive to consider the drift-kinetic limit of the gyrokinetic Dougherty collision operator in \cref{eq:Calk}. We take $ b_a \sim \epsilon_\perp \sim \epsilon$ and neglect the $O(\epsilon_\perp^2)$ terms. Using the fact that $\kernel{n}(b_a)\sim b_a^{2n}$ with $b_a \ll 1$
for $n > 1 $, while $\kernel{0}(b_a) \simeq 1 - b_a^2 / 4 $, we derive $n_a = \gyN$, $u_{\parallel a} = \gyuparallel$, $T_{\perp a} = \gyTperp$, $T_{\parallel a} = \gyTparallel$ from \cref{eq:pTperp,eq:pTparallel,eq:puparallel,eq:puperp}, in agreement with \citet{Jorge2017}. Thus, from \cref{eq:Danpqlk,eq:Panpqlk}, the field-particle collisional terms reduce to
 
 \begin{equation} \label{eq:Da0pqlk}
      D_{a0pq}^{lk}\left[ T_a , b_a\right] =  \frac{\overline{T}_a}{\overline{T}_{\parallel a}} \sqrt{l(l-1)} \delta_{p}^{l-2} \delta_q^{k} - \frac{\overline{T}_a}{\gyTperp} 2 k \delta_p^l \delta_q^{k-1},
 \end{equation}
 \\
 and
 
 \begin{equation} \label{eq:Pa0pqlk}
      P_{a0pq}^{lk}\left[\u_a ,b_a\right] =\frac{\sqrt{2 l }\gyuparallel}{\gyvthparallel} \delta_p^{l-1} \delta_q^k.
 \end{equation}
 \\
As a consequence, the drift-kinetic Doughery collision operator takes the form 
 
 \begin{align} \label{eq:DKCalk}
C_a^{lk} & = \nu_a \sum_{p,q} \left(\mathcal{C}_{apq}^{lk} +  D_{a0pq}^{lk}\left[ T_a ,b_a\right] + P_{a0pq}^{lk}\left[ \u_a,b_a\right] \right) \gyN^{*pq}(\gyR).
\end{align}
 \\
having performed the inverse Fourier transform. We remark that the absence of gyro-diffusion in the gyrocenter density in the drift-kinetic regime, since $C_a^{00}=0$ from \cref{eq:DKCalk}. 

\section{Gyrokinetic Maxwell's Equations}
\label{GyrokineticMaxwellEquations}
In this section, we derive a set of gyrokinetic Maxwell's equation that describe the temporal and spatial evolution of $\phi_0$, $\phi_1$, $A_{\parallel 0}$, $A_{\parallel 1}$. Moreover, we provide an additional field equation to self-consistently obtain the large scale magnetic vector potential $\hat{\bm A}$, and therefore $\hB$. These equations are derived self-consistently from a variational principal. While previous gyrokinetic field theories (see, e.g., \citet{Brizard2000,Madsen2010,Tronko2016}) have been used for collisionless plasmas, \corr{\citet{Sugama2015} shows that the conservation laws obtained from collisionless gyrokinetic field theories are still valid in collisional plasmas provided that the collision operator model conserves particle, momentum, and energy. Therefore, similar procedures to the ones applied in, e.g., \citet{Sugama2000,Brizard2010,Tronko2016}, and based on Noether's method, can be applied to the present model in order to derive energy, parallel and toroidal angular momentum conservation laws. These invariant properties are preserved by the Hermite-Laguerre projection in the infinite moment limit, and are altered if one considers a truncated gyro-moment hierarchy. However, alternative conservation laws associated with the truncated gyro-moment hierarchy can still be derived within the same formalism, by using the truncated distribution function to evaluate the gyrokinetic functional action \citep{Madsen2013}. The derivation of conservation laws is outside the scope of the present work.}

Contrary to \citet{Brizard2000,Tronko2016}, in our approach, we do not treat the distribution function as a dynamical field. Instead, we derive the equations for the electromagnetic fields, assuming the single particle dynamics model in \cref{Singleparticledynamicsinthetokamakperiphery}. Our field equations include the effects of strong flows and retain full FLR effects at arbitrary $\epsilon_\perp$ values to predict accurately the long and short wavelength components of the fluctuating electromagnetic fields \citep{Qin1998,Parra2008,Miyato2013}, as they might be important in the description of anomalous transport in the periphery. Within this framework, the polarization and magnetization current densities, associated with the particle and gyrocenter difference yields a classical physical interpretation of the gyrokinetic medium \citep{Qin2000,Krommes2012}. We compare our results with previous derivations and give a simple physical interpretation of the obtained equations. 

The present section is organized as follows. In \cref{GyrokineticFieldTheory}, we introduce the general formalism of our variational principle. Then, in \cref{GyrokineticPoissonEquations,GyrokineticAmpereslaw}, we derive the gyrokinetic Poisson's equations and gyrokinetic Ampere's laws from the least action principle and recover known results. Finally, in \cref{LeadingOrderEquilibriumPressureBalanceEquation}, we obtain an Ampere's law that sets the evolution of $\hat{\bm A}$, and show that our gyrokinetic formalism encompasses a leading order pressure balance equation.
\subsection{Gyrokinetic Field Theory}
\label{GyrokineticFieldTheory}

Following previous gyrokinetic literature \citep{Sugama2000,Brizard2000,Brizard2010,Squire2013,Sugama2014,Sugama2015,Tronko2016}, we introduce the gyrokinetic functional action $\mathcal{A}$,

\begin{equation} \label{eq:actionA}
    \mathcal{A}[\phi_0, \phi_1,\hat{\A},A_{\parallel 0}, A_{\parallel 1}] = \mathcal{A}_f[\phi_0, \phi_1,\hat{\A},A_{\parallel 0}, A_{\parallel 1}] + \mathcal{A}_p[\phi_0, \phi_1,\hat{\A},A_{\parallel 0}, A_{\parallel 1}] + \mathcal{A}_c,
\end{equation}
\\
where $\mathcal{A}_p$ is the gyrocenter functional action associated with the gyrocenter dynamics, $\mathcal{A}_f$ is the field functional action containing the fields contributions, and, finally, $\mathcal{A}_c$ is a collisional functional action that we assume independent of the dynamical fields. The field functional action $\mathcal{A}_f$ is given by

\begin{equation} \label{eq:actionAf}
    \mathcal{A}_f[\phi_0, \phi_1,\hat{\A},A_{\parallel 0}, A_{\parallel 1}] = \int dt \int \frac{d \x}{8 \pi} \left( \left|\E + \E_1 \right|^2  -  \left| \B +  \B_1\right|^2 \right),
\end{equation}
\\
where the total electric field is $\E + \E_1$ with $\E = -\grad \phi_0 - \partial_t(\hat{\A} + \hb A_{\parallel 0})$ and $\E_1 = - \grad \phi_1 - \partial_t (A_{\parallel 1} \hb)$, and the total magnetic field is $\B + \B_1$ where $\B = \hB + \grad \times (A_{\parallel 0}\hb)$ with $\hB = \grad \times \hat{\A}$ and $\B_1 = \grad \times (A_{\parallel 1} \hb)$ [see \cref{PeripheryOrdering}]. In \cref{eq:actionAf}, the electromagnetic fields are evaluated at the particle position $\x$. We remark that the contribution from the large scale magnetic vector potential $\hat{\A}$ is included since it is a dynamical field in our approach. We remark that the quasineutrality can be imposed by neglecting the $|\E + \E_1|^2$ term in $\mathcal{A}_{f}$ as usually applied in gyrokinetic models \citep[see, e.g.,][]{Krause2007,Madsen2013,Bottino2015,Tronko2016,Tronko2017b}. Moreover, the inductive part of the electric field is retained in \cref{eq:actionAf} since neglecting it in $\mathcal{A}_f$ can lead to spurious terms in the local conservation laws \citep{CorreaRestrepo2005}.

The gyrocenter functional action, $\mathcal{A}_p$, is defined as 

\begin{equation} \label{eq:actionAp}
    \mathcal{A}_p[\phi_0, \phi_1,\hat{\A},A_{\parallel 0}, A_{\parallel 1}] = \int d t  \sum_a \int d \gyZ \frac{B_\parallel^*}{m_a}\gyFa(\overline{\Z}) \overline{L}_a[\phi_0,\phi_1,\hat{\A},A_{\parallel 0}, A_{\parallel 1}](\gyZ,\dot \gyZ),
\end{equation}
\\
with  $d \gyZ = d \gyR d \gymu d \gyvparallel d \gytheta$, and $\overline{L}_a$ the single gyrocenter Lagrangian obtained from the gyrocenter one-form in \cref{eq:GYGamma},

\begin{equation} \label{eq:La}
    \overline{L}_a[\phi_0,\phi_1,\hat{\A},A_{\parallel 0}, A_{\parallel 1}](\gyZ,\dot \gyZ) =q_a  \overline{\A^*}[\hat{\A},A_{\parallel 0},\phi_0](\gyZ) \cdot \dot \gyR + \frac{B\gymu}{\Omega_a} \dot \gytheta - \overline{\mathcal{H}}[\phi_0,\phi_1,\hat{\A},A_{\parallel 0}, A_{\parallel 1}](\gyZ).
\end{equation}
\\
We remark that in \cref{eq:La}, the fields $\phi_0$, $\hat{\A}$ and $A_{\parallel 0}$ are evaluated at $\gyR$, while $\phi_1$ and $A_{\parallel 1}$ at $\gyR + \gyrhoa$ being gyroaveraged at constant $\gyR$ in $\overline{\mathcal{H}}$. This difference in the spatial argument between $\mathcal{A}_{f}$ and $\mathcal{A}_p$ yields polarization and magnetization effects. Since the $O(\epsilon^2,\epsilon_\delta^2)$ single gyrocenter Lagrangian $\overline{L}_a$ is gyrophase independent [see \cref{eq:GYGamma}], we can expand $\gyFa = \gyaver{\gyFa}_{\gyR} + \widetilde{\gyFa}$ and perform the $\gytheta$-integral in \cref{eq:actionAp}, showing that the contribution of $\widetilde{\gyFa}$ in $\mathcal{A}_p$ vanishes. 

We now derive a set of gyrokinetic field equations, i.e. the gyrokinetic Poisson's equations and the Ampere's laws, from the least action principle

\begin{equation}
\delta \mathcal{A} =0,
\end{equation}
\\
where the total variation of the gyrokinetic action $\mathcal{A}$, in \cref{eq:actionA}, is given by

\begin{equation} \label{eq:deltaA}
\delta \mathcal{A} =\frac{\delta \mathcal{A}}{\delta \phi_0}\circ \check{\phi}_0  + \frac{\delta \mathcal{A}}{\delta \phi_1}  \circ \check{\phi}_1 + \frac{\delta \mathcal{A}}{\delta \hat{\A}} \circ \check{\hat{\A}}+   \frac{\delta \mathcal{A}}{\delta A_{\parallel 0}} \circ \check{A}_{\parallel 0} +\frac{\delta \mathcal{A}}{\delta A_{\parallel 1}}\circ \check{A}_{\parallel 1},
\end{equation}
with $(\check{\phi}_0,\check{\phi}_1,\check{\hat{\A}},\check{A}_{\parallel 0},\check{A}_{\parallel 1})$ arbitrary test functions. The $\circ$ notation in \cref{eq:deltaA} denotes the functional derivative along an arbitrary test function, such as, e.g., 

\begin{equation} \label{eq:functionalderivative}
    \frac{\delta \mathcal{A}}{\delta \phi_0} \circ \check{\phi}_0 \equiv \frac{d}{d \epsilon} \mathcal{A}[ \phi_0 + \epsilon \check{\phi}_0, \phi_1,\hat{\A},A_{\parallel 0}, A_{\parallel 1}] \bigg \rvert_{\epsilon=0},
\end{equation}
\\
with the spatial argument of the test function $\check{\phi}_0$ being the same as $\phi_0$. Analogously, the variation with respect to a vector function is defined as the sum of the variation with respect to its component, i.e. $\delta \mathcal{A}/ \delta \hat{\A} \circ \check{\hat{\A}} \equiv \delta \mathcal{A}/ \delta \hat{\A}_i \circ \check{\hat{\A}}_i$.

We remark that, from the total variation in \cref{eq:deltaA}, two gyrokinetic Poisson's equations and three Ampere's laws are obtained. These coupled equations are necessary to determine the fields $\phi_0$, $\phi_1$, $A_{\parallel 0}$, and $A_{\parallel 1}$. We note that, by imposing the quasineutrality condition, a plasma vorticity equation is derived. One additional constraint, which corresponds to the pressure balance obtained from the evolution equation of $\hat{\A}$, can also be derived, as explained in \cref{LeadingOrderEquilibriumPressureBalanceEquation}.
\subsection{Gyrokinetic Poisson's Equations}
\label{GyrokineticPoissonEquations}
The variation of the action $\mathcal{A}$ with respect to $\phi_0$ yields the first gyrokinetic Poisson's equation, referred to as GKPI, whereas the variation with respect to $\phi_1$ produces a second gyrokinetic Poisson's equation that we denote by GKPII.

We first compute the variation of the action $\mathcal{A}$ with respect to $\phi_0$, and obtain GKPI by imposing that 

\begin{equation}\label{eq:deltaAdeltaphi0}
\frac{\delta \mathcal{A}}{\delta \phi_0} \circ \check{\phi}_0 =0.
\end{equation}
\\
The functional derivative of the field functional action, $\mathcal{A}_f$ in \cref{eq:actionAf}, is given by

\begin{equation} \label{eq:deltaAfdeltaphi0}
 \frac{\delta \mathcal{A}_f}{\delta \phi_0} \circ \check{\phi}_0   =  \int d t \int \frac{d \x}{4 \pi}\grad \cdot   \left( \E + \E_1  \right) \check{\phi}_0 - \int d t\int \frac{d \x}{4 \pi} \grad \cdot \left[   \left( \E + \E_1\right)  \check{\phi}_0 \right],
\end{equation}
\\
where the test function $\check{\phi}_0$ is evaluated at $\x$, i.e. $\check{\phi}_0 =\check{\phi}_0(\x)$. The first term in \cref{eq:deltaAfdeltaphi0} is identified as the dynamical term, i.e. a term entering in the field equation, whereas the second one is a boundary term that vanishes by carrying out the integration and using a proper choice of $\check{\phi}_0$.

We now compute the variation of the gyrocenter functional action $\mathcal{A}_p$ in \cref{eq:actionAp}. We remark that the single gyrocenter Lagrangian $\overline{L}_a$, in \cref{eq:La} depends on $\phi_0$ and on its gradient $\gygrad \phi_0$, with both quantities evaluated at $\gyR$, through the electrostatic energy, $ q_a \phi_0$, the symplectic components, $q_a \overline{\bm A^*}$, the $\E \times \B$ kinetic energy, proportional to $ |\gygrad \phi_0|^2$, and, finally, the guiding-center FLR correction term, proportional to $ ( \mu B/2 \Omega_a) \b \cdot \gygrad \times (\b \times \gygrad \phi_0/B) $. The variation of $\mathcal{A}_p$ in \cref{eq:actionAp} along $\check{\phi}_0$ is

\begin{align} \label{eq:deltaApdeltaphi0}
\frac{\delta \mathcal{A}_p}{\delta \phi_0} \circ \check{\phi}_0 &=  - \sum_a q_a \int d t\int d \x \int d \gyZ \delta(\gyR - \x) \frac{B_\parallel^*}{m_a} \gyaver{\gyFa}_{\gyR} \left[ \check{\phi}_0 \right.  \nonumber \\
& \left.  +  \frac{m_a}{B} \left(\b \times [ \dot{\gyR} - \u_E ]\right) \cdot \grad \check{\phi}_0 + \frac{\gymu B}{2 \Omega_a} \b \cdot \grad \times \frac{(\b \times \grad \check{\phi}_0)}{B} \right],
\end{align}
\\
where the spatial argument of the test function $\check{\phi}_0$ is translated from $\gyR$ to the particle position $\x$ by noticing $\check{\phi}_0(\gyR) = \int d \x  \delta(\gyR -\x) \check{\phi}_0(\x) $. This is to be consistent with the fact that the variation of field functional action, $\mathcal{A}_f$ in \cref{eq:deltaAfdeltaphi0}, is evaluated at the particle position $\x$. To evaluate the terms that provide dynamics of $\phi_0$ in \cref{eq:deltaApdeltaphi0}, we carry out successive integration by parts on terms where derivatives of $\check{\phi}_0$ appear. We obtain GKPI from \cref{eq:deltaAdeltaphi0}, \cref{eq:deltaAfdeltaphi0}, and \cref{eq:deltaApdeltaphi0}, using the arbitrariness of the test function $\check{\phi}_0$ under the assumption that it vanishes on the boundary. Since GKPI is a functional relation, we evaluate GKPI at $\gyR$ as this is where the gyro-moments are evaluated [see \cref{eq:GyromomentHierarchyEquation}], and in agreement with previous gyrokinetic models \citep{Dannert2005,Pan2018}. It yields

\begin{align} \label{eq:GKPI}
\gygrad \cdot \left( \E + \E_1 \right) =   4 \pi \sum_a \left( \overline{\varrho}_a^* -   \gygrad \cdot \bm{\mathcal{P}}^{*}_a \right).
\end{align}
\\
In \cref{eq:GKPI}, the gyrocenter charge density $\overline{\varrho}_a^*$ is 

\begin{equation} \label{eq:varrhoI}
     \overline{\varrho}_a^* = q_a \int d \gymu d \gyvparallel d \gytheta
      \frac{B_\parallel^*}{m_a} \gyaver{\gyFa}_{\gyR},
\end{equation}
\\
whereas the polarization charge density, $- \gygrad \cdot \bm{\mathcal{P}}_a^{*}$, is associated with the polarization

\begin{equation}
\label{eq:PaI}
   \bm{\mathcal{P}}_a^{*}  = \bm{\mathcal{P}}_a^{PK*} + \bm{\mathcal{P}}_a^{D*},
\end{equation} 
\\
where we have introduced the Pfirsch-Kaufman (PK) polarization \citep{Pfirsch1985,Kaufman1986a},

\begin{equation} \label{eq:PaIPK}
    \bm{\mathcal{P}}_a^{PK*} = \int d \gymu d \gyvparallel d \gytheta \frac{B_\parallel^*}{m_a} \gyaver{\gyFa}_{\gyR} \frac{m_a}{B} \b \times \left[\dot{\gyR} -\u_E\right]_\perp,
\end{equation}
\\
and the diamagnetic polarization

\begin{equation} \label{eq:PaID}
    \bm{\mathcal{P}}_a^{D*} =-  \frac{\b}{B} \times \gygrad \times \left( \int   d \gymu d \gyvparallel d \gytheta \frac{B_\parallel^*}{m_a} \gyaver{\gyFa}_{\gyR}\frac{ \gymu B}{2 \Omega_a} \b\right).
\end{equation}
\\
 We remark that the polarization $\bm{\mathcal{P}}_a^{PK*}$ in \cref{eq:PaIPK} is a known term in guiding-center theories \citep[see, e.g.,][]{Kaufman1986a,Pfirsch1985,Lee2009,Brizard2011}. This polarization term, proportional to the difference $\dot{\gyR} - \u_E$ and a consequence of writing the particle velocity $v_\perp$ in the frame moving with the background $\E \times \B$, is due to the presence of the polarization drift, $\b \times d_t^0 \U_0 /\Omega_a$, the magnetic drifts, and the small-scale drifts, $ \B \times \gygrad_\perp \gyaver{\Psi_1}_{\gyR}/B^2$. The $O(\epsilon^2)$ guiding-center FLR correction term in \cref{eq:deltaApdeltaphi0} yields the polarization $\bm{\mathcal{P}}_a^{D*}$ in \cref{eq:PaID}, which is associated with the perpendicular fluid pressure. 

We now derive the Hermite-Laguerre representation of GKPI in \cref{eq:GKPI}. We perform the velocity integral in \cref{eq:varrhoI} to find 

\begin{equation} \label{eq:varrho}
\overline{\varrho}_a^* = q_a \int d \gymu d \gyvparallel d \gytheta \frac{B_\parallel^*}{m_a} \gyaver{\gyFa}_{\gyR } = q_a  \gyN^* \gyN.
\end{equation}
\\
A similar procedure can be used to obtain the Hermite-Laguerre projections of the polarization terms, in \cref{eq:PaIPK,eq:PaID}, that are 

\begin{align} \label{eq:PaIPKHL}
    \bm{\mathcal{P}}_a^{PK*} &= \frac{m_a\gyN}{B} \b \times \left[ \U_{pa} + \frac{1}{2} \U_{\kappa} + \U_{\grad} + \U_{\omega} + \U_{\mu  a}^\perp \right. \nonumber \\
    &\left.  + \frac{\Qperp }{\Pperp \vthparallel}\left(\U_{Ba} + \U_{\mu a }^{\parallel \perp} \right)  +  \frac{1}{N_a B}\b \times \moment{}{\gygrad \gyaver{\Psi_1}_{\gyR}}\right],
\end{align}
\\
where we use \cref{eq:momentdotR} with the gyrocenter drifts defined in \cref{eq:GYdrifts}, and

\begin{equation} \label{eq:PaIDHL}
    \bm{\mathcal{P}}_a^{D* } = - \frac{\b}{B} \times \gygrad \times \left[ \frac{\Pperp}{2 \Omega_a}\corr{\left( \frac{\overline{B}_{\parallel a}^*}{B} + \frac{\Qperp}{\Pperp}\frac{\b \cdot \gygrad \times \b}{\Omega_a} \right)}\b\right],
\end{equation}
\\
respectively. 
    
    It is instructive to consider the drift-kinetic limit of GKPI to verify its consistency with previous results. For this purpose, we keep the polarization densities up to $O(\epsilon^2)$ in \cref{eq:PaI}. Therefore, only the diamagnetic polarization, $\bm{\mathcal{P}}_a^{D*}$, is retained in $\bm{\mathcal{P}}_a^{*}$ [the $\bm{\mathcal{ P}}_a^{PK*}$ polarization is $O(\epsilon^2,\epsilon_B)$, neglecting the $O(\epsilon_\delta)$ and higher corrections]. Approximating $B_\parallel^* \simeq B + O(\epsilon^2)$ in \cref{eq:PaID} and using $\b \times (\gygrad \times \b) \simeq - \gygrad_\perp B/B$, we derive the drift-kinetic Poisson's equation from \cref{eq:GKPI},

\begin{equation} \label{eq:DKP}
    \gygrad \cdot \E =  4 \pi \sum_a q_a \left[ \left( 1 + \frac{\b \cdot \gygrad \times \u_E}{\Omega_a} + \frac{\gyuparallel \b \cdot \gygrad \times \b}{\Omega_a} \right) \gyN + \frac{1}{2 m_a} \gygrad_\perp^2 \left( \frac{\Pperp}{\Omega_a^2} \right)\right].
\end{equation}
\\
 \Cref{eq:DKP} corresponds to the drift-kinetic Poisson's equation used by \citet{Jorge2017} with $\E = - \gygrad \phi_0$.

We now aim to derive the second gyrokinetic Poisson's equation, GKPII, from the variation of $\mathcal{A}$ in \cref{eq:deltaA} with respect to $\phi_1$, i.e.

\begin{equation} \label{eq:deltaAdeltaphi1}
\frac{\delta \mathcal{A}}{\delta \phi_1} \circ  \check{\phi_1} =0.
\end{equation}
\\
We first notice that the variation of the field functional action $\mathcal{A}_f$ with respect to $\phi_1$ has the same functional form as \cref{eq:deltaAfdeltaphi0}, i.e.

\begin{equation} \label{eq:deltaAfdeltaphi1}
    \frac{\delta \mathcal{A}_p}{\delta \phi_1} \circ \check{\phi}_1  = 
    \int d t \int \frac{d \x}{4 \pi}\grad \cdot   \left( \E + \E_1  \right) \check{\phi}_1 - \int d t\int \frac{d \x}{4 \pi} \grad \cdot \left[   \left( \E + \E_1\right)  \check{\phi}_1 \right],
\end{equation}
\\
with the test function $\check{\phi}_1$ evaluated at $\x$. To evaluate the variation of the gyrocenter functional action $\mathcal{A}_p$ in \cref{eq:actionAp}, we notice that the $\phi_1$ dependent terms in the gyrocenter Lagrangian $L_a$ in \cref{eq:La} are contained only in the gyrokinetic potential $\Psi_1$. Due to the complexity of the $O(\epsilon_\delta^2)$ term in $\overline{\mathcal{H}}$ contained in $\gyaver{\Psi_1}_{\gyR}$ [see \cref{eq:GYpotential}] and for numerical applications, a gyrokinetic \textit{long wavelength limit} is usually considered, neglecting the $O(\epsilon_\perp^3)$ FLR corrections \citep[see, e.g.,][]{Dubin1983,Lee1983,Hahm1988,Xu2007,Cohen2008,Hahm2009,Madsen2013,Tronko2016,Tronko2017,Tronko2017b,Shi2017}. However, it has been argued that higher order terms in $\epsilon_\perp$ are needed to correctly study also long wavelength modes and to predict the turbulent and neoclassical transport resulting from the nonlinear interactions between scales \citep{Scott2003,Parra2008,Lee2009,Miyato2013}. We therefore retain the full expression of $\gyaver{\Psi_1}$, and, as a consistency check, we show that the commonly used expression of the linear polarization is recovered in the \textit{long wavelength limit}.

The variation of the gyrocenter functional action $\mathcal{A}_p$ with respect to $\phi_1$ is

\corr{
\corrs{\begin{align} \label{eq:deltaApdeltaphi1}
    \frac{\delta \mathcal{A}_p}{\delta \phi_1} \circ \check{\phi}_1 & =- \sum_a q_a \int d t \int d \x \int d \gymu d \gyvparallel d \gytheta \left[ \gyaver{     \frac{B_\parallel^*}{m_a}  \gyaver{\gyFa}_{\gyR} }_{\x}^\dagger    \check{\phi}_1(\x)  \right.\nonumber \\
    & \left. +    \frac{  q_a^2}{ m_a \Omega_a} \left( \gyaver{ \gypmu  \left( \frac{B_\parallel^*}{m_a}  \gyaver{\gyFa}_{\gyR}\right) }_{\x}^\dagger   \Phi_1(\x)  - \gyaver{  \gyaver{ \Phi_1}_{\gyR} \gypmu \left(\frac{B_\parallel^*}{m_a}  \gyaver{\gyFa}_{\gyR} \right)  }_{\x}^\dagger  \right)  \check{\phi}_1(\x)      \right. \nonumber \\
    & \left. + \int d \gyR  \frac{B_{\parallel }^*}{m_a} \gyaver{\gyFa}_{\gyR} \frac{q_a}{2 m_a \Omega_a^2} \b \cdot \left(    \gygrad \left[\int^{\gytheta} d \gytheta'\widetilde{ \delta(\gyR + \gyrhoa -\x) \check{\phi}_1(\x)}\right] \times   \gygrad \widetilde{\Phi_1} \right. \right. \nonumber \\
&    \left. \left.+
    \gygrad \left[\int^{\gytheta} d \gytheta' \widetilde{\Phi_1}\right] \times     \gygrad \widetilde{\delta(\gyR + \gyrhoa -\x)\check{\phi}_1(\x) } \right)\right].
    \end{align}}
\\
In \cref{eq:deltaApdeltaphi1}, using that $\check{\phi}_1(\gyR + \gyrhoa) = \int d \x \delta(\gyR + \gyrhoa -\x)\check{\phi}_1(\x)$, we introduce the adjoint gyroaverage operator, $ \gyaver{\xi}_{\x}^\dagger = \gyaver{\xi}_{\x}^\dagger(\x,\gymu,\gyvparallel,t)$ acting on a function $\xi = \xi(\gyR,\gymu,\gyvparallel,t)$, defined by

\begin{align} \label{eq:adjointop}
\gyaver{\xi}_{\x}^\dagger & = \frac{1}{2 \pi}\int_0^{2 \pi} d \gytheta \int d \gyR \delta(\gyR + \gyrhoa - \x)   \xi(\gyR,\gymu,\gyvparallel,t) \nonumber \\
 & =\sum_{i\geq 0} \frac{ \grad_\perp^{2i}}{2^{2i} i! i!}  \left[(\overline{\bm \rho} \cdot \overline{\bm \rho})^i \xi(\x, \gymu, \gyvparallel,t) \right],
\end{align}
\\
and, from \cref{eq:gyaveroperator}, satisfies the property that 

\begin{equation} \label{eq:ajointpro}
\int d \gyR \gyaver{\chi}_{\gyR} \xi = \int d \x \gyaver{\xi}_{\x}^\dagger \chi.
\end{equation}
\\ 
In \cref{eq:deltaApdeltaphi1}, the adjoint gyroaverage operator, $\gyaver{\cdot}_{\x}^\dagger$, arises from the fact that the gyrocenter distribution function is gyroaveraged in the field equations giving the averaged density (or current density) at the particle position $\x$. Using \cref{eq:gyaverLaguerre}, we introduce the adjoint kernel $\kernel{n}(b_a^\dagger)$ associated with the Laguerre expansion of the adjoint gyroaverage operator $\gyaver{\cdot}_{\x}^\dagger$, where $b_a^\dagger$ is such that $(b_a^\dagger)^{2j} \to (-1)^j 2^j  \gygrad_\perp^{2}( \gyTperp / m_a\Omega_a^2)^j$ \citep{Strintzi2004}. With the adjoint kernel and \cref{eq:gyaverLaguerre}, we obtain the spatial representation of the gyroaverage adjoint operator,

\begin{align}
\gyaver{\xi}^\dagger_{\x} & = \sum_{\bm k} \sum_{n \geq 0} \kernel{n}(b_a^\dagger)  L_{n}(\sperp^2)  \xi(\bm k, \gymu, \gyvparallel,t) e^{i \bm k \cdot \x } \nonumber \\
 & =  \sum_{n \geq 0} \sum_{m \geq 0} \frac{(-1)^{n}}{n!m!}  \gygrad_\perp^{2(n+m)}   \left[  \frac{ \Tperp }{2 m_a \Omega_a^2}  L_{n}(\sperp^2)  \xi(\x, \gymu, \gyvparallel,t)\right].
\end{align}
\\
We remark that, consistently with \cref{eq:ajointpro}, the spatial dependence of fluid quantities contained in $b_a^\dagger$, such as $\gyN$ and $\gyTperp$, and contained in the argument $\sperp^2$ of the Laguerre polynomials must be retained when evaluating the adjoint gyroaverage operator. \corrs{Finally, we notice that it can be shown that the polarization associated with the last term in \cref{eq:deltaApdeltaphi1} is smaller by at least one order compared to the second term in the same equation \citep{Hahm2009}, and can therefore be neglected.}

GKPII is obtained from \cref{eq:deltaApdeltaphi1} by first performing the $\gyR$-integral, that is contained in the $\gyaver{\cdot}_{\x}^\dagger$ operator, by expanding the gyroradius, $\gyrhoa = \bm{\rho}_a(\gyR,\gymu,\gytheta)$ defined in \cref{eq:G1Rrhoa} that appears in $\delta(\gyR + \gyrhoa - \x)$}, such as

 \begin{equation} \label{eq:rhoatR}
   \bm{\rho}_a(\gyR,\gymu,\gytheta)=   \gyrhoa(\x - \gyrhoa(\x,\gymu,\gytheta),\gymu,\gytheta) \simeq \gyrhoa(\x,\gymu,\gytheta) + O\left( \left|    \gyrhoa \cdot \grad \ln B \right| \right),
 \end{equation}
\\
 where the second term can be neglected being $O(\epsilon_B)$ [see \cref{eq:epsilonB}]. We remark that, this term may become important in the case where $L_\phi \sim L_P \sim L_B$, i.e. $\epsilon \sim \epsilon_B$, typical of the tokamak core. Using the arbitrariness of the test function $\check{\phi}_1$ in \cref{eq:deltaAfdeltaphi1,eq:deltaApdeltaphi1}, \cref{eq:deltaAdeltaphi1} leads to the gyrokinetic Poisson's equation GKPII, that we evaluate at $\gyR$, 

\begin{equation} \label{eq:GKPII}
\gygrad \cdot \left( \E + \E_1 \right)  =  4 \pi \sum_a \left(\gyaver{\overline{\varrho}_a^*} +  \mathcal{P}_a^{\mu *}\right).
\end{equation}
\\
In \cref{eq:GKPII}, the gyroaveraged gyrocenter charge density $\gyaver{\overline{\varrho}_a^*}$ are given by

\begin{equation} \label{eq:gyavervarrhoGKPII}
    \gyaver{\overline{\varrho}_a^*}  = q_a \sum_{\bm k}\int d \gymu d \gyvparallel d \gytheta \frac{B_\parallel^*}{m_a} \gyaver{\gyFa}_{\gyR}(\bm k) e^{- i \bm k \cdot \gyrhoa} e^{i \bm k \cdot \gyR},
\end{equation}
\\
whereas the polarization charge density $\mathcal{P}_a^{\mu *}$ is

\begin{align} \label{eq:PaII}
\mathcal{P}_a^{\mu *} & = q_a \sum_{\bm k ,\bm k'}\int d \gymu d \gyvparallel d \gytheta \frac{B_\parallel^*}{m_a} \frac{q_a}{B} \left[\Phi_1(\bm k) - \gyaver{\Phi_1}_{\gyR}(\bm k) e^{- i \bm k \cdot \gyrhoa} \right] \nonumber \\
& \times \gypmu\gyaver{\gyFa}_{\gyR}(\bm k') e^{- \bm k' \cdot \gyrhoa} e^{i \bm K \cdot \gyR}.
\end{align}
\\
with $\bm K = \bm k + \bm k'$. We remark that the expressions of the gyroavereaged gyrocenter and polarization charge densities can be obtained by using the pull-back transformation of the gyrocenter distribution function [see \cref{eq:fTepsilonF}]. Indeed, the charge particle density $\varrho_a = q_a \int d \vi f_a$, can be written as

\begin{align} \label{eq:densitypullback}
\varrho_a & = q_a \int d \bm x' d \vi \delta(\bm x' -\bm x) f_a(\z) \nonumber \\
& = q_a \int d \R d \mu d v_\parallel d \theta \delta(\R + \rhoa - \x) \frac{B_\parallel^*}{m_a} \left[ 1 + \frac{q_a}{m_a}\widetilde{\Phi}_1  \frac{\partial }{\partial \mu}\right] \gyaver{\gyFa}_{\gyR},
\end{align}
\\
where we use that $f_a(\z(\Z) ) = T_{\epsilon_\delta} \gyFa(\Z) \simeq [1 + \gbmu_1 \partial_{\gymu}] \gyaver{\gyFa}_{\gyR} + O(\epsilon_\delta^2) $ neglecting the $\gbR_1 \cdot \grad \gyFa$ and $\gbparallel_1 \partial_{\vparallel} \gyFa$ terms since they can be shown to be higher order \citep{Hahm2009}. We identify $\gyaver{\overline{\varrho}_a^*}$ and $\mathcal{P}_a^{* \mu}$ as the first and second term in the integral of \cref{eq:densitypullback}, respectively. This transformation of the distribution function in \cref{eq:densitypullback} is a common technique to introduce the gyrocenter distribution function in the field equations in the gyrokinetic literature \citep{Dubin1983,Hahm1988,Hahm1996,Hahm2009}. 

The Hermite-Laguerre projection of GKPII is obtained by performing the velocity integrals in \cref{eq:gyavervarrhoGKPII,eq:PaII}. \corr{Using the Jacobi-Anger identity in \cref{eq:jacobianger} and expanding the Bessel functions in Laguerre polynomials using \cref{eq:Jn2Laguerre,eq:LrnLjsperp2Le,eq:LrnLj2Lf}} yields the gyro-moment expansions of the gyroaveraged gyrocenter and polarization charge densities, respectively,

\begin{equation}\label{eq:gyavervarrhoGKPIIHL}
    \gyaver{\overline{\varrho}^*_a} = q_a \sum_{\bm k} \sum_{n \geq0} \kernel{n}(\corr{b_a^\dagger}) \left( \gyN \gyN^{*0n}\right) (\bm k) e^{i \bm k \cdot \gyR},
\end{equation}
\\
and

\begin{align} \label{eq:PaIIHL}
\mathcal{P}_a^{\mu *}  & =\sum_{\bm k,\bm k'} \sum_{n > 0} \sum_{m=0}^{n-1} \kernel{n}(\corr{b'^\dagger_a}) \left[ q_a  \phi_1(\bm k)   \left( \frac{ \gyN \gyN^{*0m}  }{\gyTperp}  \right) (\bm k') - A_{\parallel 1}(\bm k) \left( \frac{\overline{J}_{\parallel a}^{*1m}}{\gyTperp} \right) \left( \bm k'\right) \right]e^{i \bm K \cdot \gyR} \nonumber \\
&-\sum_{\bm k,\bm k'} \corr{  \sum_{n } \sum_{m \geq 0} \sum_{r,s  \geq 0} \sum_{e=0}^{r+|n|} \sum_{f=0}^{e+s} \sum_{\substack{l = |m-f|\\ l \neq 0}}^{|m+f|} \sum_{l_1=0}^{l-1} K_{rsef}^{n}\alpha_{l}^{mf}  \left(\frac{b_a^\dagger}{2}\right)^{|n|}\left(\frac{b_a'^\dagger}{2}\right)^{|n|}  \kernel{r}(b_a^\dagger)  \kernel{s}(b_a'^\dagger)} \nonumber \\
& \times \kernel{m}(b_a) \left[ q_a \phi_{1 }(\bm k) \left(\frac{ \gyN \gyN^{*0l_1}}{\Tperp}\right)\left( \bm k'\right)  - A_{\parallel 1}(\bm k) \left( \frac{\overline{J}_{\parallel a }^{*1l_1}}{\Tperp}\right)(\bm k') \right]e^{i \bm K\cdot \gyR},
\end{align}
\\
where we have introduced the Fourier decomposition of $\gyaver{\Phi_1}_{\gyR}$ [see \cref{eq:gyaverLaguerre}]. Here, the generalized gyrocenter parallel current $\overline{J}_{\parallel a}^{*lm}$ is defined by $\overline{J}_{\parallel a}^{*lm} = q_a \gyN \momentstar{0m}{\gyvparallel^l}$. 

We now consider the \textit{long wavelength limit} approximation of $\mathcal{P}_a^{\mu *}$, and show that the expression for the linear polarization charge density commonly used in, e.g., gyrokinetic simulations of edge plasma dynamics \citep[see, e.g., ][]{Pan2016,Hakim2016}, is retrieved. Indeed, keeping terms up to $O(\epsilon_\perp^2)$ in \cref{eq:PaI}, we derive

\begin{equation} \label{eq:PaIILWL}
    \mathcal{P}_a^{\mu *}  = \frac{q_a^2}{m_a} \gygrad \cdot \left( \frac{\overline{B}_{\parallel a}^*\; \gyN}{B \Omega^2_a} \gygrad_\perp \phi_1 \right)  - \frac{q_a}{m_a} \gygrad \cdot \left( \frac{\overline{J}_{\parallel a}^{*10}}{\Omega_a^2} \gygrad_\perp A_{\parallel 1} \right).
\end{equation}
\\
\Cref{eq:PaIILWL} corresponds to the linear polarization charge density derived in \citet{Hahm2009} valid for arbitrary distribution function in the \textit{long wavelength limit}. 

We now consider the drift-kinetic limit of GKPII. For this purpose, we neglect the $O(\epsilon_\delta)$ terms and consider a \textit{long wavelength limit} approximation of $\gyaver{\overline{\varrho}^*}$ in \cref{eq:gyavervarrhoGKPII}. We first notice that $\mathcal{P}_a^{\mu *}$, \cref{eq:PaIIHL}, vanishes since it is $O(\epsilon_\delta)$. Expanding the lowest-order kernel up to $O(\epsilon^2)$ with $b_a \sim \epsilon$ in \cref{eq:gyavervarrhoGKPII}, in particular observing that \corr{$\kernel{0}(b_a^\dagger) \simeq 1 - (b_a^\dagger)^2/4$, and performing the inverse Fourier transform}, GKPII reduces to \cref{eq:DKP}. Thus, in the drift-kinetic limit, up to $\epsilon^2$, GKPI and GKPII are equivalent.\\
\corrIII{
As a final remark, we note that, to solve the field equations, a plasma vorticity equation can be used, which generalises the ones commonly solved in drift and long wavelength models \citep{Zeiler1997,Ricci2012,Jorge2017,Abel2018}. The plasma vorticity equation can be obtained from the quasineutrality condition in \cref{eq:GKPI}. Substituting the time derivative of the quasineutrality condition obtained from \cref{eq:GKPI} into the fluid equation of the gyrocenter density, given in \cref{eq:Na}, we obtain the plasma vorticity equation

\begin{align} \label{eq:vorticityequation}
 \gygrad \cdot \sum_a & \left\{ \frac{\partial }{\partial t} \left[ \gygrad \times \left( \frac{\Pperp}{2 \Omega_a} \frac{B_{\parallel a}^*}{B}\b\right)\times \b\right] + \frac{q_a\gyN}{\Omega_a} \b \times \frac{d_0}{d t} \U_{0a}\right\} =\sum_a q_a \gyN C_{a}^{00} \nonumber \\
& - \gygrad \cdot \sum_a \left\{ \left[ \bm{J}_a^* - q_a \gyN \U_{pa}\right] + \frac{\partial }{\partial t}  \left[  \bm{\mathcal{P}}_a^{PK*} -   \frac{\b}{B}\times \gygrad \times \left( \frac{\Qperp \b \b \cdot \gygrad \times \b}{2 \Omega_a^2 }\right) \right]\right\},
\end{align}
\\
where $\U_{0a} = \E \times \B/B^2 + \gyuparallel \b$ (with the electric field $ \E = - \gygrad \phi_0 - \partial_t \bm{A}$) and $\bm{J}_a^* = q_a \gyN \u_a^0$ is the gyrocenter current density. By comparing \cref{eq:vorticityequation} with the drift-kinetic vorticity equation found in \citet{Jorge2017}, obtained in a small mass ratio approximation, we remark the presence of collisional gyro-diffusion associated with FLR effects in the collision operator that vanish in the drift-kinetic limit. Also, the last two terms in \cref{eq:vorticityequation} are not present in \citet{Jorge2017}. These are related to the difference between gyro-center and particle densities. Finally, we remark the presence of additional currents driven by higher order drifts [see \cref{eq:ua0}]. \Cref{eq:vorticityequation} should be coupled to \cref{eq:GKPII} that generalises the gyrokinetic quasineutrality condition by including the presence of $\phi_0$ polarization effects.}

\subsection{Gyrokinetic Ampere's Laws}
\label{GyrokineticAmpereslaw}
From the variation of the action $\delta \mathcal{A}$ in \cref{eq:deltaA}, we  derive two gyrokinetic Ampere's laws, referred to as GKAI and GKAII. GKAI is obtained from the variation 

\begin{equation}\label{eq:deltaAdeltaApar0}
\frac{\delta \mathcal{A}}{ \delta A_{\parallel 0}}  \circ \check{A }_{\parallel 0}=0,
\end{equation}
\\
whereas GKAII is deduced from

\begin{equation} \label{eq:deltaAdeltaApar1}
\frac{\delta \mathcal{A}}{ \delta A_{\parallel 1}} \circ \check{A}_{\parallel 1}=0. 
\end{equation}
\\
We first consider GKAI. Noticing the presence of $A_{\parallel 0}$ in the inductive part of the electric field, the variation of the field functional action $\mathcal{A}_f$, \cref{eq:actionAf}, is

\begin{align} \label{eq:deltaAfdeltaA0}
   \frac{\delta \mathcal{A}_f}{\delta A_{\parallel 0} }\circ \check{ A}_{\parallel 0}     & = -\int d t \int \frac{d \x}{4 \pi} \left[ \left( \grad \times \left( \B + \B_1 \right)  - \frac{\partial }{\partial t} \left(\E + \E_1 \right) \right)\cdot    \hb  \check{A}_{\parallel 0}  \right. \nonumber \\
   & \left. - \grad \cdot \left( \left( \B + \B_1 \right) \times \left( \check{A}_{\parallel 0} \hb\right)  \right) + \frac{\partial }{\partial t}\left( \left(\E + \E_1 \right) \cdot \hb  \check{A}_{\parallel 0} \right)\right],
\end{align}
\\
where the test function is evaluated at $\x$, i.e. $\check{A}_{\parallel 0} =  \check{A}_{\parallel 0}(\x) $. In \cref{eq:deltaAfdeltaA0}, we identify the first term as the one contributing the field equation of $A_{\parallel 0}$, whereas the two last terms are boundary terms.

The variation of the gyrocenter functional action $\mathcal{A}_p$ in \cref{eq:actionAp} with respect to $A_{\parallel 0}$ is evaluated by noticing that $A_{\parallel 0}$ is contained in the symplectic components $q_a \bm A^*$, in the terms proportional to the magnetic field strength $B$, such as $\gymu B$, and in the magnetic vector $\B$, appearing, e.g., in $\u_E$ present in the gyrocenter Lagrangian $\overline{L}_a$, \cref{eq:La}. We remark that, while the gyrokinetic potential $\Psi_1$ also depends on $A_{\parallel 0}$ through $\b$ and the factors proportional to $1/\Omega_a$ at second order in $\epsilon_\delta$ [see \cref{eq:GYpotential}], we can neglect their contributions when evaluating the variation since they are of higher order as $\b \simeq \hb +O(\epsilon)$ and $B \simeq \hat{B} + O(\epsilon)$. The variation of $\mathcal{A}_p$ is then 

\begin{align} \label{eq:deltaApdeltaAparallel0}
\frac{\delta \mathcal{A}_p}{\delta A_{\parallel 0}} \circ \check{A}_{\parallel 0}   & = \sum_a   \int d t \int d \x  \int  d \gyZ \delta(\gyR - \x)\frac{B_\parallel^*}{m_a}  \gyaver{\gyFa}_{\gyR}  \left[ q_a \bm{\check{A}} \cdot \dot \gyR  \right. \nonumber \\
& \left. + m_a \left(  \frac{\E \times \check{\bm B}}{B^2}  - 2 \u_E \frac{\B \cdot \check{\B}}{B^2}  +\gyvparallel\frac{\check{\bm{B}}_{ \perp}}{B}  \right) \cdot \dot \gyR  \right. \nonumber \\
&   \left. -   \gymu \check{\B} \cdot \b + m_a u_E^2 \frac{\B \cdot \check{\B}}{B^2}-   \frac{m_a}{B}\left(\b \times \left[\dot \gyR -\u_E \right]\right)\cdot \frac{\partial}{\partial t}\check{\bm A} \right. \nonumber \\
&\left. -    \frac{\gymu B}{2 \Omega_a} \left(  \frac{\check{\bm{B}}_{ \perp}}{B}
      \cdot \grad \times \u_E +  \b \cdot \grad \times \left(  \frac{\E \times \check{\B}}{B^2} \right)  + \b \cdot \grad \times \left(\frac{\b \times \partial_t \check{\A}}{B}\right) \right. \right.\nonumber \\
& \left. \left.  -2  \b \cdot \grad \times\left( \u_E  \frac{\B \cdot \check{\B}}{B^2}\right) - \gyvparallel \left(  \frac{\check{\bm{B}}_{ \perp}}{B}\cdot \grad \times \b  + \b \cdot \grad \times  \left(\frac{\check{\bm{B}}_{ \perp}}{B}\right)\right)\right) \right],
\end{align}
\\
where $\check{\B} = \grad \times \check{\A} $ with $\check{\A} = \check{A}_{\parallel 0} \hb$, and $\check{\B}_\perp = \b \times \left( \check{\B} \times \b \right)$. In \cref{eq:deltaApdeltaAparallel0}, we have introduced $\check{A}_{\parallel 0}(\gyR) = \int d \x \delta(\gyR -\x) \check{A}_{\parallel 0}(\x)$, consistently with \cref{eq:deltaAfdeltaA0} where $\check{A}_{\parallel 0}$ is evaluated at $\x$. To evaluate the terms that provide dynamics of $A_{\parallel 0}$ in \cref{eq:deltaApdeltaAparallel0}, we carry out successive integrations by parts of the terms where derivatives of $\check{A}_{\parallel 0}$ appear. From \cref{eq:deltaAdeltaApar0} with the variations given in \cref{eq:deltaAfdeltaA0} and \cref{eq:deltaApdeltaAparallel0}, we then obtain GKAI, that we evaluate at $\gyR$, that is 

\begin{align} \label{eq:GKAI}
   \left[ \gygrad \times \left( \B + \B_1 \right) -\frac{\partial}{\partial t} \left( \E + \E_1 \right)
   \right] \cdot \hb 
     & =  4 \pi \sum_a \left[ \overline{\bm{J}}_{a}^* + \gygrad \times \left(  \bm{\mathcal{M}}_{a}^{ \mu *} +  \bm{\mathcal{M}}_a^{ *} + \bm{\mathcal{M}}_a^{\parallel * } + \bm{\mathcal{M}}_a^{B *}\right) \right. \nonumber \\
     & \left. + \frac{\partial }{\partial t} \bm{\mathcal{P}}^*_a \right]\cdot \hb,
\end{align}
\\
In \cref{eq:GKAI}, the gyrocenter current density $\overline{\bm{J}}_a^*$ is given by

\begin{equation} \label{eq:Ja}
    \overline{\bm{J}}_{a}^* = q_a \int  d \gymu d \gyvparallel d \gytheta\frac{B_\parallel^*}{m_a} \gyaver{\gyFa}_{\gyR}\dot \gyR,
\end{equation}
\\
whereas the magnetization current densities, due to the particle and gyrocenter difference, are due to the classical magnetization $\bm{\mathcal{M}}_a^{ \mu *}$, i.e.

\begin{equation} \label{eq:MaImu}
    \bm{\mathcal{M}}_a^{\mu *} =-   \int d \gymu d \gyvparallel d \gytheta \frac{B_\parallel^*}{m_a}  \gyaver{\gyFa}_{\gyR} \gymu  \left[ \b + \frac{(\gygrad \times \U)_\perp}{2 \Omega_a} \right],
\end{equation}
\\
to the magnetization $\bm{\mathcal{M}}_a^*$ associated with the polarization charge density $\bm{\mathcal{P}}_a^*$ [see \cref{eq:PaI}],

\begin{equation} \label{eq:MaI}
    \bm{\mathcal{M}}_a^{*} =  \b \cdot \left(  \bm{\mathcal{P}}_a^{*} \times \u_E\right) \b,
\end{equation}
\\
to the magnetization $\bm{\mathcal{M}}_a^{\parallel *}$,

\begin{equation} \label{eq:Mapara}
    \bm{\mathcal{M}}_a^{\parallel *} =  \int d \gymu d \gyvparallel d \gytheta \frac{B_\parallel^*}{m_a}  \gyaver{\gyFa}_{\gyR} \gyvparallel \frac{m_a}{B} \left[ \dot \gyR - \u_E \right]_\perp,
\end{equation}
\\
and, finally, to the Ban\~os magnetization $\bm{\mathcal{M}}_a^{B*}$,

\begin{equation} \label{eq:MB}
  \bm{\mathcal{M}}_a^{B*} = \int d \gymu d \gyvparallel d \gytheta \frac{B_\parallel^*}{m_a} \gyaver{\gyFa}_{\gyR} \frac{\gymu  B}{2 \Omega_a}\U \frac{(\b \cdot \gygrad \times \b)}{B}.
\end{equation}
We remark that the classical magnetization $\bm{\mathcal{M}}_a^{ \mu *}$ in \cref{eq:MaImu} [also referred to as the lowest-order intrinsic guiding-center magnetic dipole contribution, see \citet{Brizard2013}] is associated with the perpendicular gyrocenter pressure. Indeed, at the leading order, we have that $\bm{\mathcal{M}}_a^{\mu*} \simeq - \gyN^* \Pperp \B /B^2 $. The second term in $\bm{\mathcal{M}}_a^{\mu *}$, proportional to $-\gymu (\gygrad\times\U)_\perp/(2\Omega_a)$ [see \cref{eq:MaImu}], is a $O(\epsilon^2)$ correction to the classical magnetization. Similarly to the gyrokinetic Poisson's equation GKPI in \cref{eq:GKPI}, the magnetization $\bm{\mathcal{M}}_a^{ *}$, \cref{eq:MaI}, is due to the fact that the particle velocity $\vperp$ is written in the frame moving with the $\E \times \B$ drift. Indeed, this magnetization effect corresponds to the effective magnetization resulting from the $\bm{\mathcal{P}}_a^{ *}$ polarization \citep[see][Eq. (6.100)]{Jackson2012}. The magnetization current $\bm{\mathcal{M}}_a^{\parallel *}$, \cref{eq:Mapara}, represents the correction to the magnetic moment $\gymu$, resulting from the difference between the gyrocenter and $\E\times \B$ drifts, i.e. $\dot{\gyR} - \u_E$. \citep{Pfirsch1985,Kaufman1986a}. The magnetization $\bm{\mathcal{M}}_a^{B*}$ in \cref{eq:MB} is a $O(\epsilon^2)$ term associated with the presence of the Ban\~os drift. Finally, we remark the presence of the polarization current, $\partial_t \bm{\mathcal{P}}_a^*$, which originates from the variation of the inductive part of the $\E \times \B$ drift \citep{Brizard2007,Madsen2010,Brizard2013}.

We now obtain the Hermite-Laguerre representation of GKAI by performing the velocity integrals appearing in \cref{eq:Ja,eq:MaImu,eq:Mapara,eq:MB}. This yields 

\begin{subequations} \label{eq:GKAImagnetizations}
\begin{align}  \overline{\bm{J}}_a^* & = q_a \gyN \corr{\bm u_a^0}, \label{eq:Jastar} \\
          \bm{\mathcal{M}}_a^{ \mu *} & = - \frac{\Pperp}{B} \left[\left( \corr{\frac{\overline{B}_{\parallel a}^*}{B} + \frac{\Qperp}{\Pperp}\frac{ \b \cdot \gygrad \times \b}{\Omega_a}} \right)\left(  \b + \frac{\left[\gygrad \times \u_E\right]_\perp}{2 \Omega_a} + \uparallel \frac{[\gygrad \times \b]_\perp}{2 \Omega_a}\right)\right. \nonumber \\
    &      \left. + \left(\corr{\frac{\vthparallel}{\sqrt{2}} \frac{\b \cdot \gygrad \times \b}{\Omega_a}  + \frac{\overline{B}_{\parallel a}^*}{B}\frac{\Qperp \sqrt{2}}{\Pperp \vthparallel}} \right)\frac{\vthparallel [\gygrad \times \b]_\perp}{2 \sqrt{2} \Omega_a}\right],\\
    \bm{\mathcal{M}}_a^{\parallel *} &= \frac{m_a \gyN}{B} \left[ \frac{\gyvthparallel}{\sqrt{2}} \left(  \corr{\u_a^{\parallel 1}    - \u_E \frac{\vthparallel \b \cdot \gygrad \times \b}{\sqrt{2}\Omega_a} }\right)_\perp  +  \gyuparallel  \left(\corr{ \u_a^{0}-\frac{\overline{B}_{\parallel a}^*}{B}\u_E}\right)_\perp \right],\\
     \bm {    \mathcal{M}}_a^{B*} &= \frac{ (\b \cdot \gygrad \times \b)}{2  \Omega_a}\frac{\Pperp}{B} \left[ \frac{\vthparallel}{\sqrt{2}}\corr{ \left( \frac{\vthparallel \b \cdot \gygrad \times \b }{\sqrt{2} \Omega_a} + \frac{\overline{B}_{\parallel a}^*}{B}\frac{ \Qperp \sqrt{2} }{\Pperp \vthparallel}\right)} \b \nonumber \right. \\
&     \left.+ \U_0 \left(\corr{ \frac{\overline{B}_{\parallel a}^*}{B} + \frac{\Qperp}{\Pperp} \frac{\b \cdot \gygrad \times \b}{\Omega_a} }\right)  \right],
\end{align}
\end{subequations}
\\
where the Hermite-Laguerre projection of the magnetization $\bm{\mathcal{M}}_a^*$ is obtained from \cref{eq:PaIPKHL,eq:PaIDHL}. 

The second gyrokinetic Ampere's law, GKAII, follows from the variation of the action $\mathcal{A}$ with respect to $A_{\parallel 1}$ in \cref{eq:deltaAdeltaApar1}. We first notice that the variation of the field functional action, $\mathcal{A}_f$, has the same functional form as \cref{eq:deltaAfdeltaA0}, i.e.

\begin{align} \label{eq:deltaAfdeltaA1}
   \frac{\delta \mathcal{A}_f}{\delta A_{\parallel 1} }\circ \check{A}_{\parallel 1}     &  = -\int d t \int \frac{d \x}{4 \pi} \left[ \left( \grad \times \left( \B + \B_1 \right)  - \frac{\partial }{\partial t} \left(\E + \E_1 \right) \right)\cdot \hb \check{A}_{\parallel 1}\right. \nonumber \\
   & \left. - \grad \cdot \left( \left( \B + \B_1 \right) \times (\check{A}_{\parallel 1} \hb)  \right) + \frac{\partial }{\partial t}\left( \left(\E + \E_1 \right) \cdot (\check{A}_{\parallel 1} \hb)\right)\right],
\end{align}
\\
with the test function $\check{A}_{\parallel 1} $ evaluated at $\x$. In the evaluation of the variation of $\mathcal{A}_p$, we note that, as a result of our choice to consider the Hamiltonian gyrokinetic formalism, the symplectic components, $q_a \overline{\A^*}$ [see \cref{eq:Astar}], are independent of $A_{\parallel 1}$. Indeed, only the gyrokinetic potential $\Psi_1$, \cref{eq:GYpotential}, gives a contribution in the variation of $\mathcal{A}_f$. Therefore, the variation of the gyrocenter functional action $\mathcal{A}_p$ is given by

\corr{
\corrs{
\begin{align} \label{eq:deltaApdeltaA1}
     \frac{\delta \mathcal{A}_p}{\delta A_{\parallel 1}} \circ \check{A}_{\parallel 1} & =    \sum_a q_a
      \int d t \int d \x \int d \gymu d \gyvparallel d \gytheta \left[   \gyvparallel \gyaver{\frac{B_\parallel^*}{m_a}   \gyaver{\gyFa}_{\gyR}}_{\x}^\dagger  \check{A}_{\parallel 1}(\x)   \right. \nonumber \\
&      \left. - \frac{q_a}{m_a}  A_{\parallel 1} \gyaver{\frac{B_\parallel^*}{m_a}   \gyaver{\gyFa}_{\gyR}}_{\x}^\dagger \check{A}_{\parallel 1}(\x)  \right. \nonumber \\
    & \left. +  \gyvparallel \frac{q_a^2}{ m_a \Omega_a} \left( \gyaver{ \gypmu \left(\frac{B_\parallel^*}{m_a}   \gyaver{\gyFa}_{\gyR} \right)}_{\x}^\dagger \Phi_1 (\x)     -   \gyaver{  \gyaver{\Phi_1}_{\gyR} \gypmu \left(\frac{B_\parallel^*}{m_a}   \gyaver{\gyFa}_{\gyR} \right)}_{\x}^\dagger \right)\check{A}_{\parallel 1}  (\x)        \right. \nonumber \\
    & \left. + \int d \gyR  \frac{B_{\parallel }^*}{m_a} \gyaver{\gyFa}_{\gyR} \frac{q_a}{2 m_a \Omega_a^2} \gyvparallel \b \cdot \left(    \gygrad \left[\int^{\gytheta} d \gytheta'\widetilde{ \delta(\gyR + \gyrhoa -\x) \check{A}_{\parallel 1}(\x)}\right] \times   \gygrad \widetilde{\Phi_1} \right. \right. \nonumber \\
&    \left. \left.+
    \gygrad \left[\int^{\gytheta} d \gytheta' \widetilde{\Phi_1}\right] \times     \gygrad \widetilde{\delta(\gyR + \gyrhoa -\x)\check{A}_{\parallel 1}(\x) } \right)\right].
\end{align}}}
\\
where we introduced the adjoint gyroaverage operator, $\gyaver{\cdot}_{\x}^\dagger$ defined in \cref{eq:adjointop}, with $\check{A}_{\parallel 1}(\gyR + \gyrhoa) = \int d \x \delta(\gyR + \gyrhoa -\x) \check{A}_{\parallel 1}(\x)$. Following the similar steps and assumptions to the ones considered for the evaluation of \cref{eq:deltaApdeltaphi1}, we obtain GKAII, which we evaluate at $\gyR$,

\begin{align} \label{eq:GKAII}
 \left[ \gygrad \times \left( \B + \B_1 \right) -\frac{\partial}{\partial t} \left( \E + \E_1 \right)
   \right] \cdot \hb  =  4 \pi \sum_a \left[ \gyaver{\overline{J}_{\parallel a}^*} + \overline{J}_{\parallel a}^{\parallel *} + \overline{J}_{\parallel a}^{\mu *} \right],
\end{align}
\\ 
where the parallel gyroaveraged gyrocenter current density is 

\begin{equation} \label{eq:JaparaGKAII}
    \gyaver{\overline{J}_{\parallel a}^*} = q_a \sum_{\bm k } \int d \gymu d \gyvparallel d \gytheta \frac{B_\parallel^*}{m_a} \gyaver{\gyFa}_{\gyR}(\bm k) \gyvparallel e^{- i \bm k \cdot \gyrhoa} e^{i \bm k \cdot \gyR} ,
\end{equation}
\\
while the parallel magnetization current densities are, respectively, 

\begin{align} \label{eq:Jparapara}
    \overline{J}_{\parallel a}^{\parallel *} = - \frac{q_a^2}{m_a} \sum_{\bm k,\bm k' } \int d  \gymu d \gyvparallel d \gytheta  \frac{B_\parallel^*}{m_a} A_{\parallel 1}(\bm k) \gyaver{\gyFa}_{\gyR}(\bm k') e^{- i \bm k' \cdot \gyrhoa} e^{i \bm K \cdot \gyR},
\end{align}
\\
and 

\begin{align} \label{eq:Jparamu}
       \overline{J}_{\parallel a}^{\mu *}& =q_a \sum_{\bm k, \bm k'} \int d \gymu d \gyvparallel d \gytheta \frac{B_\parallel^*}{m_a} \gyvparallel \frac{q_a}{B}   \left[\Phi_1(\bm k) - \gyaver{\Phi_1}_{\gyR}(\bm k) e^{- i \bm k \cdot \gyrhoa} \right] \nonumber \\
&       \times \gypmu\gyaver{\gyFa}_{\gyR}(\bm k') e^{- \bm k' \cdot \gyrhoa} e^{i \bm K \cdot \gyR}.
\end{align}
\\
The Hermite-Laguerre representations of $\gyaver{\overline{J}_{\parallel a}^*}$, $ \overline{J}_{\parallel a}^{\parallel *}$, and $\overline{J}_{\parallel a}^{\mu *}$ are obtained by using the Jacobi-Anger identity in \cref{eq:jacobianger} and by performing the velocity integrals in \cref{eq:Jparapara,eq:Jparapara,eq:Jparamu}. This yields, respectively,

\begin{subequations} \label{eq:JGKAII}
\begin{align} \label{eq:gyaverJaparaHL}
  \gyaver{\overline{J}_{ \parallel a}^*}  &= \sum_{\bm k} \kernel{n}(\corr{b_a^\dagger}) \overline{J}_{ \parallel a}^{*1n}(\bm k) e^{i \bm k \cdot \gyR}, \\
 \overline{J}_{\parallel a}^{ \parallel *}  & = - \frac{q_a^2}{m_a}  \sum_{\bm k,\bm k'} \sum_{n \geq 0}A_{\parallel 1}(\bm k) \kernel{n}(\corr{b_a'^\dagger}) \left( \gyN \gyN^{*0n} \right)(\bm k') e^{i\bm K\cdot \gyR}, \\
 \overline{J}_{\parallel a}^{\mu *} & = \sum_{\bm k, \bm k'}\sum_{n > 0} \sum_{m=0}^{n-1} q_a \kernel{n}(b_a'^\dagger) \left[  \phi_1(\bm k)  \left( \frac{ \overline{J}_{\parallel a }^{*1m}  }{\gyTperp} \right)(\bm k') - A_{\parallel 1}(\bm k)  \left( \frac{\overline{J}_{\parallel a }^{*2m}}{\gyTperp} \right)\left(\bm k'\right)\right] e^{i \bm K \cdot \gyR}\nonumber \\
&-  \sum_{\bm k, \bm k'}\corr{  \sum_{n } \sum_{m \geq 0} \sum_{r,s  \geq 0} \sum_{e=0}^{r+|n|} \sum_{f=0}^{e+s} \sum_{\substack{l = |m-f|\\ l \neq 0}}^{|m+f|} \sum_{l_1=0}^{l-1} q_a K_{rsef}^{n} \alpha_{l}^{mf} \left(\frac{b_a^\dagger}{2}\right)^{|n|}\left(\frac{b_a'^\dagger}{2}\right)^{|n|}  \kernel{r}(b_a^\dagger) \kernel{s}(b_a'^\dagger) }   \nonumber \\
& \times \kernel{m}(b_a)  \left[  \phi_{1}(\bm k)  \left( \frac{\overline{J}_{\parallel a}^{*1\corr{l_1}} }{\Tperp} \right) \left( \bm k'\right) -   A_{\parallel 1}(\bm k) \left( \frac{\overline{J}_{\parallel a}^{*\corr{2l_1}}}{\Tperp}\right)(\bm k')  \right] e^{i \bm K \cdot \gyR}. \label{eq:JparamuHL}
\end{align}
\end{subequations}
\\
As a check of our expressions, we now investigate the \textit{long wavelength limit} of \cref{eq:JGKAII}, i.e. the current densities appearing in GKAII. In particular, considering $b_a \sim \epsilon_\perp \sim \epsilon$, expanding the lowest-order kernel $\kernel{0} = 1 - (b_a^{\dagger})^2/4$ up to $O(\epsilon^2)$, and performing the inverse Fourier transform, we derive

\begin{equation}
    \gyaver{\overline{J}_{\parallel a}^*}  =  \gyN^* \overline{J}_{\parallel a} + \frac{\gyvthparallel \b \cdot \gygrad \times \b}{\sqrt{2} \Omega_a}\overline{J}_{th \parallel a}  + \frac{1}{2 m_a} \gygrad_\perp^2 \left( \frac{\Pperp \overline{J}_{\parallel a}}{\gyN \Omega_a^2}\right),
    \end{equation}
\\
with $\overline{J}_{\parallel a} = q_a \gyN \gyuparallel$ and $\overline{J}_{th \parallel a} = q_a \gyN \gyvthparallel / \sqrt{2}$, while $\overline{J}_{\parallel a}^{\mu *} $ in \cref{eq:JparamuHL} reduces to

\begin{equation} \label{eq:JparamuLWL}
\overline{J}_{\parallel a}^{\mu *}     = \frac{q_a}{m_a} \gygrad \cdot \left( \frac{\overline{J}_{\parallel a}^{*10}}{\Omega_a^2} \gygrad_\perp \phi_1\right) - \frac{q_a}{m_a} \gygrad \cdot \left( \frac{\overline{J}_{\parallel a}^{*20}}{\Omega_a^2} \gygrad_\perp A_{\parallel 1}\right).
\end{equation}
\\
\Cref{eq:JparamuLWL} corresponds to the Hermite-Laguerre projection of the magnetization current densities obtained by \citet{Hahm2009} in the \textit{long wavelength limit} and valid for arbitrary distribution functions. Neglecting the $O(\epsilon_\delta)$ terms in \cref{eq:GKAI}, while keeping term up to $O(\epsilon_\perp^2)$, one obtains the drift-kinetic Ampere's law, which can be used to provide an electromagnetic extension of the drift-kinetic moment hierarchy derived in \citet{Jorge2017}. 
\subsection{Equilibrium Pressure Balance Equation}
\label{LeadingOrderEquilibriumPressureBalanceEquation}

 Since a MHD-like equilibrium pressure balance is an important element especially in the study of gradient driven modes \citep{Rogers2018}, we show how the spatial and temporal evolution of the large scale magnetic field $\hB$ can be determined such that, at leading order, it reduces to an equilibrium pressure balance. The field equation that sets the self-consistent evolution of $\hB$ can be deduced from 
 
 \begin{equation} \label{eq:deltaAdeltahatA}
\frac{\delta \mathcal{A}}{\delta \hat{\A}} \circ \check{\hat{\A}} =0,
\end{equation}
\\
that is

\begin{align} \label{eq:GKAIII}
\gygrad \times \left( \B + \B_1 \right) -\frac{\partial}{\partial t} \left( \E + \E_1 \right) & =  4 \pi \sum_a \left[ \overline{\bm{J}}_{a}^*  + \gygrad \times \left(  \bm{\mathcal{M}}_{a}^{ \mu *} +  \bm{\mathcal{M}}_a^{ *} + \bm{\mathcal{M}}_a^{\parallel * } + \bm{\mathcal{M}}_a^{B *}\right) \right. \nonumber \\
&\left. + \frac{\partial}{\partial t} \bm{\mathcal{P}}_a^* \right].
\end{align}
\\
To show that \cref{eq:GKAIII} reduces its pressure balance equation, we consider a leading order approximation by neglecting the $O(\epsilon_\delta)$ and $O(\epsilon^2)$ terms. Moreover, the magnetic fluctuations, $\delta \B$, are ignored. More precisely, we use the fact that $\left| \delta \B \right| / \hat{B} \sim \epsilon$ [see \cref{eq:deltaBoverB}], which implies that $\B + \B_1 = \hB + O(\epsilon)$. Additionally, the plasma is assumed to be quasi-neutral, i.e. $\sum_a q_a \gyN = 0$. Keeping only the leading order classical magnetization, $\bm{\mathcal{M}}_a^{*\mu} = - \Pperp \hB/\hat{B}^2 + O(\epsilon^2)$, while neglecting $\partial_t \E$ and $\partial_t \bm{\mathcal{P}}_a^*$ being higher order terms and solving for $\gygrad \times \hB/ 4 \pi$ in \cref{eq:GKAIII}, we obtain

\begin{equation} \label{eq:hatJ}
\frac{\gygrad \times \hB}{4 \pi} =  \sum_a \left[ \overline{\bm{J}}^*_a - \gygrad \times \left( \frac{\Pperp \hB}{\hat{B}^2}\right) \right],
\end{equation}
\\
where the leading order gyrocenter current is $\overline{\bm{J}}^*_a = q_a \gyN \momentstar{}{\dot \gyR}$, with the leading order gyro-moment expansion 

\begin{equation}
\momentstar{}{\dot \gyR} = \U_0  + \frac{1}{2} \U_{\bm{\kappa}} + \U_{\grad} + \U_{pa},
\end{equation}
\\
as deduced from \cref{eq:momentdotR}. Then, the leading order pressure balance is obtained by taking the cross product of \cref{eq:hatJ} with $\hB$ and using the quasi-neutrality condition,

\begin{equation} \label{eq:pressureBalance}
 \sum_a N_a m_a  \frac{d^0}{dt} \U_0 \biggr \rvert_{\perp} = \frac{1}{4 \pi} \left[ \left(\hB \cdot \gygrad \right) \hB -  \frac{\gygrad \hat{B}^2}{2}  \right]+ \left( \overline{P}_\perp - \overline{P}_\parallel \right) \bm \kappa - \gygrad_\perp \overline{P}_\perp,
\end{equation}
\\
where $\overline{P}_\parallel = \sum_a \Pparallel$ and $\overline{P}_\perp = \sum_a \Pperp$ are the parallel and perpendicular total pressures. We remark that \cref{eq:pressureBalance}, which contains the anisotropic pressure terms \citep[see, e.g.,][]{Chew1956,Lanthaler2019}, reduces to the MHD equilibrium pressure balance, $(\gygrad \times \hB ) \times \hB = 4 \pi \gygrad_\perp \overline{P}$, in the static and isotropic limit, i.e. $\overline{P}_\parallel = \overline{P}_{\perp} = \overline{P}$.

We now show that the low-$\beta$ approximation is consistent with the leading order equilibrium pressure balance equation, \cref{eq:pressureBalance}. Indeed, balancing the $\left| \gygrad \hat{B}^2 \right| \sim B^2/L_B$ and $\left| \grad_\perp \overline{P}_\perp \right| \sim 1/L_P \overline{P}_\perp$ terms yields,

\begin{equation}
\beta \sim  \frac{\epsilon_B}{\rho_s / L_P} \ll 1 ,
\end{equation}
\\
since we assumed $\rho_s / L_P \sim \epsilon$ [see \cref{eq:defepsilon}] and $\epsilon_B \sim \epsilon^3$ [see \cref{eq:epsilonB}]. Thus, steep pressure gradients on scale length $L_P$ is consistent with a low-$\beta$ plasma in the presence of an equilibrium magnetic field varying on large scales.

\section{Conclusion}
\label{conclusion}

In the present work, a gyrokinetic model is derived to evolve the turbulent plasma dynamics in the periphery of tokamak devices. \corr{This model takes the form of an infinite set of coupled fluid equations, the gyro-moment hierarchy equation given in \cref{eq:GyromomentHierarchyEquation} with a nonlinear gyrokinetic Dougherty collision operator in \cref{eq:Calk}, coupled with a set of gyrokinetic field equations, that are two Poisson's equations, GKPI in \cref{eq:GKPI} and GKPII in \cref{eq:GKPII}, and two Ampere's laws, GKAI in \cref{eq:GKAI} and GKAII in \cref{eq:GKAII}.} The model is based on second order fully electromagnetic gyrokinetic equations of motion of a charged particle. The equations are obtained by using Lie-transform perturbation theory. More precisely, by taking advantage of the scale separation between the equilibrium scales and the particle gyroscales, two changes of phase-space coordinates are performed allowing the description of the single particle motion in the presence of large and small-scale electromagnetic fluctuations and strong flows. Then, the collective behaviour is introduced by deriving a second order accurate gyrokinetic Boltzmann equation. The gyrokinetic equation is further developed into a gyro-moment hierarchy valid for far from equilibrium distribution functions, obtained by projecting the gyrokinetic Boltzmann equation onto a complete set of Hermite-Laguerre polynomials. These polynomials constitute a complete velocity-space basis used to conveniently expand the gyroaveraged distribution function. The result of the projection is an infinite set of coupled fluid equations for the temporal and spatial evolution of the Hermite-Laguerre expansion coefficients, that are referred to as gyro-moments. In the process, the linear and nonlinear coupling between the gyro-moments, associated with parallel streaming along the magnetic field lines, magnetic field gradients, and FLR effects, are analytically treated and retained at arbitrary values of the perpendicular wavenumber. In particular, a closed form of the gyroaveraged operator is derived in terms of gyro-moments. The effects of collisions in the plasma periphery dynamics are introduced by considering a gyrokinetic Dougherty collision operator. This operator is expressed as a function of the gyro-moments, is accurate at arbitrary values of the perpendicular wavenumber, and is nonlinear. We remark that the gyro-moment expansion of a full nonlinear gyrokinetic Coulomb collision operator can be derived within the same formalism as in \cref{GyrokineticCollisionOperator} to describe efficiently like and unlike species collisions in more general situations. This will be subject of a future work. Finally, a set of self-consistent gyrokinetic Poisson-Ampere equations are obtained from a variational principle. More precisely, two coupled gyrokinetic Poisson's equations and two coupled gyrokinetic Ampere's laws, with an additional Ampere's law needed to set the evolution of the large scale magnetic field, are derived to provide the necessary closure of the gyro-moment hierarchy by determining the different components of the fluctuating electromagnetic fields. Within our variational formulation, polarization and magnetization corrections, which are related to the change of phase-space coordinates at the particle Lagrangian level, appear self-consistently. The charge and current densities are expressed in terms of the gyro-moments by performing analytically the velocity integrals at arbitrary wavelengths. Thus, a complete generalization of the analytical expressions of the polarization charge and magnetization current densities, present in earlier gyrokinetic derivations, is obtained.

\begin{table}
\begin{threeparttable}
\renewcommand*{\arraystretch}{1.05}
\begin{tabular}{c|ccccccc}
&\twolinetabular{This}{Work} &\twolinetabular{Qin et \textit{al}.}{(2006)}& \twolinetabular{Hahm et \textit{al}.}{(2009)} & \twolinetabular{Dimits}{(2012)} &\twolinetabular{Madsen}{(2013)} & \twolinetabular{Jorge et \textit{al}.}{(2017)    } &\twolinetabular{Mandell et \textit{al}.}{(2018)}  \\ \hline \hline
  & Full-F & Full-F &   Full-F & Full-F &  Trunc. Full-F & Full-F &delta-F \\
DK & EM  & EM & ES & ES &ES &  ES & ES  \\
GK  & EM & EM & EM & EM &EM &  -  & ES  \\  
 \corr{CLO} & $O(\epsilon^2, \epsilon_\delta^2)$   & $O(\epsilon,\epsilon_\delta^2)$ & $O(\epsilon^2,\epsilon_\delta^2)$ & $O(\epsilon^2,\epsilon_\delta^2)$ & $O(\epsilon,\epsilon_\delta)$ & $O(\epsilon)$ & $O(\epsilon,\epsilon_\delta)$ \\
    SF & YES & YES & YES & NO & YES & YES & NO \\
  PSV & $B_\parallel^*$& $B_\parallel^*$ &  $B_\parallel^*$  & $B_\parallel^*$ & $B_\parallel^* \simeq B$ & $B_\parallel^*$ & $B_\parallel^* \simeq B$  \\ 
  $\gyaver{\cdot}$ & G-M  & NO & $J_0$  & NO  & $\moment{}{J_0} \simeq \Gamma$ & - & G-M  \\
  CO & D & -  & - & - & - & C & D  \\
  GKP &  G-M & G-M & $O(\epsilon_\perp^2)$ & $\int$ & $O(\epsilon_\perp^2)$ &   G-M &  G-M  \\
  GKA &  G-M & - & $O(\epsilon_\perp^2)$ & $\int$ & $O(\epsilon_\perp^2)$ &  - & -  \\ \hline \hline
\end{tabular}
\end{threeparttable}
\caption{Comparison between the gyrokinetic model presented herein and previous gyrokinetic theories. The present model is fully electromagnetic (EM), is in both drift-kinetic (DK) and gyrokinetic (GK) regimes, \corr{second order accurate in the collisionless part (CLO)}, in both $\epsilon$ and $\epsilon_\delta$, and includes the effects of strong flows (SF). The exact structure the phase-space volume element (PSV) is preserved. A closed and analytical form of the gyroaverage operator, $\gyaver{\cdot}$, in terms of gyro-moments (G-M) is provided, while a FLR model, $\Gamma$, is often used in other models, or a closed expression is not given. A gyrokinetic Dougherty (D) collision operator (CO) is used, while a Coulomb (C) operator will be developed in future work. The polarization and magnetization corrections in the gyrokinetic Poisson's equations (GKP) and Ampere's laws (GKA) are analytically evaluated at arbitrary wavelengths and as functions of G-M. In other theories, these terms appear as velocity integrals ($\int$), or are $O(\epsilon_\perp^2)$ accurate.}
\label{tablecomparison}
\end{table}

\Cref{tablecomparison} illustrates and summarizes the improvements and differences between the present model and previous gyrokinetic theories. In particular, we point out that the present gyro-moment hierarchy is an extension of the drift-kinetic moment hierarchy for the SOL dynamics at arbitrary collisionality developed by \citet{Jorge2017} to fully electromagnetic fluctuations allowing for perpendicular wavenumbers of the order of the ion sound Larmor radius (see \cref{tablecomparison}). 

\corr{Although a numerical implementation of the herein model is outside of the scope of the present work, we remark that closure schemes need to be applied to the gyro-moment hierarchy equation, i.e. a truncation of the series in \cref{eq:GYFadecomposition} is required. A possible closure is based on the truncation of all the gyro-moments $\gyN^{lk}$ of order higher than a given order. Then, the unresolved gyro-moments are comparable to the limitations of velocity-space grid methods to resolve fine velocity structures. However, truncations can lead to poor results especially at low Hermite-Laguerre resolution \citep{Mandell2018}. An exact asymptotic closure, known as semi-collisional closure \citep{Zocco2011,Loureiro2016}, can be rigorously applied even at small (but finite) collisionality.} With such a closure, the present hierarchy provides an ideal framework for the simulation of the dynamics of the plasma periphery. While being a rigorous asymptotic limit of the exact gyrokinetic equation [see \cref{eq:GyaverBoltzmann}], the model herein present a tuneable kinetic accuracy which depends on the number of gyro-moments retained.

\section*{Acknowledgements}

The authors acknowledge helpful discussions with A. Baillod, A. J. Brizard, S. Brunner, W. Dorland, N. F. Loureiro, A. A. Schekochihin, P. B. Snyder, M. Held and N. Tronko. This research was supported in part by the Swiss National Science Foundation, and has been carried out within the framework of the EUROfusion Consortium and has received funding from the Euratom research and training programme 2014 - 2018 and 2019 - 2020 under grant agreement No 633053. The views and opinions expressed herein do not necessarily reflect those of the European Commission.

\appendix
\section{Guiding-Center Transformation}
\label{appendixGC}
This appendix reports on the details of the second order guiding-center transformation whose results are used in \cref{GuidingCentertransformation}. Using \cref{eq:systemorderbyorder2} and computing the Lie-derivatives of $\gamma_1$ and $\Gamma_1$ given in \cref{eq:gamma1,eq:Gamma1} with the definition in \cref{eq:Lieoneform}, we obtain the $O(\epsilon^2)$ guiding-center correction $\Gamma_{2}$, expressed in the guiding-center coordinates $\Z$,  

\begin{align} \label{eq:Gamma2full}
\Gamma_{2} &=  \left[  q \g_2^{\R} \times \B + m \g^{\R}_1  \times  \left( \grad \times \U  \right) + \frac{1}{2} m \g^{\R}_1 \times \left(  \grad \times \cperp \right)  \right. \nonumber  \\
&  \left.  -  \frac{1}{2}m g_1^\theta \ptheta   \cperp -  \frac{1}{2} m g^\mu_1 \pmu \cperp  -  m  g_1^\parallel  \b + \grad S_2  \right] \cdot d \R \nonumber \\ 
  &  + \left[  m  \g_1^{\R}  \cdot \b + \pvparallel S_2  \right] d v_{\parallel} +  \left[ \frac{1}{2} m  \g^{\R}_1 \cdot \ptheta \cperp + \ptheta S_2 \right] d \theta \nonumber \\
& + \left[\frac{1}{2}  m  \g^{\R}_1  \cdot\pmu \cperp + \pmu S_2 \right] d \mu  + \left[  - q \g_{ 2 }^{\R} \cdot \E  + \frac{1}{2} m \g^{\R}_1 \cdot \grad   U^2 + \mu \g^{\R}_1 \cdot \grad  B   \right.  \nonumber  \\
&\left.  +  m g^\parallel_1 v_{\parallel} + g_1^\mu B + \frac{1}{2} m \g^{\R}_1 \cdot\grad   ( \U \cdot \cperp)  +   m \g^{\R}_1 \cdot \frac{\partial }{\partial   t}  \U +  \frac{1}{2}  m \g^{\R}_1 \cdot \frac{\partial }{\partial    t}    \cperp \right. \nonumber  \\
& \left.    +\frac{1}{2} m g^\theta_1 \U  \cdot  \ptheta\cperp +\frac{1}{2} m g^\mu_1  \U \cdot \pmu \cperp +  \frac{\partial}{\partial t} S_2 \right] d t.
 \end{align}
\\
To remove one inherent degree of freedom, we set $S_2 = 0$ and choose $\gR_1 \cdot \b =0 $. Therefore, $\gR_1$ is purely perpendicular to $\b$ and is given by \cref{eq:G1Rrhoa}. This results into $\Gamma_{2 \theta}  = B \mu / \Omega $, $\Gamma_{2 \mu} =0$ and $\Gamma_{2 \parallel} =0$. Taking into account that fact that $\cperp$ depends on $B$ (indeed, $\vperp = \sqrt{2 \mu B / m}$) results the identity

\begin{align}
\frac{1}{2}m\gR_1 \times \left(\grad \times \cperp \right) &  = - \frac{ m \mu }{q}\bm T + \frac{ m \mu }{q}\a  ( \a \cdot \bm T)+  \frac{ m\mu }{q}\b( \a \cdot \grad \c \cdot \b  )  \nonumber \\ 
& - \frac{ m}{4}  \cperp  \left( \gR_1 \cdot \grad \ln B \right) ,
\end{align}
\\
with $\bm{T} = (\grad \c) \cdot \a = \grad \e_2 \cdot \e_1$ being the first Littlejohn's gyrogauge field vector \citep{Littlejohn1988}. With $
\gR_1 \times \bm f = \left(\gR_1  \times \bm b  \right) \left( \b \cdot \bm f \right) +  \b \left(  \cperp \cdot  \bm f \right)/\Omega$ where $\bm f$ is any arbitrary vector function, and introducing the quantities $h^{\theta}$, $h^{\mu}$ and $h^\parallel$ defined by  \citep{Brizard1995}

\begin{subequations} \label{eq:hquantities}
\begin{align} 
h^{\mu} & = \gmu_1  + \mu \gR_1 \cdot \grad \ln B, \label{eq:hmu} \\
h^{\theta} &= \gtheta_1 - \gR_1 \cdot \bm{T}, \label{eq:htheta}\\
h^\parallel & =  g_1^\parallel - \gR_1 \cdot \grad \b \cdot \cperp, \label{eq:hparallel} 
\end{align}
\end{subequations}
\\
 into \cref{eq:Gamma2full}, the perpendicular and parallel components to $\bm b$ of the symplectic components $\Gamma_{2 \R}$ can be separated as 

\begin{align}\label{eq:Gamma2Rfull} 
\Gamma_{2 \R}   & = - \frac{ m \mu }{q}\bm T   - \frac{m}{2} \left( h^\mu \pmu \cperp + h^\theta \ptheta \cperp \right)  - \B \times \left( q \gR_{2 }  + m \gR_1 \frac{ \left( \b \cdot \grad \times \U \right)}{B}\right) \nonumber \\ 
& -  \b \left(  m h^\parallel + \frac{m}{2}  \gR_1 \cdot \grad \b \cdot \cperp -  m  \cperp \cdot \frac{\left( \grad \times \U \right)}{\Omega}  \right) .
\end{align}
\\
Introducing $h^\parallel = \gyaver{h^\parallel} + \widetilde{h^\parallel}$ and $ \gR_1 \cdot \grad \b \cdot \cperp  = \gyaver{ \gR_1 \cdot \grad \b \cdot \cperp} + \widetilde{ \gR_1 \cdot \grad \b \cdot \cperp}$ where $\gyaver{\chi} \equiv \int_0^{2\pi} d \theta \chi$ and $\widetilde{\chi} = \chi - \gyaver{\chi}$ being the gyrophase dependent part of any gyrophase dependent function $\chi = \chi(\theta)$, we cancel the gyrophase dependent terms parallel to $\b$ by setting

\begin{align}
\widetilde{h^\parallel} =  \cperp \cdot \frac{\left( \grad \times \U \right)}{\Omega}  - \frac{1}{2}\widetilde{ \gR_1 \cdot \grad \b \cdot \cperp},
\end{align}
\\
and in the perpendicular direction by taking the vector product of \cref{eq:Gamma2Rfull} with $\b$ to obtain an expression for $\g_{2 \perp }^{\R}$,

\begin{equation} \label{eq:G2Rperp}
\g_{2 \perp}^{\R} = \frac{1}{2 \Omega}   \b \times \left(  h^{\theta} \ptheta   \cperp +h^{\mu} \pmu
  \cperp \right)  - \gR_1  \frac{\b \cdot  \left(  \grad \times \U \right) }{\Omega} .
\end{equation}
\\
Thus, $\Gamma_{2 \R}$ reduces to

\begin{equation}
\Gamma_{2 \R} = - \frac{B\mu}{\Omega} \bm T  - \b m \left[ \gyaver{h^\parallel} + \frac{1}{2} \gyaver{\gR_1 \cdot \grad \b \cdot \cperp} \right].
\end{equation}
\\
At this stage, we have the freedom to choose $\gyaver{h^\parallel} = -   \gyaver{\gR_1 \cdot \grad \b \cdot \cperp}/2$. This particular choice of $\gyaver{h^\parallel}$ has the effect of transferring the latter term in the Hamiltonian component. An alternative possibility is to set $\gyaver{h^\parallel} =0$, and so keeping the $m \gyaver{\gR_1 \cdot \grad \b \cdot \cperp}/2$ in the symplectic component \citep{Brizard1995,Madsen2010}. Thus, at second order, we find

\begin{equation} \label{eq:Gamma2Rfinal}
\Gamma_{2 \R} = - \frac{B \mu }{\Omega} \bm T,
\end{equation}
\\
while, $\gyaver{h^\parallel} =0$ yields $\Gamma_{2 \R} = - B \mu  \bm T^*/ \Omega $ where $\bm T^* = \bm T - \bm b \bm b \cdot \grad \times \bm b /2$.

The Hamiltonian component of $\Gamma_{2}$, namely $\Gamma_{2t}$, can be developed as 

\begin{align}  \label{eq:Gamma2tfull}
\Gamma_{2 t} &=   - q \gR_2 \cdot \E + \frac{1}{2} m \g^{\R}_1 \cdot \grad   U^2  + B h^\mu + m v_\parallel h^\parallel + \frac{1}{2} m v_\parallel \gR_1 \cdot \grad \b \cdot \cperp \nonumber \\
& + \frac{1}{2} m \gR_1 \cdot \grad \U \cdot \cperp + m \gR_1 \cdot \frac{\partial}{\partial t} \U \nonumber \\
& +  \frac{1}{2}  m \g^{\R}_1 \cdot \frac{\partial }{\partial    t}    \cperp + \frac{1}{2} m h^\theta \U \cdot  \ptheta \cperp + \frac{1}{2} m h^\mu \U \cdot \pmu \cperp.
\end{align}
\\
Using the expression for $\gR_{2 \perp}$ given in \cref{eq:G2Rperp} into \cref{eq:Gamma2tfull}, terms combine, and $\Gamma_{2t}$ reduces to

\begin{align}
\Gamma_{2t} & =q \frac{\b \cdot \grad \times \U}{\Omega} \gR_1 \cdot \E + \frac{1}{2} m \g^{\R}_1 \cdot \grad   U^2  + B h^\mu + m v_\parallel h^\parallel + \frac{1}{2} m v_\parallel \gR_1 \cdot \grad \b \cdot \cperp \nonumber \\
& + \frac{1}{2} m \gR_1 \cdot \grad \U \cdot \cperp + m \gR_1 \cdot \frac{\partial}{\partial t} \U + \frac{1}{2} m \gR_1 \cdot \frac{\partial }{\partial t} \cperp.
\end{align}
\\
Introducing $h^\mu = \gyaver{h^\mu} + \widetilde{h^\mu}$, $h^\parallel = \gyaver{h^\parallel} + \widetilde{h^\parallel}$, and $ \gR_1 \cdot \grad \b \cdot \cperp = \gyaver{ \gR_1 \cdot \grad \b \cdot \cperp} + \widetilde{ \gR_1 \cdot \grad \b \cdot \cperp}$ in $\Gamma_{2t}$, allows to separate the gyrophase dependent and independent parts by noticing that 

\begin{equation}
\gR_1 \cdot \frac{\partial }{\partial t} \cperp = - \frac{2 \mu B}{m \Omega}  S,
\end{equation}
\\
where $S = \a \cdot \partial_t \c$ being the second gyrogauge field vector introduced by \citet{Littlejohn1988}, such that

\begin{align} \label{eq:Gamma2t}
\Gamma_{2t} = - \frac{\mu B}{\Omega}  S + B \gyaver{h^\mu} + m v_\parallel \left( \gyaver{h^\parallel}  + \frac{1}{2} \gyaver{\gR_1 \cdot \grad \b \cdot \cperp} \right)+ \frac{1}{2} m \gyaver{\gR_1 \cdot \grad \U \cdot \cperp},
\end{align}
\\
where the gyrophase dependent term are cancelled by setting

\begin{align}
 - B \widetilde{h^\mu}  & =   q \frac{\b \cdot \grad \times \U}{\Omega} \gR_1 \cdot \E + \frac{1}{2} m \g^{\R}_1 \cdot \grad   U^2    + m v_\parallel \widetilde{h^\parallel} + \frac{1}{2} m v_\parallel \widetilde{\gR_1 \cdot \grad \b \cdot \cperp} \nonumber \\
& + \frac{1}{2} m \widetilde{\gR_1 \cdot \grad \U \cdot \cperp} + m \gR_1 \cdot \frac{\partial}{\partial t} \U.
\end{align}
\\
Finally, with $\gyaver{h^\parallel} = - \gyaver{\gR_1 \cdot \grad \b \cdot \cperp}/2$ into \cref{eq:Gamma2t}, the second order guiding-center correction $\Gamma_2$ is given by  

\begin{equation} \label{eq:Gamma2secondorder}
\Gamma_{2} = - \frac{B\mu}{\Omega} \bm{T} \cdot d \R + \frac{\mu B}{\Omega} d \theta + \left[   \frac{B \mu }{\Omega} S + \frac{B
  \mu}{2 \Omega} \b \cdot \grad \times \U    +     B \gyaver{{h}^{\mu}} \right]  d t.
\end{equation} 

The analytical expression for $\gyaver{h^\mu}$ in $\Gamma_2$ given in \cref{eq:Gamma2secondorder} must be obtained by considering a third-order calculation that ensures $\dot{\mu} = O(\epsilon^4)$. A sufficient condition is to impose that $\Gamma_{3 \theta} =0 $, $\Gamma_{3 \mu} =0 $ and $\Gamma_{3 \parallel} =0$. As a results, we also obtain the closed expressions for $\b \cdot \gR_2$ and $h^\theta$. From \cref{eq:systemorderbyorder3}, we remark that only the Lie-derivatives $\L_1^2 \gamma_1$, $\L_1^2 \Gamma_1$ and $\L_2 \Gamma_1$ contribute in $\Gamma_{3\theta}$, $\Gamma_{3 \mu}$ and $\Gamma_{3 \parallel}$. In fact, while $\gamma_3$ and $\gamma_2$ are identically zero [see \cref{eq:gamma0,eq:gamma1}], $\L_3 \gamma_0$ has only a $\R$ and a $t$ component. Thus, we derive that

\begin{subequations} \label{eq:L12gamma1}
\begin{align}
\left( \L_1^2 \gamma_1 \right)_{\theta} & = m \gR_1 \cdot \ptheta \left(   \gR_1 \times \grad \times ( \U  + \cperp)   - m \gtheta_1 \ptheta \cperp - m \gmu_1 \pmu \cperp -  g_1^\parallel \b  \right) \nonumber  \\
&  - m \gR_1  \cdot \grad \left( \gR_1  \cdot \ptheta  \cperp \right) - m \gmu_1 \pmu \left( \gR_1 \cdot \ptheta \cperp \right)  +   m \gmu_1 \ptheta \left(  \gR_1 \cdot \pmu \cperp \right), \\
\left(  \L_1^2 \gamma_1 \right)_{\mu} & = m \gR_1 \cdot \pmu \left(   \left( \gR_1 \times \grad \times ( \U  + \cperp) \right)  -  \gtheta_1 \ptheta \cperp - \gmu_1 \pmu \cperp    -  g_1^\parallel \b\right) \nonumber  \\
&    -m \gR_1 \cdot \grad \left( \gR_1 \cdot \pmu \cperp  \right) - m \gtheta \ptheta  \left( \gR_1 \cdot \pmu \cperp  \right)   + m \gtheta_1 \pmu \left( \gR_1 \cdot \ptheta \cperp \right)   , \\
\left(  \L_1^2 \gamma_1  \right)_{\parallel} &  =  - m \gR_1 \cdot \pvparallel \left( \gtheta_1 \cdot \ptheta \cperp \right),
\end{align}
\end{subequations}
\\
and 

\begin{equation} \label{eq:L12Gamma1}
\L_1^2 \Gamma_1 =   m \gR_1 \cdot \ptheta \left(  \gR_1 \times  \grad \times \U \right) d \theta,
\end{equation}
\\
while

\begin{equation} \label{eq:L2Gamma1}
\L_2 \Gamma_1 = - m \b \cdot \gR_2  d v_\parallel.
\end{equation}
\\
By explicit evaluation of the different terms, and introducing $h^{\mu}$ given in \cref{eq:hmu}, we derive 

\begin{align}
\Gamma_{3 \theta} =- \frac{ \mu B}{\Omega^2 } \b \cdot \grad \times \U + \frac{2}{3} \frac{m \mu}{q} \gR_1 \cdot \grad \ln B  - \frac{B}{\Omega} \left( \gyaver{h^\mu} + \widetilde{h^\mu}\right)  + \ptheta s_3.
\end{align}
\\
Setting $\Gamma_{3 \theta} =0$ yields

\begin{align} \label{eq:closedhmu}
\gyaver{h^\mu} &=- \frac{\mu}{\Omega} \b \cdot \grad \times \U, 
\end{align}
\\
and

\begin{align}
s_3 & = \frac{B}{\Omega} \int d  \theta' \left[ \widetilde{h^\mu} - \frac{2}{3} \mu \gR_1 \cdot \grad \ln B\right],
\end{align}
\\
where we used the identity $\gyaver{\gR_1 \cdot \grad \U \cdot \cperp} = \mu \b \cdot \grad \times \U / 2$. An closed expression for $h^\theta$ can be derived from $\Gamma_{3 \mu} =0$, i.e.

\begin{equation}
h^\theta =-  \frac{\Omega}{B} \pmu s_3,
\end{equation}
\\
while from $\Gamma_{3 \parallel}$ one obtains

\begin{equation}
\gR_2 \cdot \b = - \frac{1}{m } \frac{\partial}{\partial v_\parallel} s_3.
\end{equation}
\\
\corr{Thus, with \cref{eq:closedhmu}, the second order accurate guiding-center one-form, $\Gamma = \Gamma_0 + \Gamma_1 + \Gamma_2$, is given by

\begin{align} \label{eq:GammaGC}
\Gamma &=  \left[q \A^* - \frac{B\mu}{\Omega}\bm T\right] \cdot d \R + \frac{B \mu}{\Omega} d \theta \nonumber \\
& - \left[\frac{\mu B}{\Omega} S + q \phi + \frac{m}{2}v_{\parallel }^2 + \frac{m}{2} u_E^2
  +    \mu B  + \frac{\mu B}{2 \Omega} \b \cdot \grad \times \U \right] dt,
\end{align}
\\
with $\A^* = \A + m \U/q$. We remark that the Ba\~nos term, proportional to $\mu v_\parallel B \b \cdot \grad \times \b / 2 \Omega$ contained in the last term of \cref{eq:GammaGC}, can be removed from the Hamiltonian component with the choice $\gyaver{h^\parallel} =0$. Indeed, one finds

\begin{align} \label{eq:GammaBrizard}
\Gamma & =  \left[q \A^* - \frac{B\mu}{\Omega}\bm T^*  \right] \cdot d \R  + \frac{B \mu}{\Omega} d \theta \nonumber \\
& - \left[\frac{\mu B}{\Omega}S + q \phi + \frac{m}{2}v_{\parallel }^2 + \frac{m}{2} u_E^2
  +    \mu B  + \frac{\mu B}{2 \Omega} \b \cdot \grad \times \u_E \right] dt,
\end{align}
\\
where $\bm T^* = \bm T + \b \b \cdot \grad \times \b / 2$ \citep{Brizard1995}.}
\section{Gyrocenter Transformation}
\label{appendixGY}
We describe the second order gyrocenter transformation in $\epsilon_\delta$ whose results are presented in \cref{GyrocenterTransformation}. We start to derive $\overline{\Gamma}_1$ expressed in the gyrocenter coordinates $\overline{\Z}$. Within the choice of a Hamiltonian formulation, the symplectic components of $\overline{\Gamma}_1$ vanish. Therefore, from \cref{eq:systemorderbyorder1}, we obtain that 

\begin{subequations}
\begin{align}
q \gbR_1 \times \B^*- m \gbparallel_1 \b + q \A_1 + \gygrad S_1 & =0, \label{eq:equationG1R} \\
m \gbR_1 \cdot \b  + \frac{\partial S_1}{\partial \gyvparallel} & = 0 , \label{eq:equationG1parallel} \\
q \A_1   \cdot   \frac{\partial \bm \rho}{\partial \gytheta} - \frac{m}{q } \gbmu_1 + \frac{\partial S_1}{\partial \gytheta} & =0, \label{eq:equationG1mu}\\
  q \A_1 \cdot \frac{\partial \bm \rho}{\partial \gymu} + \frac{m  }{q } \gbtheta_1   + \frac{\partial S_1}{\partial \gymu}  & =0,\label{eq:equationG1theta}
\end{align}
\end{subequations}
\\
The first order gyrocenter correction, $\overline{\Gamma}_1 = \overline{\Gamma}_0 - \L_1 \delta \Gamma$, reduces to

\begin{equation} \label{eq:overlineGamma1tfull}
\begin{aligned}
  \overline{\Gamma}_{1} & = \left[ - q \phi_1 + \gbmu_1 \gypmu  \overline{\mathcal{H}}_0 + \gbparallel_1 \gypvparallel \overline{\mathcal{H}}_0 +  \gbR_1  \cdot \left( \gygrad \: \overline{\mathcal{H}}_0 + \frac{\partial }{\partial t} \overline{\bm A}^* \right) + \frac{\partial S_1 }{\partial t}\right] d t.
\end{aligned}
\end{equation}
\\
 The closed analytical expressions of the first order generating functions $\gbparallel_1  $ and $\gbR_1$ can be obtained by taking the scalar and vector product of \cref{eq:equationG1R} with $\B^*$ and $\b$, respectively, while the one for $\gbmu_1$ and $\gbtheta_1$ follows \cref{eq:equationG1mu,eq:equationG1theta}. Thus, we obtain

\begin{subequations}
\label{eq:G1}
\begin{align}
\gbR_1 &= -\frac{1}{q B^*_{\parallel}} \b \times \left( q
  \A_1 +  \gygrad S_1 \right) - \frac{\B^*}{m B^*_{\parallel}} \frac{\partial S_1}{\partial \gyvparallel}, \\
\gbparallel_1  &= \frac{\B^*}{m B^*_{\parallel}} \cdot \left(   q \A_1 + \gygrad S_1 \right), \\
\gbmu_1 &=  \frac{q}{m } \left(q  \A_1 \cdot \frac{\partial \bm \rho}{\partial \gytheta}   +\frac{\partial S_1}{\partial \gytheta} \right).\\
\gbtheta_1  &= - \frac{q}{m } \left( q \A_1 \cdot \frac{\partial \bm \rho}{\partial \gymu}  +\frac{\partial S_1}{\partial \gymu} \right).
\end{align}
\end{subequations}
\\
With $\A_1 = A_{\parallel 1} \hb$, \cref{eq:G1} reduces to \cref{eq:GYG1}. We remark that the contribution of $\A_{1}$ can be transferred to the symplectic components of $\overline{\Gamma}_1$ via the transformation $\bm A_{1} \to \widetilde{\bm A_{1}}$ in $\gbR_1$ and in $\gbparallel_1$ in \cref{eq:G1}, yielding $\overline{\Gamma}_{1\gyR} = q \gyaver{\bm A_{1}}_{\gyR}$ \citep{Brizard2007,Madsen2013,Tronko2016}.
\corr{Isolating the gyrophase dependent terms in the electromagnetic fluctuations by using \cref{eq:gyaverandwidetilde}, approximating $\partial_{\gyvparallel} \overline{\mathcal{H}}_0 \simeq m \gyvparallel + O(\epsilon^2)$, $\partial_{\gymu}  \overline{\mathcal{H}}_0 \simeq B +O(\epsilon^2)$ and $B_{\parallel}^* \simeq B + O(\epsilon^2)$ and, finally, neglecting the time derivative of the effective magnetic potential, being $\partial_t \A^* \sim \epsilon^2 \Omega \A^*$, the first order gauge function $S_1$ satisfies,

 \begin{align} \label{eq:pthetaS1}
 \Omega \frac{\partial S_1}{\partial \gytheta} +  \frac{d_{gc}}{dt} S_1 - \frac{1}{m} \frac{\partial S_1}{\partial \gyvparallel} \gygrad_\parallel \;\overline{\mathcal{H}}_0 \simeq q \left( \widetilde{ \phi_1 } - \widetilde{ \A_1 \cdot \c_{\perp}} -  \left( \gyvparallel \b + \bm D_\perp \right) \cdot  \widetilde{ \A_1 } \right) +O(\epsilon^2),
 \end{align}
 \\
 where we introduce the guiding-center convective derivative $d_{tgc} \equiv \partial_t + \left( \gyvparallel \b  + \bm D_\perp \right) \cdot \gygrad $ with $ \bm D_\perp = \b \times \gygrad \; \overline{\mathcal{H}}_0 /(q B)$. We remark that $\bm D_\perp$ reduces to $\u_E$ at the leading order. Ordering the terms on the left-hand side of \cref{eq:pthetaS1} according to \cref{Singleparticledynamicsinthetokamakperiphery}, i.e. $\Omega \partial_{\gytheta} S_1 \sim \Omega S_1$, $d_{tgc} S_1 \sim \epsilon \Omega S_1$ while $\partial_{\gyvparallel}S_1 \gygrad_\parallel \overline{\mathcal{H}}_0/m \sim \epsilon^2 \Omega S_1$, shows that a solution for $S_1$ can be obtained iteratively by expanding $S_1 = S_{10} +  S_{11} + \dots$ with $S_{11} \sim O(\epsilon S_{10})$. We remark that an analytical solution of the first order gauge function, $S_1$, is an important element for the second order analysis. With $\A_{1} = A_{\parallel 1} \b$, the leading order solution, $S_{10}$, satisfies $\Omega \partial_{\gytheta} S_{10} = q \widetilde{\Phi_1}$ where we introduce the first order gyrokinetic potential, 
 
 \begin{equation}
 \Phi_1 = \phi_1 - \gyvparallel A_{\parallel 1},
 \end{equation}
 \\
 being $\phi_1 \sim \gyvparallel A_{\parallel 1}$, yielding
 
\begin{equation} \label{eq:S1integral}
     S_{10}  = \frac{q}{\Omega} \int d \gytheta' \left[ \Phi_1 - \gyaver{\Phi_1}_{\gyR}\right] \equiv \frac{q}{\Omega} \overline {\widetilde{\Phi_1}}.
 \end{equation}
 \\
Then, the first order correction, $S_{11}$, is given by $\Omega \partial_{\gytheta} S_{11} = - d_{tgc}S_{10}$, such that $S_{11} = - d_{tgc}\overline{ S_{10}}/\Omega$.} \corr{A closed analytical expression for $S_{10}$ can be derived, using \cref{eq:S1integral}, performing the $\gytheta$-integral explicitly in Fourier-space. Indeed, expanding the fluctuation $\widetilde{\Phi_1}$ in Fourier harmonics, i.e. $\widetilde{\Phi_1}(\bm k)  = \sum_{n \neq 0} i^n J_n(b) \Phi_1(\bm k) e^{i n \gytheta}$ (where $\Phi_1(\bm k)$ is the Fourier harmonics of $\Phi_1(\gyR)$). Here, $J_n(x)$ is the $n$th-order Bessel function [see \cref{HermiteLaguerreRepresentationOfGyroaverageOperator}]. Then, inserting the latter expression for $\widetilde{\Phi_1}$ yields

\begin{equation} \label{eq:S10closed}
S_{10} = \sum_{\bm k} \sum_{n \neq 0 } \frac{i^{n}}{in} J_n\left( b\right) e^{i n \gytheta} \Phi_1(\bm k) e^{i \bm k \cdot \gyR}.
\end{equation}
}
\\
With \cref{eq:pthetaS1}, the first order gyrocenter one-form correction $\overline{\Gamma}_1$ in \cref{eq:overlineGamma1tfull} reduces to
 
 \begin{equation} \label{eq:overlineGamma1tfinal}
 \overline{\Gamma}_{1} = - q \gyaver{\Phi_1}_{\gyR} dt \equiv  - \overline{\mathcal{H}}_1 dt.
 \end{equation}

 We now perform the second order analysis in $\epsilon_\delta$. From \cref{eq:systemorderbyorder2}, the second order gyrocenter one-form correction, $\overline{\Gamma}_2$, reads

\begin{align} \label{eq:overlineGamma2full}
\overline{\Gamma}_2 & = \left[  q \gbR_2 \times \B^* + \frac{q}{2} \gbR_1 \times (\gygrad \times \A_1 ) - \frac{q}{2}  \gbmu_1 \frac{\partial }{\partial \gymu} \A_1 -  \frac{q}{2}  \gbtheta_1 \frac{\partial }{\partial \gytheta}\A_1
 - m \gbparallel_2 \b +
\gygrad S_2 \right] \cdot d \gyR  \nonumber \\
& + \left[ m\gbR_2 \cdot \b + \frac{\partial S_2}{\partial \gyvparallel} 
 \right] d \gyvparallel + \left[  \frac{q}{2} \gbR_1  \cdot   \frac{\partial }{\partial \gytheta}  \A_1 - \frac{q}{2 \Omega}  \gbR_1 \cdot \gygrad   (\A_1 \cdot \cperp) \right. \nonumber \\
 & \left.   - \frac{q}{2 \Omega}  \gbmu_1     \gypmu (\A_1 \cdot \cperp) -  \frac{q}{2 \Omega}  \gbtheta_1 \ptheta (\A_1 \cdot \cperp)  +  \frac{q}{4   \gymu} \gbmu_1   \gyptheta (\A_1 \cdot \bm \rho)
 -  \frac{m}{q } \gbmu_2  + \frac{\partial S_2}{\partial \gytheta} \right] d \gytheta \nonumber  \\
& + \left[  \frac{q}{2} \gbR_1 \cdot \frac{\partial }{\partial \gymu} \A_1  - \frac{q}{2} \gbR_1 \cdot \gygrad  ( \A_1 \cdot \bm \rho )  - \frac{q}{2} \gbmu_1 \gypmu  ( \A_1 \cdot \bm \rho )\right. \nonumber \\
& \left. - \frac{q}{2} \gbtheta_1  \gyptheta ( \A_1 \cdot \bm \rho )   +  \frac{q}{2 \Omega} \gbtheta_1  \gypmu (\A_1 \cdot \cperp)  +  \frac{m  }{q } \gbtheta_2  + \frac{\partial S_2}{\partial \gymu} \right] d \gymu - \overline{\mathcal{H}}_{2 } d t.
\end{align}
\\
with 

\begin{align} \label{eq:H2}
    - \overline{\mathcal{H}}_{2 }&  =   \frac{q}{2} \gbR_1\cdot \gygrad \phi_1 + \frac{q}{2} \gbmu_1 \frac{\partial }{\partial \gymu} \phi_1   + \frac{q}{2} \gbtheta_1 \frac{\partial }{\partial \gytheta} \phi_1 + \frac{q}{2} \gbR_1 \cdot \gygrad  \left<  \Phi_1 \right>_{\gyR}  + \frac{q}{2} \gbmu_1 \gypmu  \left<  \Phi_1 \right>_{\gyR}  \nonumber \\
    & + \frac{q}{2} \gbparallel_1 \gypvparallel \gyaver{\Phi_1}_{\gyR}+      \gbmu_2  \gypmu \overline{\mathcal{H}}_0 +  \gbparallel_2 \gypvparallel \overline{\mathcal{H}}_0 + \frac{q}{2} \gbR_1 \cdot \frac{\partial }{\partial t} \A_1  \nonumber \\
 & + \frac{q}{2}\gbmu_1 \frac{\partial}{\partial t}\left( \A_1 \cdot \frac{\partial \bm \rho}{\partial \gymu} \right) + \frac{q}{2} \gbtheta_1 \frac{\partial}{\partial t}\left( \A_1 \cdot \frac{\partial \bm \rho}{\partial \gytheta}  \right) +  \gbR_2 \cdot \left( \gygrad  \:\mathcal{\overline{H}}_0  + \frac{\partial}{\partial t} \overline{\bm A^*}\right) + \frac{\partial S_2}{\partial t}.
\end{align}
\\
Applying the gyrocenter transformation rules of our Hamiltonian formulation, we find the second order generating functions,

\begin{subequations} \label{eq:GY2}
\begin{align} 
\gbR_2& = \frac{1}{q} \frac{\b}{B^*_{\parallel}} \times \left(  \frac{q}{2} \gbmu_1 \gypmu \A_1 + \frac{q}{2} \gyptheta \A_1 - \frac{q}{2} \gbR_1 \times ( \gygrad \times \A_1 ) -  \gygrad  S_2\right) \nonumber  \\ 
& - \frac{\B^*}{m B^*_{\parallel}} \frac{\partial S_2}{\partial \gyvparallel}  , \label{eq:GYG2R}\\
\gbparallel_2& =  \frac{ \B^*}{m B^*_{\parallel}} \cdot\left( \frac{q}{2 } \gbR_1 \times \left( \gygrad \times \A_1 \right) - \frac{q}{2} \gbmu_1 \gypmu \A_1 - \frac{q}{2} \gbtheta_1 \gyptheta \A_1 + \gygrad S_2  \right), \label{eq:GYG2parallel}\\
\gbmu_2& = \frac{q}{m  } \left(    \frac{q}{2} \gbR_1 \cdot \gyptheta \A_1  - \frac{q}{2 \Omega}  \gbR_1 \cdot \gygrad  (\A_1 \cdot \cperp) -  \frac{q}{2 \Omega} \gbmu_1 \gypmu  (\A_1 \cdot \cperp)   \right. \nonumber \label{eq:GYG2mu}\\
& \left.  - \frac{q}{2 \Omega} \gbtheta_1 \gyptheta   (\A_1 \cdot \cperp)  +  \frac{q}{4   \gymu} \gbmu_1 \gyptheta (\A_1 \cdot \bm \rho)    +  \frac{\partial  S_2}{\partial \gytheta} \right),  \\
\gbtheta_2& =   \frac{q}{m  }\left(   \frac{q}{2} \gbR_1 \cdot \gygrad ( \A_1 \cdot \bm \rho ) + \frac{q}{2} \gbmu_1 \gypmu ( \A_1 \cdot \bm \rho )  + \frac{q}{2} \gbtheta_1 \gyptheta  (\A_1 \cdot \bm \rho) \right.  \nonumber \\ 
&\left.  -  \frac{q}{2} \gbR_1 \cdot \gypmu \A_1  -  \frac{q}{2 \Omega} \gbtheta_1 \gypmu (\A_1 \cdot \cperp)  - \gypmu S_2 \right). \label{eq:GYG2theta}
\end{align}
\end{subequations}
\\
\Cref{eq:GY2} reduces to \cref{eq:GYG2} at the leading-order with $\A_1  =  A_{\parallel 1} \hb$. \corr{Using \cref{eq:GY2} and the closed expressions for the first order generating functions, given in \cref{eq:G1} with $\A_1 = A_{\parallel 1} \hb$ and within the same approximations made in the first order analysis, $- \overline{\mathcal{H}}_2$ reduces to 

\begin{align} \label{eq:fullgyH2}
\overline{\mathcal{H}}_2 & =  \frac{1}{2 B} \b \times \gygrad S_1 \cdot \gygrad \Phi_1  + \frac{q}{2 m } \frac{\partial S_1}{\partial \gyvparallel} \gygrad_\parallel \Phi_1+ \frac{q}{2 m} \frac{\partial S_1}{\partial \gyvparallel} \left( \frac{d_{gc}}{dt} A_{\parallel 1}  + \Omega \gyptheta A_{\parallel 1}\right) \nonumber \\ 
& - \frac{q^2}{2 m }\left(  \frac{\partial S_1}{\partial \gytheta}\frac{\partial \Phi_1}{\partial \gymu} - \frac{\partial S_1}{\partial \gymu} \frac{\partial \Phi_1 }{\partial \gytheta}\right) - \frac{1}{2 B} \b \times \gygrad S_1 \cdot \gygrad \gyaver{\Phi_1}_{\gyR} \nonumber \\
&  + \frac{q}{2m} \frac{\partial S_1}{\partial \gyvparallel}  \gygrad_\parallel \gyaver{\Phi_1}_{\gyR}  - \frac{q^2}{2 m} \frac{\partial S_1}{\partial \gytheta} \gypmu \gyaver{\Phi_1}_{\gyR} + \frac{q^2}{2 m} A_{\parallel 1} \gyaver{A_{\parallel 1}}_{\gyR}\nonumber \\
&+ \frac{q}{2m}  \gygrad_\parallel S_1 \gyaver{A_{\parallel 1}}_{\gyR} - \Omega \frac{\partial S_2}{\partial \gytheta} \frac{d_{gc} S_2}{dt}  + \frac{1}{m} \frac{\partial S_2}{\partial \gyvparallel}  \gygrad_\parallel \;\overline{\mathcal{H}}_0.
\end{align}
\\
Noticing that $\gyaver{S_1}_{\gyR} =0$ and isolating the gyrophase dependent terms, we obtain the expression given in \cref{eq:fullH2}. In particular, we cancel the gyrophase dependent terms by choosing $S_2$ such that it satisfies

\begin{align} \label{eq:S2}
\Omega \frac{\partial S_2}{\partial \gytheta} + \frac{d_{gc}}{dt}S_2- \frac{1}{m} \frac{\partial S_2}{\partial \gyvparallel}  \gygrad_\parallel \;\overline{\mathcal{H}}_0 &=  \frac{1}{2B} \b \times \gygrad S_1 \cdot \gygrad\gyaver{ \Phi_1}_{\gyR} - \frac{q^2}{2m} \frac{\partial S_1}{\partial \gytheta} \frac{\partial }{\partial \gymu} \gyaver{\Phi_1}_{\gyR} \nonumber \\
& + \frac{q}{2 m} \frac{\partial S_1}{\partial \gyvparallel}  \left(   \gygrad_\parallel \gyaver{\Phi_1}_{\gyR} + \frac{d_{gc}}{dt} \gyaver{A_{\parallel 1}}_{\gyR} \right) \nonumber \\
&  + \frac{q^2}{2m} \widetilde{A_{\parallel 1}}\gyaver{A_{\parallel 1}} + \frac{q}{2 m } \gygrad_\parallel S_1 \gyaver{A_{\parallel 1}}.
\end{align}
\\
We remark that we do not solve explicitly $S_2$ from \cref{eq:overlineGamma2}. Approximating with $S_{1} \simeq S_{10}$ and neglecting terms proportional to $d_{tgc} \sim \epsilon \Omega$ and parallel gradients in \cref{eq:fullH2}, we derive the leading order second order gyrocenter correction one-form $\overline{\Gamma}_2$,

\begin{align} \label{eq:overlineGamma2}
\overline{\Gamma}_2    &=    \left[  \frac{q^3}{2  m \Omega } \pmu \left( \gyaver{\Phi_1^2}_{\gyR} - \gyaver{\Phi_1}^2_{\gyR} \right)    - \frac{q^2}{2 m} \gyaver{A_{\parallel 1}^2}_{\gyR} - \frac{q} {2\Omega m} \gyaver{ \left(\b \times \gygrad S_{10} \right) \cdot \gygrad \widetilde{\Phi_1}}_{\gyR} \right] d t \nonumber \\
&  = - \gyaver{\overline{\mathcal{H}}_{2 }}_{\gyR}  d t.
\end{align}
\\
}
\corrs{
\section{Hermite-Laguerre Expansion of $\gyaver{(\b \times \gygrad S_1) \cdot \gygrad \widetilde{ \Phi_1} }_{\gyR}$}
\label{appendixHLnonlinearpot}

This appendix reports the leading order Hermite-Laguerre projection of the last term in  \cref{eq:momentstarlkgygradPsi1}. Using the expression of $\Phi_1$ given in \cref{eq:Phi1} with $S_{1}\simeq S_{10}$ in \cref{eq:S1integral}, the considered term can be written as 

\begin{align}\label{eq:momentlkS1Phi}
& \moment{lk}{\gygrad \gyaver{ \left(\b \times \gygrad \overline{\widetilde{\Phi_1}}\right) \cdot \gygrad \widetilde{\Phi_1}}_{\gyR}}  =\sum_{p,j}\left[ \delta_p^l \delta_j^k \moment{pj}{\gygrad \gyaver{ \left(\b \times \gygrad \overline{\widetilde{\phi_1}}\right) \cdot \gygrad \widetilde{\phi_1}}_{\gyR}}\right. \nonumber \\
& \left. - \phaseV_{apj}^{lk}  \moment{pj}{\gygrad \gyaver{ \left(\b \times \gygrad \overline{\widetilde{A_{\parallel 1}}}\right) \cdot \gygrad \widetilde{\phi_1}}_{\gyR}}  - \phaseV_{apj}^{lk} \moment{pj}{\gygrad \gyaver{ \left(\b \times \gygrad \overline{\widetilde{\phi_1}}\right) \cdot \gygrad \widetilde{A_{\parallel 1 }}}_{\gyR}}\right. \nonumber \\
& \left. + \phaseV_{apj}^{2lk} \moment{pj}{\gygrad \gyaver{ \left(\b \times \gygrad \overline{\widetilde{A_{\parallel 1}}}\right) \cdot \gygrad \widetilde{A_{\parallel 1 }}}_{\gyR}} \right].
\end{align}
\\
In the evaluation of \cref{eq:momentlkS1Phi}, we pursue a procedure similar to the one applied in \cref{HermiteLaguerreRepresentationOfGyroaverageOperator} to evaluate the nonlinear terms. In particular, we derive the Hermite-projection of the first term in \cref{eq:momentlkS1Phi} which can then be used to evaluate the remaining terms. Using the closed analytical expression of $S_{10}$ given in \cref{eq:S10closed}, expanding $\widetilde{\phi_1}$ in Fourier harmonics, applying the gradient operator and, finally, the gyroaverage operator yields

 \begin{align} \label{eq:gyaverS1phi1}
 \gyaver{\left(\b \times \gygrad \overline{\widetilde{\phi_1}}\right) \cdot \gygrad \widetilde{\phi_1}}_{\gyR} = \sum_{\bm k,\bm k'} \sum_{n\neq 0} \frac{(-1)^{n}e^{in \alpha}}{i n}  J_{n}(b)J_{n}(b')(\bm k \times \b)\cdot \bm k' \phi_1(\bm k)\phi_1(\bm k') e^{i \bm K \cdot \gyR},
\end{align}
\\
where we neglected the terms proportional to $|\gygrad J_n(b)| \sim \lvert \gygrad_\perp B/B \rvert \lvert J_n'(b)\rvert$ in \cref{eq:gyaverS1phi1} since they are smaller by a least a factor $\epsilon$. Expressing the Bessel functions in terms of associated Laguerre polynomials using \cref{eq:Jn2Laguerre}, and applying consecutively the two following identities

\begin{equation}
L_r^{n}(\sperp^2) = \sum_{r_1=0}^r \overline{L}_{rr_1}^n L_{r_1}(\sperp^2),
\end{equation}
\\
where 

\begin{equation}
\overline{L}_{rr_1}^n =\binom{n + r - r_1 -1 }{r - r_1 },
\end{equation}
\\
and 

\begin{equation}
L^m_r(\sperp^2) L_j(\sperp^2) \sperp^{2m} = \sum_{s=0}^{m+r+j} d_{rjs}^m L_s(\sperp^2),
\end{equation}
\\
where 

\begin{equation}
d_{rjs}^m  = \sum_{r_1 = 0}^r \sum_{j_1 =0}^j   \sum_{s_1=0}^s L_{jj_1}^{-1/2}  L_{rr_1}^{m-1/2}  L_{s s_1}^{-1/2} (r_1 + j_1 + s_1 + m)!,
\end{equation}
\\
we derive that the Hermite-Laguerre projection of the last term appearing in \cref{eq:momentstarlkgygradPsi1}, i.e.

\begin{align} \label{eq:momentS1phitildepj}
\moment{pj}{\gygrad \gyaver{ \left(\b \times \gygrad \overline{\widetilde{\phi_1}}\right) \cdot \gygrad \widetilde{\phi_1}}_{\gyR}}& = \sum_{\bm k,\bm k'} \sum_{n \neq 0} \sum_{r,s \geq 0}\sum_{r_1 =0}^r \sum_{s_1=0}^{|n| + r_1 + s} \frac{\overline{K}^n_{rsr_1s_1} }{in} \left( \frac{b_a}{2}\right)^{|n|} \left( \frac{b_a'}{2}\right)^{|n|}  \nonumber \\
& \times \kernel{r}(b_a) \kernel{s}(b_a') \bm{\overline{\mathcal{D}}}_{anrs}^{pjs_1}(b_a,b_a',\bm K)   (\bm k \times \b) \cdot \bm k' \nonumber \\
&
\times \phi_{1}(\bm k) \phi_1(\bm k') e^{i \bm K \cdot \gyR}
\end{align}
\\
with $\overline{K}^n_{rsr_1s_1} = (-1)^n e^{in \alpha}\overline{L}_{rr_1}^{|n|} d_{sr_1s_1}^{|n|}  r! s!/[(|n| +r)!(|n|+s)!]$, and where the operator $\bm{\overline{\mathcal{D}}}_{anrs}^{pjs_1}(b_a,b_a',\bm K) $ is defined in \cref{eq:Danrslkg}. \Cref{eq:momentS1phitildepj} can be generalized to evaluate the remaining terms in \cref{eq:momentlkS1Phi}.}

\bibliographystyle{jpp}
\bibliography{biblio}

\end{document}